\documentclass[12pt]{iopart}
\usepackage{iopams}
\usepackage{graphicx}
\usepackage{lscape}

\usepackage{yfonts}
\usepackage{hyperref}

\usepackage{natbib}
\newcommand\newblock{\hskip .11em\@plus.33em\@minus.07em}

\newcommand{\be}{\begin{equation}}
\newcommand{\ee}{\end{equation}}
\newcommand{\bea}{\begin{eqnarray}}
\newcommand{\eea}{\end{eqnarray}}
\newcommand{\nn}{\nonumber}
\newcommand{\pp}{\varphi}
\newcommand{\dd}{\mathrm{d}}

\usepackage{xspace}
\usepackage{url}
\newcommand{\gy}{\textsc{Gyoto}\xspace}

\hypersetup{pdftitle={GYOTO: a polarized relativistic ray-tracing code},
  pdfkeywords={LaTeX},
  colorlinks=true
}

\begin{document}

\title{GYOTO 2.0: a polarized relativistic ray-tracing code}

\author{N Aimar$^{1}$\footnote{The three first authors in the list contributed equally to the article.},
  T Paumard$^{1}$\footnotemark[\value{footnote}],
  F H Vincent$^{1}$\footnotemark[\value{footnote}],
  E Gourgoulhon$^{2,3}$, and
  G Perrin$^{1}$}
  %}
\address{$^{1}$ LESIA, Observatoire de Paris, CNRS,  Universit\'e Pierre et Marie Curie, Universit\'e  Paris Diderot, 5 place Jules Janssen, 92190 Meudon, France\\
$^{2}$ Laboratoire Univers et Théories, Observatoire de Paris, CNRS, Université PSL, Université Paris Cité, 5 place Jules Janssen, 92190 Meudon, France\\
$^{3}$ Laboratoire de Mathématiques de Bretagne Atlantique, CNRS, Université de Bretagne Occidentale,
6 avenue Le Gorgeu, 29238 Brest, France
}
\ead{nicolas.aimar@obspm.fr}

\begin{abstract}
  Polarized general-relativistic radiative transfer in the vicinity of black holes and other compact objects has become a crucial tool for probing the properties of relativistic astrophysics plasmas. Instruments like GRAVITY, the Event Horizon telescope, ALMA, or IXPE make it very timely to develop such numerical frameworks. In this article, we present the polarized extension of the public ray-tracing code \gy, and offer a python notebook allowing to easily perform a first realistic computation. The code is very modular and allows to conveniently add extensions for the specific needs of the user. It is agnostic about the spacetime and can be used for arbitrary compact objects. We demonstrate the validity of the code by providing tests, and show in particular a perfect agreement with the \textsc{ipole} code. Our article also aims at pedagogically introducing all the relevant formalism in a self-contained manner.
\end{abstract}

\section{Introduction}

Generating synthetic images of black hole environments is tricky because of relativistic effects such as aberration, Doppler beaming, gravitational redshift and light bending. General-Relativistic Radiative Transfer (GRRT) computation through ray-tracing method, i.e.~integration of photons trajectories (null geodesics, assuming no impact other than gravitation on the trajectory), allows to naturally account for all relativistic effects to generate synthetic observables of such environments that can be compared to observational data.
% This technic is used to model accretion flows and derive physical parameters of the flow as properties of the black hole (like mass or spin), or as post-processing of general relativistic MHD (GRMHD) simulations of black hole accretion to study their variability.
This technique is key to obtain meaningful constraints on the parameters
describing the emitting accretion flow or the spacetime geometry.
It can be used either on analytically described accretion flows,
or in post-processing of general relativistic magnetohydrodynamical
(GRMHD) simulations.
For example, the Numerical Observatory of Violent Accreting system~\citep[\textsc{NOVAs};][]{varniere18,NOVAs} combines GRMHD simulations of accretion flows around black holes as computed by the code GR-AMRVAC%\footnote{Freely available at \url{https://github.com/amrvac/amrvac}}
~\citep{casse17}, with the ray-tracing code
\gy~\citep{vincent11} to produce various synthetic observables.
%light curves, spectra.

The recent development of instruments now allows to measure the polarization of light coming from the extremely close environment of black holes. The Event Horizon Telescope (EHT) released the millimetric polarized image of M87*~\citep{EHT2022}. GRAVITY~\citep{gravity18,Gravity2023} in the near infrared, and the Atacama Large Millimeter/submillimeter Array~\citep[ALMA,][]{Wielgus2022} have detected a polarized signature of the radiation flares associated with the supermassive black hole at the center of the Galaxy, Sagittarius~A* (Sgr~A*).
% The dominant emission process of these sources at these observing frequencies is synchrotron radiation which is expected to be highly polarized~\citep{RL79}.
Moreover, the Imaging X-ray Polarimetry Explorer (IXPE) has obtained important
constraints on the geometry of the accretion flow of an X-ray binary by
measuring its polarized radiation~\citep{krawczynski22}.

Thus, polarized ray-tracing codes are of particular interest to generate synthetic observables from accretion models, be they analytic~\citep[see e.g.][]{Broderick2016,Gralla2018, Gralla2019, Vincent2019, Gralla2020, Gravity2020, Nalewajko2020, Dovciak2022, Vos2022, Aimar2023, CardenasAvendano2023}, or numeric~\citep[see e.g.][]{Chan2015, Moscibrodzka2016, Chael2018, Davelaar2018, Chael2019, EHT2019, Anantua2020, Dexter2020, Porth2021}.
%that should be even more constrained by polarized observables.
In particular, polarization signatures might allow probing the nature of spacetime close to the event horizon of black hole candidates, making GRRT a crucial tool for constraining general relativity (GR) in the strong-field regime~\citep{himwich20,Jimenez2021, vincent23}.

%Dovciak?

The need for ray tracing when computing images in GR led to the development of multiple codes. Many of them were developed for unpolarized light~\citep{Noble2007, Dexter2009, Dauser2010, vincent11, Pu2016, Chan2018, Bronzwaer2018, Younsi2020}. Some codes also keep track of the electric vector position angle and polarization degree, assuming
a geometrically thin equatorial accretion flow~\citep{Dovciak2008,Gelles2021,AART},
the latter two being specialized for highly-lensed features by implementing
adaptive ray tracing.
Only a handful of codes are able of treating the most demanding problem
of integrating
the full polarized radiative transfer: \textsc{grtrans}
%\footnote{\url{https://github.com/jadexter/grtrans}}
~\citep{Dexter2016},
\textsc{ipole}
%\footnote{\url{https://github.com/moscibrodzka/ipole}}
~\citep{Moscibrodzka2018},
\textsc{Arcmancer}
~\citep{Pihajoki18},
%\textsc{Odyssey}
%\footnote{\url{https://github.com/hungyipu/Odyssey}}
%~\citep{Pu2016},
\textsc{Bhoss}~\citep{Younsi2020}, \textsc{Raptor}
%\footnote{\url{https://github.com/jordydavelaar/raptor}}
~\citep{Bronzwaer2020},
\textsc{Blacklight}
~\citep{White2022},
as well as \textsc{Lemon}~\citep{Yang21} which specializes on polarized
radiative transfer with scattering.
Some of these polarized GRRT codes were recently compared by~\cite{Prather2023}.
% The first two, \textsc{grtrans} and \textsc{Odyssey}, are limited to Kerr spacetime due to its analytical solution for integrating the polarization along null geodesics while the other are independent of the spacetime metric (more details of the comparison of these codes in~\cite{Prather2023}).
% \textsc{BHOSS} and \textsc{RAPTOR} use Runge-Kutta-Fehlberg method, of fourth-order for the first one and adaptive for the second, to integrate the null geodesics.

\gy~\citep{vincent11} is %, as the codes mentioned above,
a backwards ray-tracing code (i.e.~integrating from the observer to the source), operating in any given (analytically or numerically computed) metric, and solving the radiative transfer equation.
It can also integrate timelike geodesics for computing e.g.~stellar orbits~\citep{grould17}. % equations for massive objects.
The code is publicly available
\footnote{\url{https://github.com/gyoto/Gyoto/blob/master/INSTALL.Gyoto.md}}
, built to be modular so that extensions are easy to integrate and user-friendly with XML and python interfaces.

%Explain why polarization in curved spacetime matters. Focus on
%SgrA*/flares and M87, GRAVITY and EHT as our main scientific interest.
%Read in particular the papers of Jimenez, Dexter+ on polarization
%study of flares and BH images and ph rings.
%Note that there is also interest for Xrays with IXPE, eXTP. So there is
%clearly a very fertile observational ground in which the tool
%that we provide will be useful.

%Review of past results. Code of Czech colleagues, Jason, Monika, Zach,
%certainly others (Eckart group?). Try to cover everything that was done and
%highlight the specificities of each. Note that
%there are not so many codes doing full GR polarized radiative transfer.
%I guess only Jason and Monika (TBC).

%It would probably be interesting to also review the past results
%regarding the formalism of GR polarization. Eg the work of
%Shcherbakov+, Gammie+, ...

%Note that in particular in the context of Gravity and EHT, the question
%of testing the nature of spacetime is at the center of the attention.
%Insist that GR polarized raytracing might be a crucial tool in this quest.

The goals of this paper are the following: (i) providing the new
version of \gy, with full polarized GRRT included,
publicly available at the same address as the older version,
%open-source,
%GR polarized
%radiative transfer code,
%agnostic about spacetime,
%user-friendly
with the addition of a python snippet
\footnote{\url{https://github.com/gyoto/Gyoto/blob/master/doc/examples/Gyoto_Polar_example.ipynb}}
to allow the interested
reader to immediately be able to compute a non-trivial setup;
(ii) providing a pedagogical, in-depth presentation of the formalism
of GR polarization,
%with a lot of room dedicated to
%by discussing all the relevant equations and assumptions in detail,
%re-deriving
%everything from first principles
as well as a detailed description of the technical
%dirty
implementation.
%in section~\ref{sec:formalism}.
In the following discussion, we will focus on polarized synchrotron radiation,
as it is the dominant emission mechanism for our sources of primary scientific
interest (Sgr~A* and M87).
But the code is able to compute polarized observables for other emission mechanism as soon as the electric field is provided in the model.
Section~\ref{sec:formalism} presents the formalism of GR polarized
radiative transfer.
In section~\ref{sec:tests}, we present various tests that we made to validate our code. The last section is dedicated to discussion and conclusion.

%No astrophysical application here, only tests. I will prepare
%in parallel a QU-loop non-Kerr paper that might be the first
%astrophysical application paper.

% \vspace{1cm}
% \textbf{MEMO LIST:}
% \begin{itemize}
% \item Rereads and comments:
%   ask Frederic Marin (Xray view, what should we highlight
%   to also attract the attention of Xray IXPE/eXTP people); ask Jason
%   and/or Monika; of course Maciek; ask Alex Lupsasca;
%   ask Benoît Cerutti for putative
%   future use with his code.
% \end{itemize}

\section{Formalism}\label{sec:formalism}

We will discuss the problem of polarized GRRT taking the usual
point of view of ray tracing. We thus consider a light ray
(mathematically speaking, a null geodesic) integrated backwards
from a distant observer's screen
towards some source of radiation. The problem can then be divided into
three main parts that will be discussed hereafter:
\begin{itemize}
\item the definition of a wave vector at the distant observer's screen,
  tangent to the considered null geodesic, together with a pair of
  spacelike vectors forming an orthonormal basis
  of the observer's screen;
\item the backwards parallel propagation along the considered
  null geodesic of the wave vector together with the screen basis,
  until a source of radiation is reached;
\item the integration of polarized radiative transfer
  within the source.
\end{itemize}

Before describing these three steps in detail, we will start
by providing important definitions in the next section.

\subsection{Geometric optics, light ray, covariant and observer-specific polarization vectors}
\label{sec:geom_optics}

We consider a monochromatic plane electromagnetic wave propagating
in a given spacetime. The geometrical optics
approximation of Maxwell's equations under Lorenz gauge condition
allows to describe this wave as follows.

% In this section, we consider only geometric optics, so we consider
% a photon traveling along a null geodesic, that will be described
% in terms of its Faraday tensor and associated electric and magnetic fields
% as measured in the frame comoving with the emitter of the photon.
% The electric and magnetic fields
% that are discussed below are that corresponding to this photon,
% observed by the emitter. It should not be confused with
% the ambient magnetic field close to Sgr~A*.

The complex 4-potential 1-form reads
\be
\label{eq:defpot}
\mathbf{\hat{A}} = \mathbf{\hat{a}} \, e^{i\Phi},
\ee
where $\mathbf{\hat{a}}$ is the amplitude 1-form, assumed to vary
much slower than the phase $\Phi$ (this is the basic idea of
the geometrical optics approximation). The hat reminds that we are
dealing with complex quantities.
The Faraday electromagnetic 2-form,
\be
\boldsymbol{\hat{\mathcal{F}}} = \mathbf{d \hat{A}},
\ee
then reads in components
\bea
\hat{\mathcal{F}}_{\alpha \beta} &=& \nabla_\alpha \hat{A}_\beta - \nabla_\beta \hat{A}_\alpha \\ \nn
&=& e^{i\Phi} \nabla_\alpha \hat{a}_\beta + \hat{a}_\beta\,i \,e^{i\Phi}\,\nabla_\alpha \Phi - \left( e^{i\Phi} \nabla_\beta \hat{a}_\alpha + \hat{a}_\alpha\,i \,e^{i\Phi}\,\nabla_\beta \Phi\right) \\ \nn
&\approx& i\left(\hat{a}_\beta\,k_\alpha - \hat{a}_\alpha\,k_\beta \right)\,e^{i \Phi},
\eea
where we introduce the \textit{wave vector}
\be
\mathbf{k} \equiv \boldsymbol{\nabla}{\Phi},
\ee
and use the geometric optics approximation to neglect the variations
of the amplitude, i.e.~the $\nabla_\alpha \hat{a}_\beta$ terms.

Plugging this into Maxwell's equations and assuming Lorenz gauge
(that is, the divergence of $\mathbf{\hat{A}}$ should vanish) leads to the following
results:
\begin{itemize}
\item $\mathbf{k}$ is a null vector parallel propagated along itself,
  \be
  \mathbf{k} \cdot \mathbf{k} = 0, \quad \boldsymbol{\nabla}_\mathbf{k}\, \mathbf{k} = \mathbf{0},
  \ee
  so that it defines a null geodesic, which we define as a \textit{light ray};
\item we can introduce the unit spacelike \textit{covariant polarization vector}
  \be
  \label{eq:polarvec}
  \mathbf{\hat{f}} \equiv \frac{\mathbf{\hat{a}}}{a},
  \ee
  where $\mathbf{\hat{a}}$ is the complex vector corresponding to the
  amplitude 1-form introduced in Eq.~\ref{eq:defpot} by metric duality
  (we use the same notation for both quantities in order to
  simplify the notation).
  The scalar quantity $a$ is the modulus of the complex
  vector $\mathbf{\hat{a}}$,
  \be
  a = \sqrt{\hat{a}_\mu \hat{a}^{\mu *}}.
  \ee
  Our naming convention specifies that this vector is covariant in order
  to differentiate it from the polarization vector as observed by a specific observer,
  which will be our quantity of prime interest in the following.
  The covariant polarization vector satisfies the two following important properties:
  (i) it is perpendicular to the wave vector $\mathbf{k}$ (this is a consequence
  of the Lorenz gauge choice),
  and (ii) it is parallel transported along $\mathbf{k}$ in vacuum
  (this is a consequence of Maxwell's equations):
  \be
  \label{eq:kprop}
  \mathbf{\hat{f}} \cdot \mathbf{k} = 0, \quad \boldsymbol{\nabla}_\mathbf{k}\, \mathbf{\hat{f}} = \mathbf{0}.
  \ee
\end{itemize}

We can thus reexpress the Faraday tensor in terms of the polarization
and wave vectors as follows
\be
{\hat{\mathcal{F}}_{\alpha\beta} = i\,a\,\left(\hat{f}_\beta\,k_\alpha - \hat{f}_\alpha\,k_\beta \right)\,e^{i \Phi}.}
\ee
From this expression we deduce the important property that it
is possible to add any multiple of the wave vector $\mathbf{k}$
to the polarization vector $\mathbf{\hat{f}}$ without altering
the Faraday tensor. So the polarization vector can be arbitrarily
transformed under
\be
\label{eq:fddl}
\mathbf{\hat{f}} \mapsto \mathbf{\hat{f}} + q \mathbf{k}
\ee
for any scalar field $q$ (note that $q$ is not necessarily a constant,
it is an arbitrary scalar field).

So far, we have only used global quantities that are not defined
with respect to any particular observer. We now want to introduce
the electric and magnetic fields as observed by the distant observer,
$\mathcal{O}$,
the oscillations of which define the observed electromagnetic wave.
Let us denote by $\mathbf{u_0}$ the 4-velocity of observer $\mathcal{O}$.
By definition, the electric linear form and the magnetic vector
as measured by a generic observer with 4-velocity $\mathbf{u}$
read
\footnote{We highlight that the electric and magnetic fields
  discussed here are the electromagnetic fields describing the
  monochromatic wave that reaches the observer's screen. They should
  not be confused with the electromagnetic fields that might exist
  at the source location, for instance the magnetic field of the
  accretion flow surrounding the black hole.}
\bea
\hat{E}_\alpha = \hat{\mathcal{F}}_{\alpha\mu} \,u^\mu, \\ \nn
\hat{B}^\alpha = -\frac{1}{2} \epsilon^{\alpha\mu\nu}_{\:\:\:\:\:\:\: \rho} \, \hat{\mathcal{F}}_{\mu\nu}\, u^\rho
\eea
where $\boldsymbol{\epsilon}$ is the Levi-Civita tensor.
The electric field vector $\mathbf{\hat{E}_0}$
as observed by the distant observer $\mathcal{O}$ thus reads
\bea
\hat{E}^\rho &=& g^{\rho \alpha} \, \hat{E}_\alpha \\ \nn
&=& g^{\rho \alpha} \,\hat{\mathcal{F}}_{\alpha \beta}\,u^\beta \\ \nn
&=& i\,a\,e^{i \Phi} \, g^{\rho \alpha} \left(\hat{f}_\beta\,k_\alpha - \hat{f}_\alpha\,k_\beta \right) u^\beta \\ \nn
&=& i\,a\,e^{i \Phi} \,\left(\hat{f}_\beta\,k^\rho - \hat{f}^\rho\,k_\beta \right) u^\beta \\ \nn
&=& i\,a\,e^{i \Phi} \, \left( (\mathbf{\hat{f}_0} \cdot \mathbf{u_0}) \,k^\rho + \omega_0 \hat{f}^\rho \right)
\eea
where we drop the lower index $0$ for the components of the various
tensors for simplicity (all of them being evaluated at the distant
observer's location),
and we introduce $\omega_0 \equiv -\mathbf{k_0} \cdot \mathbf{u_0}$,
where $\mathbf{k_0}$ is the wave vector at the distant observer's
location. This quantity  $\omega_0$ is
the pulsation of
the photon as measured by $\mathcal{O}$. All vectors with a lower index $0$ are
defined at the distant observer's screen.

Let us decompose the vectors
$\mathbf{\hat{f}_0}$ and $\mathbf{k_0}$ in parts parallel and orthogonal to the
observer's 4-velocity:
\bea
\label{eq:kfdecomp}
\mathbf{k_0} &=& \omega_0\,\mathbf{u_0} + \mathbf{K_0}, \qquad\qquad \mathbf{K_0} \perp \mathbf{u_0}, \\ \nn
\mathbf{\hat{f}_0} &=& -(\mathbf{\hat{f}_0} \cdot \mathbf{u_0})\,\mathbf{u_0} + \mathbf{\hat{f}_0^\perp}, \quad \mathbf{\hat{f}_0^\perp} \perp \mathbf{u_0}. \\ \nn
\eea
Note that
\be
\label{eq:fperp_norm}
\mathbf{\hat{f}_0^\perp} \cdot \mathbf{\hat{f}_0^\perp} = 1 + \left(\mathbf{\hat{f}_0} \cdot \mathbf{u_0} \right)^2
\ee
so that $\mathbf{\hat{f}_0^\perp}$ is not a unit vector in general.
It is normalized only if $\mathbf{\hat{f}_0} \cdot \mathbf{u_0}=0$,
in which case we simply have $\mathbf{\hat{f}_0^\perp} = \mathbf{\hat{f}_0}$.
Similary, $\mathbf{K_0}$ is not a unit vector and it is easy to show that
\be
\mathbf{K_0} \cdot \mathbf{K_0} = \omega_0^2.
\ee
The vector $\mathbf{K_0}$ coincides with the incident wave vector as measured
by observer $\mathcal{O}$.

In terms of the vectors normal to $\mathbf{u_0}$ we immediately
obtain the final expression of the electric vector as observed
by $\mathcal{O}$:
\be
\label{eq:E}
{\mathbf{\hat{E}_0} = i\,a\,e^{i \Phi} \left( \omega_0 \, \mathbf{\hat{f}_0^\perp} - \frac{\mathbf{K_0} \cdot \mathbf{\hat{f}_0^\perp}}{\omega_0}\, \mathbf{K_0} \right).}
\ee
This vector is clearly orthogonal to the direction of propagation $\mathbf{K_0}$,
and it is also orthogonal to the 4-velocity $\mathbf{u_0}$, as it should for a vector
living in the local rest space of the observer.

Then, the magnetic field vector $\mathbf{\hat{B}_0}$ as observed by $\mathcal{O}$ reads
\bea
\hat{B}^\rho &=& -\frac{1}{2}   \,\epsilon^{\rho \mu \nu}_{\:\:\:\:\:\:\alpha} \hat{\mathcal{F}}_{\mu\nu} u^\alpha \\ \nn
&=& -\frac{1}{2} i\,a\,e^{i \Phi} \, \epsilon^{\rho \mu \nu}_{\:\:\:\:\:\:\alpha} \left(\hat{f}_\nu k_\mu - \hat{f}_\mu k_\nu \right) u^\alpha \\ \nn
&=& - i\,a\,e^{i \Phi} \, \epsilon^{\rho \mu \nu}_{\:\:\:\:\:\:\alpha} \hat{f}_\nu k_\mu u^\alpha \\ \nn
&=& - i\,a\,e^{i \Phi} \, \epsilon^{\rho}_{\:\:\mu\nu\alpha} u^\alpha k^\mu \hat{f}^\nu \\ \nn
&=& i\,a\,e^{i \Phi} \, \epsilon^{\:\:\:\:\:\:\:\:\rho}_{\alpha\mu\nu} u^\alpha k^\mu \hat{f}^\nu \\ \nn
&=& i\,a\,e^{i \Phi} \, \epsilon^{\:\:\:\:\:\:\:\:\rho}_{\alpha\mu\nu} u^\alpha K^\mu \hat{f}^{\perp \nu}, \\ \nn
\eea
where we have used extensively the antisymmetric nature
of the Levi-Civita tensor.
The last expression exactly coincides with the
definition of the cross product in the vector space
orthogonal to $\mathbf{u_0}$ (which we label by $\times_\mathbf{u_0}$),
such that finally the magnetic
field vector as measured by $\mathcal{O}$ reads
\be
\label{eq:BKf}
{\mathbf{\hat{B}_0} = i\,a\,e^{i \Phi} \,\mathbf{K_0} \times_\mathbf{u_0} \mathbf{\hat{f}_0^\perp}.}
\ee
This vector is also obviously orthogonal to the direction
of propagation $\mathbf{K_0}$, and to the electric vector.

We can now define the notion of
\textit{polarization vector as measured by observer $\mathcal{O}$}:
\be
\label{eq:locpolar}
\mathbf{\hat{F}_0} = \mathbf{K_0} \times_\mathbf{u_0} \mathbf{\hat{B}_0},
\ee
which is by construction normal to the direction of propagation
and to the magnetic field vector. We note that this quantity
depends on the observer, just as the electric and magnetic fields,
while the covariant polarization vector $\mathbf{\hat{f}_0}$, defined in
Eq.~\ref{eq:polarvec}, is a covariant quantity.
They obviously differ given that by construction $\mathbf{\hat{F}_0}$
is orthogonal to the observer's 4-velocity $\mathbf{u_0}$,
while $\mathbf{\hat{f}_0}$ is defined independently from $\mathbf{u_0}$.
A natural question is to investigate the relation between
$\mathbf{\hat{F}_0}$ and $\mathbf{\hat{f}_0^\perp}$ that both live
in the vector space orthogonal to $\mathbf{u_0}$. These two vectors
are completely independent a priori, because $\mathbf{\hat{F}_0}$ is
by construction orthogonal to both $\mathbf{K_0}$ and $\mathbf{\hat{B}_0}$
(see Eq.~\ref{eq:locpolar}), while $\mathbf{\hat{f}_0^\perp}$ is only
orthogonal to $\mathbf{\hat{B}_0}$ (see Eq.~\ref{eq:BKf}),
but not to $\mathbf{K_0}$. Indeed, from Eq.~\ref{eq:kprop}
and~\ref{eq:kfdecomp}, we have
\bea
\label{eq:fFlink}
&&\mathbf{\hat{f}_0} \cdot \mathbf{k_0} = 0 \\ \nn
\Leftrightarrow &&\left(\mathbf{\hat{f}_0} \cdot \mathbf{u_0} \right) \, \omega_0 + \mathbf{\hat{f}_0^\perp} \cdot \mathbf{K_0} = 0 \\ \nn
\eea
so that only if $\mathbf{\hat{f}_0} \cdot \mathbf{u_0} = 0$ (which has no
reason to hold in general) is $\mathbf{\hat{f}_0^\perp}$ orthogonal
to $\mathbf{K_0}$. In this special case, we saw in Eq.~\ref{eq:kfdecomp}
that $\mathbf{\hat{f}_0^\perp} = \mathbf{\hat{f}_0}$, so that when
$\mathbf{\hat{f}_0} \cdot \mathbf{u_0} = 0$, and only then, $\mathbf{\hat{f}_0}$
is the unit vector along $\mathbf{\hat{F}_0}$.
However, we have seen in Eq.~\ref{eq:fddl} that the covariant
polarization vector is defined up to a term proportional to
the wavevector, the proportionality coefficient being a scalar field.
We can thus choose to work with a covariant polarization vector
$\mathbf{f'}$ such that, at the distant observer's location
\be
\mathbf{\hat{f}'_0} = \mathbf{\hat{f}_0} +  \frac{\mathbf{\hat{f}_0} \cdot \mathbf{u_0}}{\omega_0} \,\mathbf{k_0}.
\ee
This vector is such that
\be
\mathbf{\hat{f}'_0} \cdot \mathbf{u_0} = 0,
\ee
and we saw just above that this implies that
$\mathbf{\hat{f}'_0}$ is then the unit vector along
$\mathbf{\hat{F}_0}$. Thanks to the degree of liberty in the
definition of the covariant polarization vector expressed
by Eq.~\ref{eq:fddl}, we can thus confuse the covariant and
non-covariant polarization vectors at the observer,
$\mathbf{\hat{f}_0}$ and
$\mathbf{\hat{F}_0}$.

\begin{figure}[htbp]
\centering
\includegraphics[width=0.6\textwidth]{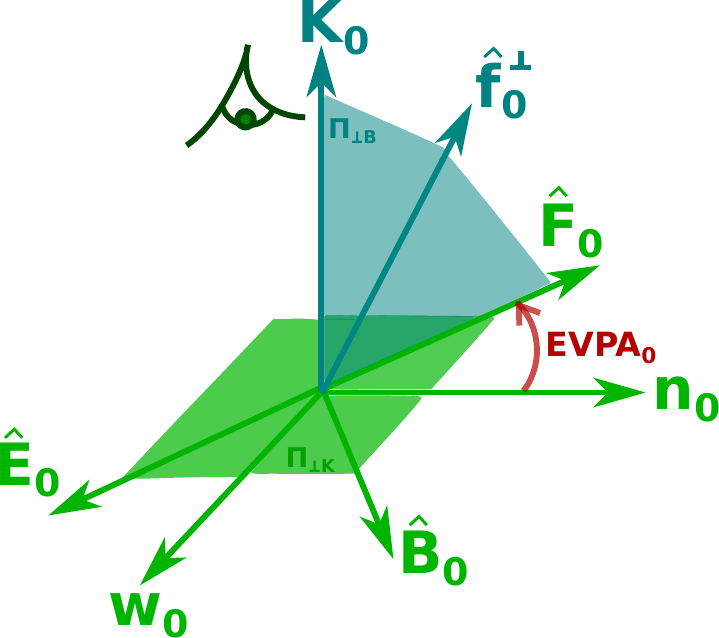}
\caption{Vectors lying in the distant observer $\mathcal{O}$'s
  local rest space (the vector space orthogonal to $\mathcal{O}$'s
  4-velocity $\mathbf{u_0}$).
  $\mathbf{K_0}$ is the wave vector
  projected orthogonal to $\mathbf{u_0}$. The vectors $\mathbf{\bar{w}_0}$ and
  $\mathbf{\bar{n}_0}$ are unit vectors pointing towards the local West and North directions,
  so that $(\mathbf{K_0},\mathbf{\bar{w}_0},\mathbf{\bar{n}_0})$
  forms a direct orthogonal triad.
  $\mathbf{\hat{E}_0}$ and $\mathbf{\hat{B}_0}$ are the electric and magnetic
  vectors as measured by $\mathcal{O}$,
  associated with the incident light wave, so that
  $(\mathbf{K_0},\mathbf{\hat{E}_0},\mathbf{\hat{B}_0})$
  is a direct orthogonal triad.
  $\mathbf{\hat{F}_0}$
  is the polarization vector as measured by $\mathcal{O}$
  (defined in Eq.~\ref{eq:locpolar}),
  which is antiparallel to
  $\mathbf{\hat{E}_0}$ as stated by Eq.~\ref{eq:FandE}.
  $\mathbf{\hat{f}_0^\perp}$ is the covariant polarization vector
  (defined in Eq.~\ref{eq:polarvec})
  projected orthogonal to $\mathbf{u_0}$.
  The plane orthogonal to $\mathbf{K_0}$ (observer's screen plane)
  is drawn in green. It contains the electric and magnetic field vectors,
  and the polarization vector as measured by $\mathcal{O}$.
  The plane orthogonal to $\mathbf{\hat{B}_0}$
  is drawn in blue-green. It contains the photon's wave vector
  $\mathbf{K_0}$, the electric vector $\mathbf{\hat{E}_0}$,
  and both the covariant and $\mathcal{O}$'s specific
  polarization vectors. The observed Electric Vector Position Angle (EVPA$_0$),
  measured East of North in the screen's plane, is shown in dark red.
}
\label{fig:EB}
\end{figure}

By virtue of the double vector product law we have
\bea
(\mathbf{K_0} \times_\mathbf{u_0} \mathbf{\hat{B}_0})^\alpha &=& i \, a \, e^{i \phi} \left[ (\mathbf{K_0}\cdot\mathbf{\hat{f}_0^\perp}) K^\alpha - (\mathbf{K_0}\cdot\mathbf{K_0}) \hat{f}^{\perp\alpha} \right] \\ \nn
&=& i \, a \, e^{i \phi}\,\omega_0 \, \left[ \frac{\mathbf{K_0}\cdot\mathbf{\hat{f}_0^\perp}}{\omega_0} K^\alpha - \omega_0 \hat{f}^{\perp\alpha} \right] \\ \nn
\eea
so that finally
\be
\label{eq:FandE}
{\mathbf{\hat{F}_0} = (\mathbf{K_0} \times_\mathbf{u_0} \mathbf{\hat{B}_0}) = - \omega_0 \,\mathbf{\hat{E}_0}}.
\ee
We thus conclude that the polarization vector as measured by $\mathcal{O}$
coincides, up to a normalization factor, with the electric field
as measured by $\mathcal{O}$. The various vectors lying in observer
$\mathcal{O}$'s local rest space (that is, the vector space orthogonal to
the 4-velocity $\mathbf{u_0}$) are depicted in Fig.~\ref{fig:EB}.
We introduce in this figure the \textit{electric vector position angle} (EVPA),
defined in the local frame of the distant observer,
which is the angle between a reference direction (the local North of the
distant observer) and the polarization vector $\mathbf{\hat{F}_0}$.

%In the following, we will never discuss the covariant polarization vector,
%and will only focus on the observer-specific polarization vector $\mathbf{\hat{F}_0} $.

%which makes sense given that $(\mathbf{K},\mathbf{E},\mathbf{B})$
%should be a right-handed triad (note that the three vectors are not
%normalized here).

\subsection{Polarization basis defined at the distant observer's screen}

\begin{figure}[htbp]
\centering
\includegraphics[width=\textwidth]{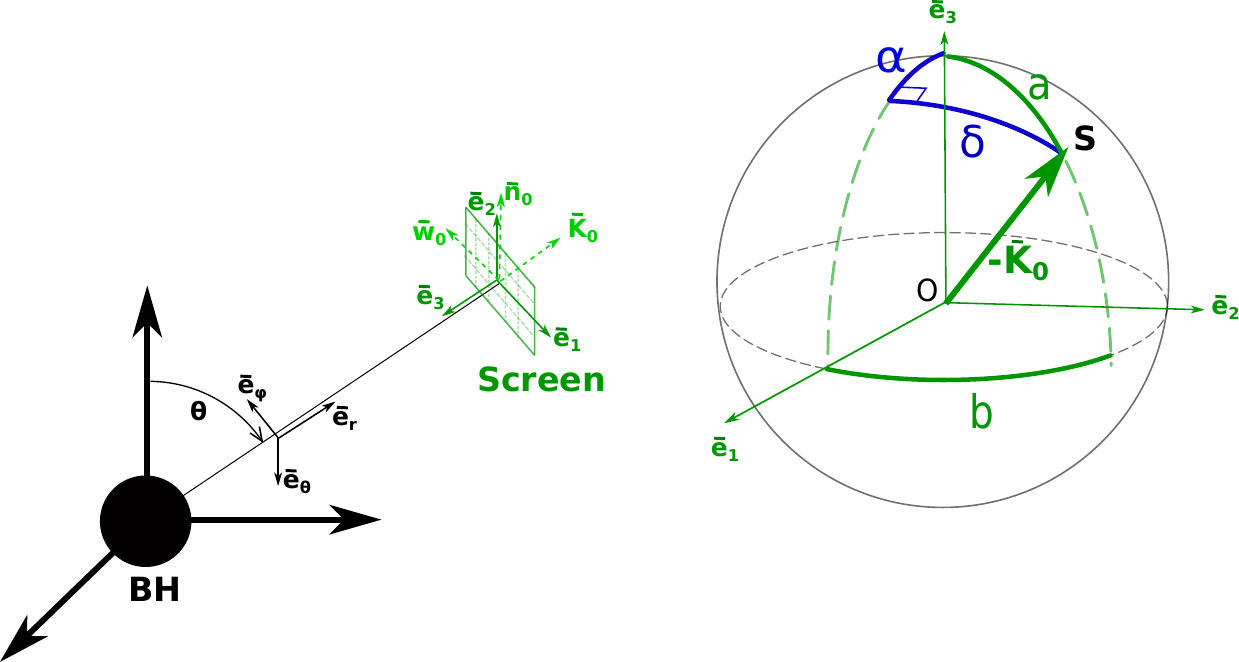}
\caption{Initial condition of the polarized ray-tracing problem
  at the distant observer's screen. \textbf{Left:} A black hole (BH)
  spacetime is represented for being specific, but the figure is very
  general and applies to any kind of spacetime. We consider spherical
  coordinates and the spacelike orthonormal basis associated with
  these coordinates are labeled $(\mathbf{\bar{e}_r},\mathbf{\bar{e}_\theta},
  \mathbf{\bar{e}_\pp})$.
  The observer's local rest
  space is described by a direct orthonormal triad,
  $(\mathbf{\bar{e}_1},\mathbf{\bar{e}_2},\mathbf{\bar{e}_3})$,
  where $\mathbf{\bar{e}_3} = -\mathbf{\bar{e}_r}$, and where the screen's plane
  is contained in $\mathrm{Span}(\mathbf{\bar{e}_1},\mathbf{\bar{e}_2})$.
  We consider here that
  the ``upwards'' direction of the observer's screen coincides
  with the projection of the black hole's angular momentum
  on the screen, i.e. that $\mathbf{\bar{e}_2}=-\mathbf{\bar{e}_\theta}$.
  The unit direction of photon reception at the observer's screen is $\mathbf{\bar{K}_0}$.
  The local polarization basis at the observer's screen,
  $(\mathbf{\bar{K}_0},\mathbf{\bar{w}_0},\mathbf{\bar{n}_0})$, is shown, and corresponds
  to the central pixel of the screen, that is, to the purely radial
  incoming direction of the photon.
  \textbf{Right:} zoom on the local
  rest space $(\mathbf{\bar{e}_1},\mathbf{\bar{e}_2},\mathbf{\bar{e}_3})$ and
  local celestial sphere of the observer. For a source of radiation
  located at S on the local celestial sphere, the unit direction
  of incidence is $\mathbf{\bar{K}_0}$ (it is $-\mathbf{\bar{K}_0}$ on the
  figure because $\mathbf{\bar{e}_3}$ points towards
  the source, while the incidence direction is of course in the
  opposite direction). The vector $\mathbf{\bar{K}_0}$ is here not purely
  along the radial direction $\mathbf{\bar{e}_3}$ (in contrast with the left panel):
  it thus corresponds to a pixel
  that is not located at the center of the screen.
  The corresponding equatorial angles labeling
  the source, $(\alpha,\delta)$, are shown, together with the
  corresponding spherical angles on the observer local sky, $(a,b)$.
  For typical ray-tracing problems where the observer is far away,
  we have $a \ll 1$.
}
\label{fig:obsframe}
\end{figure}

We take here the typical point of view of a ray-tracing problem
where the initial conditions are fixed at the far-away observer's
screen, and the integration is performed backwards in time from
the screen towards the source. This allows to save a lot of computing time
by integrating only those geodesics that will approach the source
by shooting light rays only within a small solid angle subtending
the source. The aim of this subsection is to explicitly describe
our initial conditions at the observer's screen, which is
illustrated in Fig.~\ref{fig:obsframe}. In order to be specific,
we will consider a black hole spacetime, but the discussion is
very general and is not restricted to this particular case.

The observer's screen is considered to be a pin-hole camera,
with the various pixels corresponding to different directions
on sky. The local rest space of the observer is spanned by
a direct orthonormal triad $(\mathbf{\bar{e}_1},\mathbf{\bar{e}_2},\mathbf{\bar{e}_3})$.
Here and in the following, a bar on top of a vector denotes a spacelike unit vector.
The vector $\mathbf{\bar{e}_3}$ is along the line of sight, normal to the screen, towards the black hole.
If we consider spherical coordinates centered on the black hole (e.g. Boyer-Lindquist coordinates) and a direct orthonormal triad
$(\mathbf{\bar{e}_r},\mathbf{\bar{e}_\theta},\mathbf{\bar{e}_\pp})$ associated with
these coordinates, then $\mathbf{\bar{e}_3}=-\mathbf{\bar{e}_r}$.
The screen's plane is spanned by $(\mathbf{\bar{e}_1},\mathbf{\bar{e}_2})$,
and we consider the special case $\mathbf{\bar{e}_2} = -\mathbf{\bar{e}_\theta}$,
which boils down to assuming that the projection of the black hole's
angular momentum on the screen is along $\mathbf{\bar{e}_2}$.
For a $N \times N$ pixels screen, one pixel with
indices $(i,j)$, with $i,j=1..N$, corresponds to a pair of
equatorial angles (see Fig.~\ref{fig:obsframe})
\bea
\alpha &=& \frac{f}{N} \left(i - \frac{N+1}{2} \right), \\ \nn
\delta &=& \frac{f}{N} \left(j - \frac{N+1}{2} \right), \\ \nn
\eea
where $f$ is the field of view of the observer. The corresponding
spherical angles (see Fig.~\ref{fig:obsframe})
on the local sky of the observer are given by standard spherical
trigonometry relations:
\bea
\cos a &=& \cos \alpha \, \cos \delta, \\ \nn
\tan b &=& \frac{\tan \alpha}{\sin \delta}. \\ \nn
\eea
The local unit direction of photon incidence then reads
\bea
\label{eq:K0def}
\mathbf{\bar{K}_0} &=& -\sin a \cos b \,\boldsymbol{\bar{e}_1} - \sin a \sin b \,\boldsymbol{\bar{e}_2} - \cos a  \,\boldsymbol{\bar{e}_3} \\ \nn
&=& \frac{\cos a}{\sqrt{g_{rr}}} \,\boldsymbol{\partial_r}
+ \frac{\sin a \sin b}{\sqrt{g_{\theta\theta}}} \,\boldsymbol{\partial_\theta}  + \frac{\sin a \cos b}{\sqrt{g_{\varphi\varphi}}} \,\boldsymbol{\partial_\varphi}, \\ \nn
\eea
where we have used the relations
\be
\mathbf{\bar{e}_r} = \frac{\boldsymbol{\partial_r}}{\sqrt{g_{rr}}}, \quad \mathbf{\bar{e}_\theta} = \frac{\boldsymbol{\partial_\theta}}{\sqrt{g_{\theta\theta}}}, \quad \mathbf{\bar{e}_\pp} = \frac{\boldsymbol{\partial_\pp}}{\sqrt{g_{\pp\pp}}},
\ee
where $\boldsymbol{\partial_i}$
are spherical coordinate basis vectors, and $g_{ii} = \boldsymbol{\partial_i} \cdot \boldsymbol{\partial_i}$ are the corresponding
metric coefficients.
The null 4-vector tangent to the null geodesic when incident on the
observer's screen then reads
\be
\mathbf{k_0} = \boldsymbol{\partial_t} + \mathbf{\bar{K}_0}
\ee
where we consider that the observer's 4-velocity is
$\mathbf{u_0} = \boldsymbol{\partial_t}$ (assuming a
static observer, and that $g_{tt} \to -1$ at the
observer's location) and where we
have assumed that $- \mathbf{k_0} \cdot \mathbf{u_0} = 1$,
so that the spacelike vector $\mathbf{\bar{K}_0}$ in the last
equation is normalized.
This last assumption means that the photon's energy as measured by the far-away observer
is unity,
which does not change anything to the problem, it simply scales the
energy, and physical values of energies can be easily retrieved
when radiative transfer calculations
are performed.

Once the photon arrival direction $\mathbf{\bar{K}_0}$ has been
defined, it must be completed by two other vectors to form
the local orthonormal polarization basis $(\mathbf{\bar{K}_0},\mathbf{\bar{w}_0},
\mathbf{\bar{n}_0})$. The vector $\mathbf{\bar{K}_0}$ corresponding to the
central pixel of the screen coincides with a purely radial
direction of arrival, $\mathbf{\bar{K}_0^\mathrm{cen}} = -\mathbf{\bar{e}_3}$ (the superscript 'cen' refers to the central pixel of the screen),
see the left panel of Fig.~\ref{fig:obsframe}.
This particular vector can be easily
completed by $\mathbf{\bar{w}_0^\mathrm{cen}} = -\mathbf{\bar{e}_1}$ and
$\mathbf{\bar{n}_0^\mathrm{cen}} = \mathbf{\bar{e}_2}$, see the left panel of
Fig.~\ref{fig:obsframe}.
In the general case of a vector $\mathbf{\bar{K}_0}$
defined by the two spherical angles
$(a,b)$ (see the right panel of Fig.~\ref{fig:obsframe}),
\ref{app:observertriad} shows that the observer's screen polarization basis
reads
\bea
\mathbf{\bar{w}_0} &=& \left[ - \sin^2 b \left(1 - \cos a \right) - \cos a\right] \mathbf{\bar{e}_1} + \sin b \cos b \left(1 - \cos a \right) \mathbf{\bar{e}_2} \\ \nn
&& + \cos b \sin a \, \mathbf{\bar{e}_3}, \\ \nn
\mathbf{\bar{n}_0} &=&  -\sin b \cos b \left(1 - \cos a \right)\mathbf{\bar{e}_1} + \left[  \cos^2 b \left(1 - \cos a \right) + \cos a\right] \mathbf{\bar{e}_2} \\ \nn
&&- \sin b \sin a \,\mathbf{\bar{e}_3}. \\ \nn
\eea
It is straightforward to check that these vectors are unit vectors,
orthogonal to each other, and to $\mathbf{\bar{K}_0}$. Moreover, in a typical
ray-tracing problem where $a \ll 1$, we have as we should $\mathbf{\bar{w}_0} \approx \mathbf{\bar{w}_0^\mathrm{cen}} = -\mathbf{\bar{e}_1}$, and $\mathbf{\bar{n}_0} \approx \mathbf{\bar{n}_0^\mathrm{cen}} = \mathbf{\bar{e}_2}$. We note that the plane
$(\mathbf{\bar{w}_0},\mathbf{\bar{n}_0})$ (where the polarization angle will be defined)
strictly speaking only coincides with the screen's plane
$(\mathbf{\bar{e}_1},\mathbf{\bar{e}_2})$ for the central pixel of the screen (with $a=0$).
We will neglect this small difference between the polarization plane
and the screen's plane, which is perfectly valid as long as
$a \ll 1$, that is, as long as the field of view is sufficiently small.

At this point, we have fully defined our initial condition by specifying
the triad $(\mathbf{k_0}, \mathbf{\bar{w}_0}, \mathbf{\bar{n}_0})$ at the observer's
screen. The next step is to parallel transport these vectors along
the light ray towards
the source.

\subsection{Relevant frames, parallel transport of polarization basis, EVPA} \label{sec:parallel_transport_EVPA}

Let us start by introducing the three relevant frames for describing the
polarized GRRT problem. We will focus on synchrotron radiation, which is our
primary science interest, but most of the discussion is rather general.
The frames of interest are:
\begin{itemize}
\item the \textit{observer frame}, described in detail in the previous section,
  defined by the 4-velocity $\mathbf{u_0}$,
\item the \textit{fluid frame}, defined by the 4-velocity $\mathbf{u}$ describing the
  bulk motion of the emitting fluid (for instance, Keplerian motion around
  a black hole),
\item the \textit{particle frame}, which follows the helical motion of the
  synchrotron-emitting electron around the magnetic field lines described
  in the fluid frame.
\end{itemize}
This section will mostly deal with the fluid frame, and the link with the
particle frame is further discussed in~\ref{app:PolarVec}.

We consider a light ray, modeled by a null geodesic,
joining the far-away observer to the emitting
accretion flow surrounding a black hole.
We want to parallel-transport the null 4-vector $\mathbf{k}$ tangent to the
null geodesic, backwards from the observer towards the emitter.
We will also parallel-transport the local West and North
unit spacelike directions,
$\mathbf{\bar{w}}$ and $\mathbf{\bar{n}}$.
Note that the index $0$ used for these three vectors in the previous
section meant that they were considered at the screen position.
We now consider their evolution along the ray and drop the index.
So we must integrate the following
equations
\bea
\label{eq:paralleltp}
\boldsymbol{\nabla_\mathbf{k}} \mathbf{k} &=& \mathbf{0}, \\ \nn
\boldsymbol{\nabla_\mathbf{k}} \mathbf{\bar{w}} &=& \mathbf{0}, \\ \nn
\boldsymbol{\nabla_\mathbf{k}} \mathbf{\bar{n}} &=& \mathbf{0}, \\ \nn
\eea
with the initial conditions that $(\mathbf{k}, \mathbf{\bar{w}}, \mathbf{\bar{n}}) = (\mathbf{k_0}, \mathbf{\bar{w}_0}, \mathbf{\bar{n}_0})$ at the screen.
Given that the parallel transport preserves the scalar product
between vectors
\footnote{This is an obvious property: let us consider two
  vectors $a^\mu$ and $b^\mu$ parallel-transported along $k^\mu$,
  then $\nabla_{\mathbf{k}} \left( \mathbf{a} \cdot \mathbf{b}\right) = k^\mu \nabla_\mu \left( g_{\alpha\beta} a^\alpha b^\beta\right) = g_{\alpha\beta} \left(a^\alpha k^\mu \nabla_\mu b^\beta + b^\beta k^\mu \nabla_\mu a^\alpha \right) = 0$, because of the parallel-transport relations $k^\mu \nabla_\mu a^\alpha = 0$ and $k^\mu \nabla_\mu b^\beta = 0$. We used the fact that the connexion
  $\nabla$ is compatible with the metric to get $\nabla_\mu g_{\alpha \beta} = 0$
  and take the metric tensor out of the covariant derivative.}
,
and given that $(\mathbf{k},\mathbf{\bar{w}},\mathbf{\bar{n}})$
are mutually orthogonal at the observer, they remain mutually
orthogonal when parallel-transported at the emitter, and $\mathbf{\bar{w}}$
and $\mathbf{\bar{n}}$ remain unit vectors.

It is useful at this point to note that, in vacuum, the EVPA is a conserved
quantity along a geodesic. Let us demonstrate this result.
We have seen that, at the distant observer's location,
we might confuse the covariant and non-covariant polarization vectors,
$\mathbf{\hat{f}_0}$ and
$\mathbf{\hat{F}_0}$. Let us consider the point along a photon's geodesic
corresponding to the exit from the emitting source region, meaning
that the part of the geodesic located in between this point and the
distant observer is in vacuum. We hereafter call this point the exit point.
We can make the exact same reasoning
at the exit point as we made at the distant observer's location, and
conclude that we can confuse the covariant and non-covariant polarization vectors
at the exit point,
$\mathbf{\hat{f}_\mathrm{exit}}$ and
$\mathbf{\hat{F}_\mathrm{exit}}$, where $\mathbf{\hat{F}_\mathrm{exit}}$
is the polarization vector as measured by the emitter
at the exit point. This implies more generally that $\mathbf{\hat{f}}$ and
$\mathbf{\hat{F}}$ can be confused at any point along the part of the
geodesic located in vacuum.
Given that $\mathbf{\hat{f}}$ and the screen
basis $(\mathbf{\bar{w}},\mathbf{\bar{n}})$ are parallel propagated along the
geodesic in vacuum (see Eqs.~\ref{eq:kprop} and~\ref{eq:paralleltp}), the angle between $\mathbf{\hat{f}}$ and the basis vectors
is conserved in vacuum,
hence the EVPA is conserved along the part of the
geodesic located in vacuum.
This is of course no longer valid in the source region,
where the covariant polarization vector is no longer
parallel propagated (the parallel propagation of $\mathbf{\hat{f}}$
is a consequence of Maxwell's equations in vacuum).

\begin{figure}[htbp]
\centering
\includegraphics[width=\textwidth]{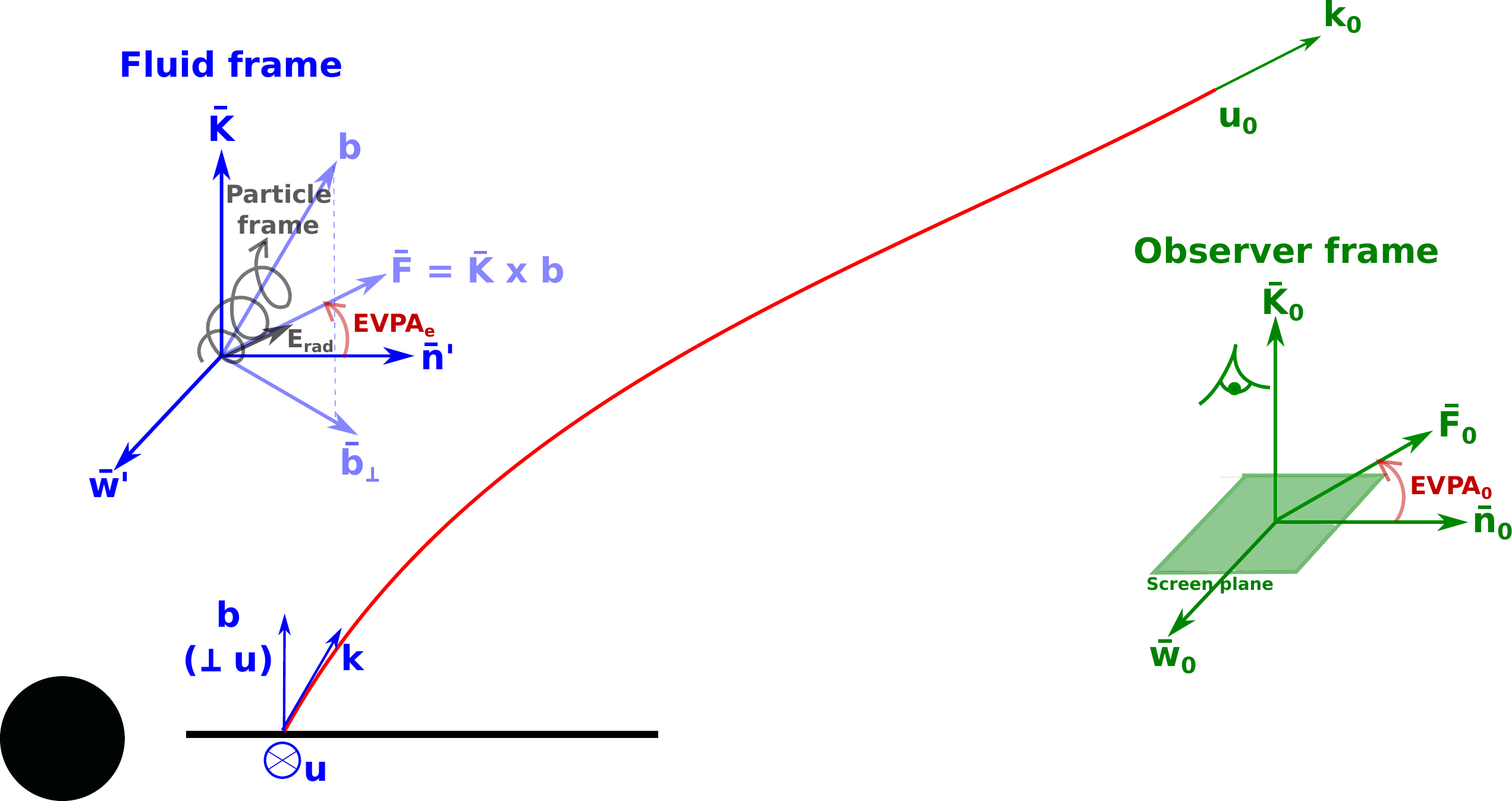}
\caption{Frames of interest for the polarized ray-tracing problem.
  The far-away observer's rest frame (orthogonal to the
  observer's 4-velocity $\mathbf{u_0}$)
  is shown in green,
  it is a simplified version of Fig.~\ref{fig:EB} and shows
  the local incidence direction $\mathbf{\bar{K}_0}$ (spacelike vector)
  of the observed
  light ray, together with the null 4-vector tangent to the
  incident null geodesic $\mathbf{k_0}$,
  the local North and West spacelike directions $(\mathbf{\bar{w}_0},\mathbf{\bar{n}_0})$,
  such that $(\mathbf{\bar{K}_0},\mathbf{\bar{w}_0},\mathbf{\bar{n}_0})$ is a direct
  orthonormal triad,
  and the polarization vector as measured by the far-away observer,
  $\mathbf{\bar{F}_0}$. The observed EVPA is labelled EVPA$_0$,
  lying between the screen's
  North direction and the observed polarization vector $\mathbf{\bar{F}_0}$.
  Starting from the far-away observer, a null
  geodesic is integrated backwards towards the source (red line),
  until it reaches the accretion flow (black line) surrounding the
  black hole (black disk). The fluid frame (orthogonal to the
  emitter's 4-velocity $\mathbf{u}$) is shown
  in blue. The null 4-vector tangent to the null geodesic at the
  emission point is called $\mathbf{k}$. The magnetic field spacelike 4-vector
  as measured in the fluid frame (thus, orthogonal to $\mathbf{u}$)
  is called $\mathbf{b}$. The synchrotron-emitting electron's trajectory
  is represented by the pale black helix. The local direction of photon emission,
  as measured in the fluid frame, is the spacelike unit vector $\mathbf{\bar{K}}$.
  The pair of vectors $(\mathbf{\bar{n}'},\mathbf{\bar{w}'})$ is related to
  the pair $(\mathbf{\bar{n}_0},\mathbf{\bar{w}_0})$, parallel-propagated along
  the null geodesic (see text for details), such that
  $(\mathbf{\bar{K}},\mathbf{\bar{w}'},\mathbf{\bar{n}'})$ is a direct orthonormal triad.
  The unit polarization vector as measured by the emitter is
  called $\mathbf{\bar{F}}$. The radiation field $\mathbf{E_\mathrm{rad}}$
  associated with the helical motion of the electron is shown in
  pale black, and lies along $\mathbf{\bar{F}}$ for a relativistic
  electron (see~\ref{app:PolarVec}
  for a demonstration). The unit projection of the magnetic
  4-vector orthogonal to $\mathbf{\bar{K}}$ is called $\boldsymbol{\bar{b}_\perp}$.
  Thus $(\mathbf{\bar{K}},\boldsymbol{\bar{b}_\perp},\mathbf{\bar{F}})$ is also a direct
  orthonormal triad,
  rotated with respect to $(\mathbf{\bar{K}},\mathbf{\bar{w}'},\mathbf{\bar{n}'})$
  by the emission EVPA, labeled EVPA$_\mathrm{e}$, lying between the parallel-transported
  North direction and the fluid-frame polarization vector $\mathbf{\bar{F}}$.
}
\label{fig:setup}
\end{figure}

We want to project the parallel-transported basis vectors
$(\mathbf{k}, \mathbf{\bar{w}}, \mathbf{\bar{n}})$
orthogonally to the 4-velocity $\mathbf{u}$ of the emitting fluid, that is,
project them in the local
rest space of the fluid.
By doing so without further precaution,
we would of course lose the mutual orthogonality between these
vectors, which is not preserved in a projection.
Let us consider
\bea
\label{eq:np}
\mathbf{\bar{w}'} = \mathbf{\bar{w}} - \frac{\mathbf{\bar{w}} \cdot \mathbf{u}}{\mathbf{k} \cdot \
\mathbf{u}} \, \mathbf{k}, \\ \nn
\mathbf{\bar{n}'} = \mathbf{\bar{n}} - \frac{\mathbf{\bar{n}} \cdot \mathbf{u}}{\mathbf{k} \cdot \
\mathbf{u}} \, \mathbf{k}, \\ \nn
\eea
where the denominator, $\mathbf{k} \cdot \mathbf{u}$, is minus the energy
of the photon as measured in the fluid frame, and as such, non zero, so that
these expressions are well defined.
It is easy to check that these two vectors are spacelike unit vectors,
orthogonal to $\mathbf{u}$,
to each other, and to
\be
\label{eq:kproju}
\mathbf{\bar{K}} = \frac{\mathbf{k} + \left(\mathbf{k} \cdot \mathbf{u} \right)\,\mathbf{u}}{\vert \mathbf{k} \cdot \mathbf{u} \vert},
\ee
the normalized projection of $\mathbf{k}$ orthogonal to $\mathbf{u}$,
which coincides with the unit direction of emission of the photon in the
fluid frame.
We thus obtain a well-defined
orthonormal direct triad $(\mathbf{\bar{K}},\mathbf{\bar{w}'}, \mathbf{\bar{n}'})$
of the fluid frame, illustrated in Fig.~\ref{fig:setup}.
We note that if we consider a vector $\mathbf{F}$ in the fluid frame,
normal to $\mathbf{\bar{K}}$, then our definition leads to
\be
\label{eq:Fdotn}
\mathbf{F} \cdot \mathbf{\bar{n}} = \mathbf{F} \cdot \mathbf{\bar{n}'},
\ee
and similarly for $\mathbf{w}$, so that our definition allows to keep unchanged
the angles between such a vector $\mathbf{F}$ and the reference directions, be
they primed or unprimed. This will be important later.

Let us now consider the magnetic field 4-vector $\mathbf{b}$ of the
accretion flow, as measured in the fluid frame. By construction, this
vector lies in the local rest frame of the fluid, so it is orthogonal
to $\mathbf{u}$. We are also interested in its normalized
projection orthogonally
to $\mathbf{\bar{K}}$, which reads
\be
\boldsymbol{\bar{b}_\perp} = \frac{\mathbf{b} - \left(\mathbf{b} \cdot \mathbf{\bar{K}}\right) \mathbf{\bar{K}}}{\vert \vert \mathbf{b} - \left(\mathbf{b} \cdot \mathbf{\bar{K}}\right) \mathbf{\bar{K}} \vert \vert},
\ee
(note that the minus sign in the numerator and denominator of the rhs,
compared to the plus sign in the numerator of the
rhs of Eq.~\ref{eq:kproju}, comes from the fact that $\mathbf{\bar{K}}$
is spacelike while $\mathbf{u}$
is timelike),
and in the unit polarization vector as measured in the fluid frame
\be
\label{eq:Femit}
\mathbf{\bar{F}} = \frac{\mathbf{\bar{K}} \times \mathbf{b}}{\vert \vert \mathbf{\bar{K}} \times \mathbf{b} \vert \vert} = \mathbf{\bar{K}} \times \boldsymbol{\bar{b}_\perp}.
\ee
We thus have constructed a second orthonormal direct triad of
the fluid rest space, $(\mathbf{\bar{K}},\boldsymbol{\bar{b}_\perp},\mathbf{\bar{F}} )$.
We note that it is not obvious that the vector defined by Eq.~\ref{eq:Femit}
coincides with the emission polarization vector for synchrotron radiation,
that is, with the direction of the
radiation electric field emitted by an electron moving around the $\mathbf{b}$
field lines, given that we have never discussed the emitting electron motion so far.
In~\ref{app:PolarVec}, by relating the particle frame and the fluid frame,
we demonstrate that, provided the emitting electron
is relativistic, this is indeed so. %It is important to keep in mind that this
%is only so for relativistic electrons (with Lorentz factor in the fluid frame $\gamma \gg 1$).

We have thus at hand two orthonormal triads of the fluid frame,
the observer-related $(\mathbf{\bar{K}},\mathbf{\bar{w}'}, \mathbf{\bar{n}'})$,
and the magnetic-field-related $(\mathbf{\bar{K}},\boldsymbol{\bar{b}_\perp},\mathbf{\bar{F}} )$.
These two frames are rotated with respect to each other by the
angle
\be
\label{eq:EVPAem}
\chi \equiv (\mathbf{\bar{n}'},\mathbf{\bar{F}}) = (\mathbf{\bar{n}},\mathbf{\bar{F}}) \equiv \mathrm{EVPA_e},
\ee
the EVPA in the fluid frame, where the index e reminds that we
are dealing with an emission EVPA, as compared to the observed EVPA
of Fig.~\ref{fig:EB}. Both angles are illustrated in Fig.~\ref{fig:setup}.
We note that the emission EVPA evolves as the light ray evolves through
the emitting fluid; the EVPA is only conserved in vacuum as demonstrated above.
So a sequence of emission EVPAs corresponds to a
unique observed EVPA. Note that the second equality in Eq.~\ref{eq:EVPAem}
is a consequence of Eq.~\ref{eq:Fdotn}.
This emission EVPA will be crucial in the polarized radiative transfer
formalism that we introduce in the next section.
A practical, code-friendly expression for the emission EVPA is the following
\be
\mathrm{EVPA_e} = \frac{\pi}{2} - \mathrm{atan2} \left(\mathbf{\bar{b}_\perp}\
 \cdot \mathbf{\bar{w}'}, \mathbf{\bar{b}_\perp} \cdot \mathbf{\bar{n}'} \right).
\ee

After having discussed the parallel transport of the vectors of
interest along the null geodesic, the last step of the
polarized GRRT problem is to integrate the polarized
radiative transfer within the accretion flow surrounding the
compact object.

\subsection{Polarized radiative transfer}

\subsubsection{Stokes parameters}

\begin{figure}[htbp]
\centering
\includegraphics[width=0.5\textwidth]{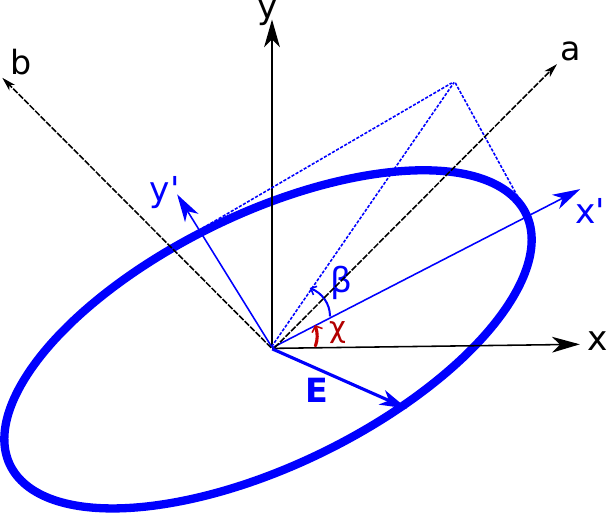}
\caption{ Polarization ellipse of a general monochromatic wave (in blue).
  The black axes $(x,y)$ label the frame of interest where we want to
  formulate the problem, while the blue axes
  $(x',y')$ are along the major and minor axes of the ellipse
  and are therefore naturally adapted to the elliptically
  polarized wave.
  The angle between the two bases is called $\chi$ (it would coincide
  with the notion of EVPA for a linearly polarized wave along
  the $x$ axis),
  and the quantity $\tan \beta$ encodes the ellipse axes ratio.
  The dashed black axes $(a,b)$ are tilted by $45^\circ$ relative
  to $(x,y)$ and are useful for defining the U Stokes parameter.
}
\label{fig:ellipsepolar}
\end{figure}

The most general monochromatic electromagnetic wave has
an elliptical polarization, in the sense that the electric
field vector describing the wave draws an ellipse during
its time evolution in the plane normal to the direction
of propagation, see Fig.~\ref{fig:ellipsepolar}.

Let us introduce the electric field complex vector of the
monochromatic wave
\be
\mathbf{\hat{E}} = \hat{E}_x \mathbf{\bar{e}_x} + \hat{E}_y \mathbf{\bar{e}_y},
\ee
decomposed in an arbitrary orthonormal basis $(\mathbf{\bar{e}_x},\mathbf{\bar{e}_y})$ of the plane orthogonal to the direction of propagation,
where $\hat{E}_x$ and $\hat{E}_y$ are the complex components along
the axes.
The polarization ellipse described by the vector $\mathbf{\hat{E}}$
can be equivalently described by two sets of parameters.
From a geometrical point of view, it is very natural to provide
the total intensity $|\mathbf{\hat{E}}|^2$, together with two angles
$\chi$ and $\beta$. The angle $\chi$ lies between the
basis $(\mathbf{\bar{e}_x},\mathbf{\bar{e}_y})$ and the basis corresponding to the
axes of the ellipse, while $\tan \beta$ encodes the ellipse axes ratio
(see Fig.~\ref{fig:ellipsepolar}). However, this parametrization is not
practical from a physical point of view given that the two angles
are not directly observable. So from a physical point of view, it is more
natural to consider the following set of four Stokes
parameters~\citep{RL79}
\bea
\label{eq:Stokes}
I &=& |\hat{E}_x|^2 + |\hat{E}_y|^2, \\ \nn
Q &=& |\hat{E}_x|^2 - |\hat{E}_y|^2 = I \, \cos 2\chi \, \cos 2\beta, \\ \nn
U &=& |\hat{E}_a|^2 - |\hat{E}_b|^2 = I \, \sin 2\chi \, \cos 2\beta, \\ \nn
V &=& |\hat{E}_r|^2 - |\hat{E}_l|^2 = I \, \sin 2\beta, \\ \nn
\eea
where $\hat{E}_a$ and $\hat{E}_b$ are the complex components of
the electric field in an orthonormal basis $(\mathbf{\bar{e}_a},\mathbf{\bar{e}_b})$
rotated by $45^\circ$ compared to $(\mathbf{\bar{e}_x},\mathbf{\bar{e}_y})$,
see the dashed black axes in Fig.~\ref{fig:ellipsepolar}.
The quantities $\hat{E}_r$ and $\hat{E}_l$ are the complex
components of the field in an orthonormal complex
basis, $\mathbf{\bar{e}_{l,r}} = \sqrt{2}/2 (\mathbf{\bar{e}_x} \pm i \mathbf{\bar{e}_y})$.
These relations clearly show that the Stokes parameters
are all sums or differences of intensities along specific directions,
and as such are directly observable and well adapted to being evolved in
a radiative transfer problem. Equations~\ref{eq:Stokes} show how
to construct the Stokes parameters from the geometrical
angular parameters $\chi$, $\beta$ of the polarization ellipse.
The reverse expression is easy to find and reads
\bea
\tan 2\chi &=& \frac{U}{Q}, \\ \nn
\sin 2\beta &=& \frac{V}{I}. \\ \nn
\eea

For a circular polarization, $\beta=\pi/4$, so
\be
Q=0, \quad U=0, \quad V=I \quad (\mathrm{circular \: polarization}),
\ee
while for a linear polarization, $\beta=0$, and
\be
Q = I \cos 2\chi, \quad U = I \sin 2 \chi, \quad V=0 \quad (\mathrm{linear \: polarization}),
\ee
and if the wave is polarized along the $x$ axis of Fig.~\ref{fig:ellipsepolar},
then $Q=I$ and $U=0$, while if the wave is polarized at $45^\circ$
from the $x$ axis, then $Q=0$ and $U=I$.
So $Q$ and $U$ encode linear polarization along the directions $\mathbf{\bar{e}_x}$ or $\mathbf{\bar{e}_y}$
and $\mathbf{\bar{e}_a}$ or $\mathbf{\bar{e}_b}$, respectively, and $V$ encodes circular polarization.

Let us consider the Stokes parameters $(I,Q,U,V)$ defined in a
basis $(\mathbf{\bar{e}_x},\mathbf{\bar{e}_y})$, and the parameters
$(I',Q',U',V')$ defined in a basis $(\mathbf{e'_x},\mathbf{e'_y})$,
rotated by an angle $\chi$ with respect to $(\mathbf{\bar{e}_x},\mathbf{\bar{e}_y})$,
see Fig.~\ref{fig:ellipsepolar}. It is easy to show that the $Q$ and $U$
Stokes parameters transform following
\be
\label{eq:StokesRot}
 \left( \begin{array}{c}
     Q    \\
     U   \end{array} \right)
   =
   \left( \begin{array}{cc}
     \cos 2\chi        &   -\sin 2\chi   \\
     \sin 2\chi          &   \cos 2\chi  \end{array} \right)
  \left( \begin{array}{c}
     Q'    \\
     U'      \end{array} \right),
 \ee
 while $I$ and $V$ are invariant.

For a monochromatic radiation, there must exist a relation between
the four Stokes parameters, that are equivalent to the
set of three parameters $(I,\chi,\beta)$, and thus cannot
be independent. This relation reads
\be
I^2 = Q^2 + U^2 + V^2 \quad (\mathrm{monochromatic \: / \: fully \: polarized})
\ee
and the radiation is then said to be fully polarized.
For a superimposition of waves at different frequencies,
the radiation is only partially polarized and the resulting
Stokes parameters verify
\be
I^2 \geq Q^2 + U^2 + V^2 \quad (\mathrm{partially \: polarized}).
\ee
It is then useful to introduce the degree of polarization
\be
\mathrm{d_p} = \frac{\sqrt{Q^2 + U^2 + V^2}}{I},
\ee
and the degree of linear polarization
\be
\mathrm{d_{lp}} = \frac{\sqrt{Q^2 + U^2}}{I}.
\ee

\subsubsection{Stokes parameters for synchrotron radiation, conventions}

\begin{figure}[htbp]
\centering
\includegraphics[width=\textwidth]{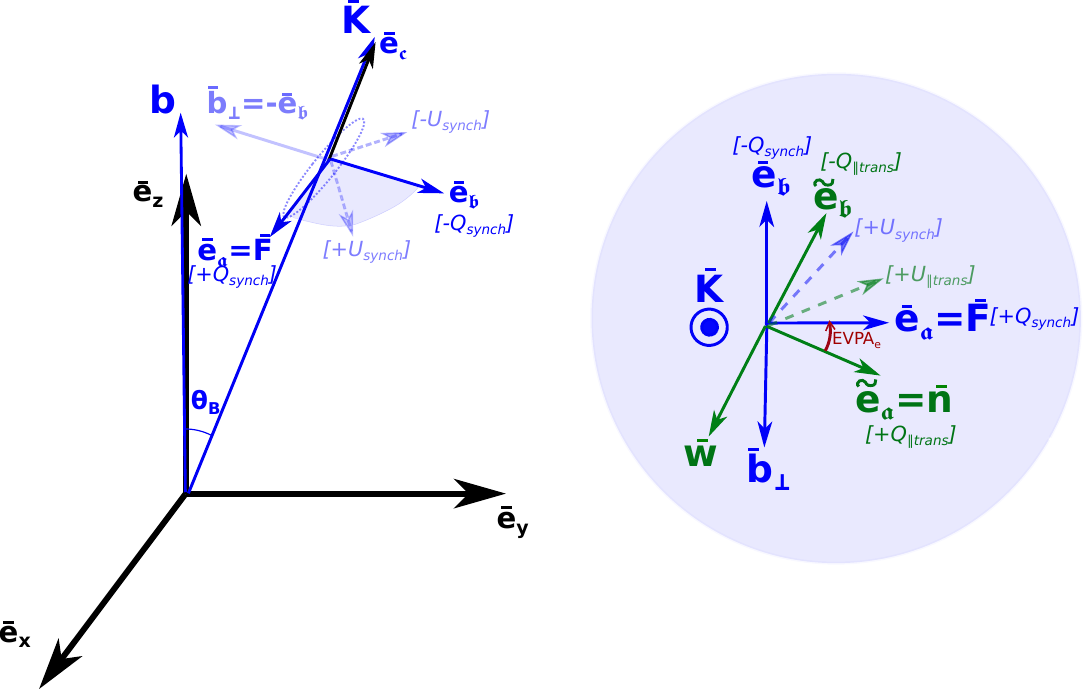}
\caption{Geometry of the polarized synchrotron problem and Stokes
  parameter illustration. All quantities depicted here are defined
  in the fluid frame.
  The magnetic field is $\mathbf{b}$,
  which lies along the $\mathbf{\bar{e}_z}$ unit vector.
  The direction of emission in this frame
  is $\mathbf{\bar{K}}$, which lies along the $\mathbf{\bar{e}_\textfrak{c}}$ unit vector.
  We define the $\mathbf{\bar{e}_\textfrak{a}}$ unit vector as lying along the major axis
  of the synchrotron polarization ellipse (shown in dashed pale blue).
  This vector is defined up to an unimportant sign convention.
  The $\mathbf{\bar{e}_\textfrak{b}}$ unit vector is such that $(\mathbf{\bar{e}_\textfrak{a}},\mathbf{\bar{e}_\textfrak{b}},\mathbf{\bar{e}_\textfrak{c}})$
  is a direct orthonormal triad of the fluid frame. The vector $\mathbf{\bar{e}_x}$ is
  parallel to $\mathbf{\bar{e}_\textfrak{a}}$. The $\mathbf{\bar{e}_y}$ unit vector is
  such that $(\mathbf{\bar{e}_x},\mathbf{\bar{e}_y},\mathbf{\bar{e}_z})$
  is a direct orthonormal triad of the fluid frame. The angle between $\mathbf{b}$
  and $\mathbf{\bar{K}}$ is called $\theta_B$. The Stokes parameters Q and U
  are defined in the $(\mathbf{\bar{e}_\textfrak{a}} = \mathbf{F},\mathbf{\bar{e}_\textfrak{b}}=-\mathbf{\bar{b}_\perp})$ basis,
  that we call the synchrotron polarization basis,
  illustrated
  by the zoom on the right of the figure. This zoom shows
  the synchrotron polarization basis (in blue; subscript 'synch')
  as well as the parallel-transported polarization basis (in green;
  subscript '$\parallel$trans'),
  $(\mathbf{\tilde{e}_\textfrak{a}}=\mathbf{\bar{n}},\mathbf{\tilde{e}_\textfrak{b}}=-\mathbf{\bar{w}})$.
  These two bases are rotated
  by the emission EVPA angle.
  The sign conventions of the Stokes parameters are
  as shown in this zoom.
  Note that the orientation convention used in this figure is the same as
  that used by e.g.~\citet{Dexter2016,Huang2011}, which results in a positive
  emission coefficient for Stokes Q. Some authors use an
  alternative orientation convention,
  taking $\mathbf{\bar{e}_\textfrak{a}}$ along the minor axis of the polarization ellipse,
  see~\citet{Shcherbakov08,Pandya2021}. This simply leads to Stokes Q being
  multiplied by $-1$, and to a negative emission coefficient for Stokes Q.
  % Note: the full ellipse is described by the Erad vector of App. B as the electron
  % gyrates around the mf. But for a relativistic electron, it is only when K is along
  % v_elec that there is radiation observed at infinity, so this is when F is as in
  % Fig. 5, orthogonal to b. So the synchro radiation becomes nearly fully linearly
  % polarized along the direction Kxb when the electron is relativistic.
  %
}
\label{fig:polarframes}
\end{figure}

We are primarily interested in polarized synchrotron
radiation given that our main science interest is the
millimeter and infrared radiation emitted by nearby
supermassive black hole environments.
Let us consider a single electron following a helical motion
around the field lines of a magnetic field $\mathbf{b}$ as
measured in the fluid frame. The emitted synchrotron radiation
is elliptically polarized, with the minor axis of the polarization ellipse aligned
along the direction of the magnetic field projected
orthogonally to the direction of propagation, and major axis
along the fluid-frame polarization vector~\citep{Huang09}.
The Stokes parameters are thus naturally expressed in
a basis aligned with the axes of this polarization ellipse,
that is, the $(\mathbf{\bar{F}},-\mathbf{\bar{b}_\perp})$ basis
(see the illustration in Fig.~\ref{fig:polarframes}).
We call this basis the \textit{synchrotron polarization basis}.
This basis is rotated by the emission EVPA with respect
to the observer-related $(\mathbf{\bar{n}'},-\mathbf{\bar{w}'})$ basis.
This last basis is called the \textit{parallel-transported polarization basis}.
%We note that the observer polarization basis is indeed a basis
%of the fluid frame. But its basis vectors are parallel propagated
%from the observer frame, hence the name.
For integrating the radiative transfer in the observer's
frame, we will need to take care of this rotation between
the synchrotron and the parallel-transported polarization bases.
This is described in the next section.

% This is illustrated in Fig.~\ref{fig:polarframes}.
% For a distribution of electrons emitting synchrotron radiation,
% the resulting radiation is linearly polarized along the
% main axis of the polarization ellipse, provided that the
% distribution of pitch angle (the angle between the
% magnetic field direction and the velocity of the
% electron) is smooth~\citep{Huang09}. So a typical
% synchrotron radiation will lead to a high degree of
% linear polarization.

Our sign conventions for the Stokes parameters are illustrated
in Fig.~\ref{fig:polarframes}. It complies with
the convention of the International
Astronomical Union~\citep[see Fig.~1 of the
second reference]{IAU74,hamaker96}.

% \begin{figure}[htbp]
% \centering
% \includegraphics[width=0.5\textwidth]{./Figure/SynchPolar.pdf}
% \caption{\textbf{Probably no longer useful?}. Polarization ellipse (black solid contour)
%   in the plane $(\mathbf{\bar{F}},\boldsymbol{\bar{b}_\perp})$,
%   orthogonal to the direction of propagation $\mathbf{\bar{K}}$,
%   of the
%   single-electron synchrotron radiation. The major axis is aligned
%   with the local polarization vector direction, while the minor
%   axis is aligned with the projection of the magnetic field
%   normal to the direction of propagation. The dashed black line
%   shows the linear polarization direction resulting from the
%   emission of an assembly of electrons with a smooth pitch
%   angle distribution. The rest of the figure is taken from
%   Fig.~\ref{fig:setup}.
% }
% \label{fig:synchpolar}
% \end{figure}

\subsubsection{Transfer equation}

Just like in the unpolarized version of \gy~\citep{vincent11}, we will integrate
the radiative transfer equation in the fluid frame, and
then transform the quantities to the observer's frame. The unpolarized
radiative transfer equation used by \gy reads
\be
\frac{\dd I^\mathrm{em}_\nu}{\dd s} = j_\nu - \alpha_\nu \, I^\mathrm{em}_\nu ,
\ee
where $I^\mathrm{em}_\nu$ is the specific intensity (the index $\nu$ means
that we are considering an intensity per unit of frequency),
$\dd s$ is the element
of optical path length,
$j_\nu$ and $\alpha_\nu$ are the specific emission and absorption
coefficients, all these quantities being measured in the fluid frame
(hence the superscript 'em' for the intensity, referring to the emitting fluid; we discard it for the
other quantities for simplicity).
The intensity in the observer frame (superscript 'obs') then follows from
\be
\label{eq:Liouville}
I^\mathrm{obs}_\nu = I^\mathrm{em}_\nu \, \left( \frac{\nu^\mathrm{obs}}{\nu^\mathrm{em}}\right)^3,
\ee
which is a consequence of Liouville's theorem~\citep{MTW}.

The polarized radiative transfer equation is naturally written
in the synchrotron polarization basis of the fluid frame.
%meaning that the
%Stokes parameters $Q_\nu$ and $U_\nu$ (the index $\nu$ again means
%that we are dealing with specific quantities, per unit frequency)
%are associated with the directions
%aligned and at $45^\circ$ with respect to the last two vectors
%of the tetrad. More precisely, the Stokes parameters are
%expressed in the $(\mathbf{\bar{F}},-\mathbf{\bar{b}_\perp})$
%basis, see Fig.~\ref{fig:polarframes}.
In this basis,
the transfer equation reads
\be
 \frac{\dd}{\dd s}\left( \begin{array}{c}
     I_\nu^\mathrm{em;synch}    \\
     Q_\nu^\mathrm{em;synch}    \\
    U_\nu^\mathrm{em;synch}       \\
   V_\nu^\mathrm{em;synch}       \end{array} \right)
   =
   \left( \begin{array}{c}
     j_{\nu,I}    \\
     j_{\nu,Q}    \\
    j_{\nu,U}       \\
   j_{\nu,V}        \end{array} \right)
   -
   \left( \begin{array}{cccc}
     \alpha_{\nu,I}        &   \alpha_{\nu,Q}  &  \alpha_{\nu,U}  &  \alpha_{\nu,V}   \\
      \alpha_{\nu,Q}          &   \alpha_{\nu,I}  &r_{\nu,V} & -r_{\nu,U}  \\
     \alpha_{\nu,U}          &  -r_{\nu,V} & \alpha_{\nu,I} & r_{\nu,Q}  \\
    \alpha_{\nu,V}          &  r_{\nu,U} & -r_{\nu,Q} &  \alpha_{\nu,I}   \end{array} \right)
  \left( \begin{array}{c}
     I_\nu^\mathrm{em;synch}    \\
     Q_\nu^\mathrm{em;synch}    \\
    U_\nu^\mathrm{em;synch}       \\
   V_\nu^\mathrm{em;synch}        \end{array} \right),
\ee
where the 'em;synch' label is here to remind that we are
dealing with Stokes parameters expressed
in the synchrotron polarization basis of the emitting fluid frame;
moreover,
$j_{\nu,X}$ and $\alpha_{\nu,X}$ are emission and absorption
coefficients for the Stokes parameter X, $r_{\nu,Q}$ and $r_{\nu,U}$
are Faraday conversion parameters, and $r_{\nu,V}$ is the
Faraday rotation parameter. All these transfer coefficients
are defined in the synchrotron polarization basis of the fluid frame
(we discard
the superscript for simplicity). We refer to Fig.~\ref{fig:polarframes}
for the details of the sign conventions.
%emitter's
%$(\mathbf{\bar{b}_\perp},\mathbf{\bar{F}})$ basis,
%with sign conventions
%as illustrated in Fig.~\ref{fig:polarframes}.
%The superscript em for the
%Stokes parameters highlights that we are dealing with quantities
%in this particular basis.

However, we rather want to integrate this equation in
the parallel-transported polarization basis,
$(\mathbf{\bar{n}'},-\mathbf{\bar{w}'})$,
which is rotated by the emission
EVPA with respect to the synchrotron polarization
basis, see Fig.~\ref{fig:polarframes}.
%so that the $Q,U$ Stokes parameters will be aligned with
%the observer-related axes $(\mathbf{\bar{w}'},\mathbf{\bar{n}'})$, rather
%than with the magnetic-field-related axes of the Stokes basis,
%$(\mathbf{\bar{b}_\perp},\mathbf{\bar{F}})$,
In the parallel-transported polarization basis of the fluid frame,
the transfer equation reads
\bea
\label{eq:polareqgyoto}
\frac{\dd}{\dd s}
\left( \begin{array}{c}
         I_\nu^\mathrm{em;\parallel trans}    \\
         Q_\nu^\mathrm{em;\parallel trans}    \\
         U_\nu^\mathrm{em;\parallel trans}       \\
         V_\nu^\mathrm{em;\parallel trans}
       \end{array} \right)
     &=&
     \mathbf{R}\left(\chi_\mathrm{e}\right)
     \left( \begin{array}{c}
              j_{\nu,I}    \\
              j_{\nu,Q}    \\
              j_{\nu,U}       \\
              j_{\nu,V}        \end{array} \right)
          \\ \nn
          &&-
          \mathbf{R}\left(\chi_\mathrm{e}\right)
          \left( \begin{array}{cccc}
                   \alpha_{\nu,I}        &   \alpha_{\nu,Q}  &  \alpha_{\nu,U}  &  \alpha_{\nu,V}   \\
                   \alpha_{\nu,Q}          &   \alpha_{\nu,I}  &r_{\nu,V} & -r_{\nu,U}  \\
                   \alpha_{\nu,U}          &  -r_{\nu,V} & \alpha_{\nu,I} & r_{\nu,Q}  \\
                   \alpha_{\nu,V}          &  r_{\nu,U} & -r_{\nu,Q} &  \alpha_{\nu,I}   \end{array} \right)
               \mathbf{R}\left(-\chi_\mathrm{e}\right)
               \left( \begin{array}{c}
                        I_\nu^\mathrm{em;\parallel trans}    \\
                        Q_\nu^\mathrm{em;\parallel trans}    \\
                        U_\nu^\mathrm{em;\parallel trans}       \\
                        V_\nu^\mathrm{em;\parallel trans}        \end{array} \right),
\eea
where the superscript 'em;$\parallel$trans' reminds that we are dealing with
Stokes parameters defined in the parallel-transported polarization
basis of the emitting fluid frame, and
\be
\label{eq:rotmatrix}
\mathbf{R}\left(\chi_\mathrm{e}\right) = \left( \begin{array}{cccc}
     1        &   0  &  0 &  0   \\
      0          &   \cos 2\chi_\mathrm{e}  & -\sin 2\chi_\mathrm{e} & 0  \\
     0          &  \sin 2\chi_\mathrm{e}  & \cos 2 \chi_\mathrm{e} & 0  \\
    0          &  0 & 0 &  1   \end{array} \right)
\ee
is a rotation matrix describing the rotation by the
angle $\chi_\mathrm{e} \equiv EVPA_\mathrm{e}$, the emission EVPA,
between the synchrotron
and the parallel-transported polarization bases. This is the exact same
transformation as that described by Eq.~\ref{eq:StokesRot}.
%,
%where the 'obs' and 'em' frames considered here correspond to the unprimed
%and primed bases, respectively, of Eq.~\ref{eq:StokesRot}.

Solving Eq.~\ref{eq:polareqgyoto} is a well-known problem that we briefly
discuss in~\ref{app:solve_RT}.
The corresponding Stokes parameters in the observer frame then follow from
\be
X^\mathrm{obs}_\nu = X^\mathrm{em;\parallel trans}_\nu \, \left( \frac{\nu^\mathrm{obs}}{\nu^\mathrm{em}}\right)^3,
\ee
similarly as in Eq.~\ref{eq:Liouville}, where $X$ is either of the Stokes parameters.

\subsubsection{Polarized synchrotron coefficients}

We now need to express the synchrotron coefficients in the synchrotron polarization basis. In this basis, the transfer coefficients for the Stokes parameter U, i.e. $j_{\nu,U}$, $\alpha_{\nu,U}$ and $r_{\nu,U}$ are zero by definition. However, the computation of emission, absorption and rotation synchrotron coefficients for the others Stokes parameters, from an arbitrary distribution of electrons could be quite heavy. Indeed, even for isotropic distributions, the computation of the emission coefficients require a double integral~\citep{RL79} and the others are even more complex using the susceptibility tensor~\citep[see Appendix B of ][]{Pandya2021}. Fortunately, \citet{Pandya2021, Dexter2016, Huang2011} derived fitted formulae for the emissivities, absorptivities and rotativities for well-defined isotropic distributions of electrons : Thermal (Maxwell-Jüttner), Power Law or Kappa (thermal core with a power law tail). We choose to implement the formulae of~\citet{Pandya2021} in \gy to compute the synchrotron coefficients as they are the only one who provides formulae for a kappa distribution. These formulae are valid for a specific range of parameters.

For a thermal distribution, parametrized by the dimensionless temperature $\Theta_e = k_B T / m_e c^2$, the fits are accurate for $3 < \Theta_e < 40$ and for $\nu / \nu_c \gg 1$ with $\nu_c=eB/(2\pi m_e c)$ the cyclotron frequencies~\citep{Pandya2021}.

For a Power Law distribution, parametrized by a minimum and maximum Lorentz factor, $\gamma_\mathrm{min}$ and $\gamma_\mathrm{max}$ respectively, and by a power law index $p$, the fits are accurate for $\gamma_\mathrm{min} < 10^2$, $1.5 < p < 6.5$ and, as before, for $\nu / \nu_c \gg 1$~\citep{Pandya2021}.

The Kappa distribution is characterized by two parameters $w$ (equivalent to the dimensionless temperature) and $\kappa = p + 1$. Contrary to the other distributions where the fits are continuous functions of the parameters, the fits for rotation coefficients for the Kappa distribution have been done for four specific values of $\kappa = (3.5, 4.0, 4.5, 5.0)$ and are not defined for any other value. The fits are valid while $3 < w < 40$, $\nu / \nu_c \gg 1$ and $X_\kappa \gg 10^{-1}$ where $X_\kappa=\nu/ \nu_c (w \kappa)^2 \sin \theta$ with $\theta$ the angle between the magnetic field vector and the photon tangent vector~\citep{Pandya2021}.

For some tests in section~\ref{sec:tests}, we will use the formulae from~\citet{Dexter2016}, especially for the comparison with the ray-tracing code \textsc{ipole} which use the formulae of~\citet{Dexter2016} (as the code \textsc{grtrans}). The order of the maximum relative error between all the fits in~\citet{Pandya2021} and the true values is of 30\%. For the typical parameters of the accretion flow of Sgr~A*, the difference between the formulae of~\citet{Pandya2021} and~\citet{Dexter2016} is lower or equal to $10\%$.

\section{Tests}\label{sec:tests}

%Could we think of providing online some jupyter notebook where a polarized
%computation would be done wit
%I can think of:h plenty of comments?
%NA: Yes could be interesting, if i have time, I can do it.

\subsection{Test of the parallel transport}
The first test that we have to make is to check that the observer polarization basis' vectors, i.e. $\mathbf{\bar{n}}$ and $\mathbf{\bar{w}}$, are well parallel transported along the null geodesics.
The parallel transport equation given by Eq.~\ref{eq:paralleltp}
is fully general and agnostic about the particular spacetime considered. However, in the
special case of the Kerr spacetime, it is well known that the
special algebraic kind of the spacetime allows the existence
of the Walker-Penrose constant~\citep{WalkerPenrose70}, defined as follows.
If $\mathbf{k}$ is the tangent vector to a null geodesic,
and if $\mathbf{f}$ is a vector orthogonal to $\mathbf{k}$
and parallel-transported along $\mathbf{k}$, then
the following complex quantity
\bea
\label{eq:WPcst}
K_1 - i K_2 =&& (r-ia \cos\theta) \bigg[ (k^t f^r - k^r f^t) + a \sin^2\theta (k^r f^\pp - k^\pp f^r)\bigg. \\ \nn
&& \bigg. - i \sin\theta \left\{(r^2+a^2) (k^\pp f^\theta - k^\theta f^\pp) -a(k^t f^\theta - k^\theta f^t)\right\}\bigg], \\ \nn
\eea
here expressed in Boyer-Lindquist coordinates,
is conserved along the null geodesic.
Thus, knowing the evolution of $\mathbf{k}$ along the null geodesic,
the vectors $\mathbf{\bar{n}}$ and $\mathbf{\bar{w}}$ (that obviously fulfill
the conditions on $\mathbf{f}$ in Eq.~\ref{eq:WPcst})
can be immediately obtained
without further computation by using this constant.
This Kerr-specific result is only used for testing purposes in \gy.
The code is agnostic about the spacetime and does not use this property.

We thus check the conservation of $K_1$ and $K_2$ for specific geodesics and obtain a conservation to within $10^{-5}$ for default integration parameters of \gy. This can be improved by setting a lower tolerance value for the integration steps, but at the cost of a longer calculation time.

\subsection{EVPA calculation test}
Parallel transport having been tested, the observer polarization basis is well defined in the rest frame of the emitter $(\mathbf{\bar{K}},\mathbf{\bar{w}'}, \mathbf{\bar{n}'})$, through Eq.~(\ref{eq:np}). As said in section~\ref{sec:parallel_transport_EVPA}, the natural basis to express the synchrotron coefficients is $(\mathbf{\bar{K}},\boldsymbol{\bar{b}_\perp}, \mathbf{\bar{F}})$ with $\mathbf{\bar{K}}$ a common vector between the two bases. Thus, to
%express the coefficients in observer basis from the natural synchrotron
express the radiative transfer in the observer-related basis,
rather than in the synchrotron basis, we just need to apply the rotation matrix
defined in Eq.~(\ref{eq:rotmatrix}). The angle between these two bases corresponds to the emission EVPA (see section~\ref{sec:parallel_transport_EVPA}).

% To check that the computation of this angle works correctly, we define a basic setup for which the result is trivial.
We define a simple setup to check the computation of this crucial angle.
We consider a Page-Thorne disk~\citep[geometrically thin, optically thick;][]{PT1974} in a Minkowski metric (to avoid GR effects), seen face-on.
%(inclination of $0.01$° as it should not be zero).
We consider two magnetic field configurations $\mathbf{b}$, toroidal and radial. % A vertical magnetic field would result in
% a pitch
%an emission
%angle $\theta_B$ (angle between the direction of the photon's emission and the magnetic field) close to zero as seen face-on, and to a close to vanishing polarization vector.
%It is more interesting to consider a toroidal or radial magnetic field,
%The results are thus less trivial compared to the two previous cases
%which will result in an emission angle close to 90°.
Expressed in Boyer-Lindquist coordinates, the two magnetic field configurations read, in the rest frame of the emitter% as follow
\begin{equation}
    \mathrm{\textbf{Toroidal:}} \:
    \label{eq:B_toroidal}
    b^\alpha = \left\{
    \begin{array}{l}
        b^t = \sqrt{\frac{g_{\varphi \varphi}}{g_{tt}} \frac{\Omega^2}{g_{tt} +g_{\varphi \varphi}\Omega^2}}, \\
        b^r = 0, \\
        b^\theta = 0, \\
        b^\varphi = \sqrt{\frac{g_{tt}}{g_{\varphi \varphi}} \frac{1}{g_{tt} + g_{\varphi \varphi}\Omega^2}},
    \end{array} \right.
\end{equation}
where $\Omega = u^\varphi/u^t$ and $\mathbf{u}$ is the 4-velocity of the emitting fluid assumed to be Keplerian, and
\begin{equation}
    \mathrm{\textbf{Radial: }} \:
    \label{eq:B_radial}
    b^\alpha = \left\{
    \begin{array}{l}
        b^t = 0, \\
        b^r = \sqrt{\frac{1}{g_{rr}}}, \\
        b^\theta = 0, \\
        b^\varphi = 0.
    \end{array} \right.
\end{equation}

In the toroidal case, for all azimuthal angles of the disk, the wave-vector $\mathbf{\bar{K}}$ is made of two components, one almost vertical (face-on view), and one azimuthal component resulting from special-relativistic aberration~\citep{vincent23}. The magnetic field $\mathbf{b}$ is
%orthogonal to $\mathbf{\bar{K}}$
in the toroidal direction. Thus, the resulting polarization vector $\mathbf{\bar{F}} = \mathbf{\bar{K}} \times \boldsymbol{\bar{b}_\perp}$ is in the radial direction. Similarly, for the radial magnetic field, the resulting polarization vector is in the toroidal direction.

We consider an arbitrary $I_\nu=1$ emission for unpolarized intensity. \gy computes images for the four Stokes parameters from which we can compute the orientation of the polarization vector, i.e.~the EVPA. As we are only interested by the computation of the EVPA for the moment, i.e.~without radiative transfer, we assume a fully linearly polarized radiation by taking $Q_\nu =  I_\nu \, \cos\left(2*EVPA\right)$ and $U_\nu =  I_\nu \, \sin\left(2*EVPA\right)$. This means that we do not take into account any absorption nor Faraday rotation.

\begin{figure}[htbp]
    \centering
    \includegraphics[width=\textwidth]{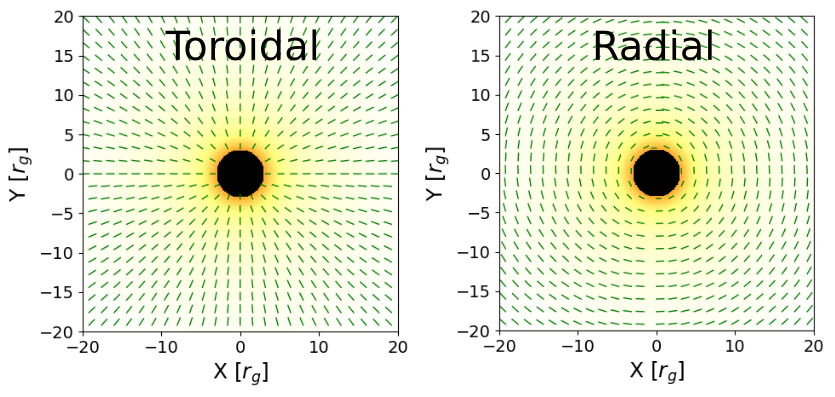}
    \caption{Image of total (unpolarized) intensity of a Page-Thorne disk with Keplerian orbit in a Minkowski space-time seen face-on for two magnetic field configurations (toroidal left, radial right). The inner radius is at $r=3\, r_g$. The green lines represent the orientation of polarization vectors.}
    \label{fig:basic_test_EVPA}
\end{figure}

We show in the left panel of Fig.~\ref{fig:basic_test_EVPA} in background the unpolarized images of the setup described above and the polarization vector with the green ticks.
The position angle of these vectors is the observed EVPA. As expected, the polarization vectors are radial for a toroidal magnetic field and toroidal for radial magnetic field.

\begin{figure}[htbp]
    \centering
    \includegraphics[width=\textwidth]{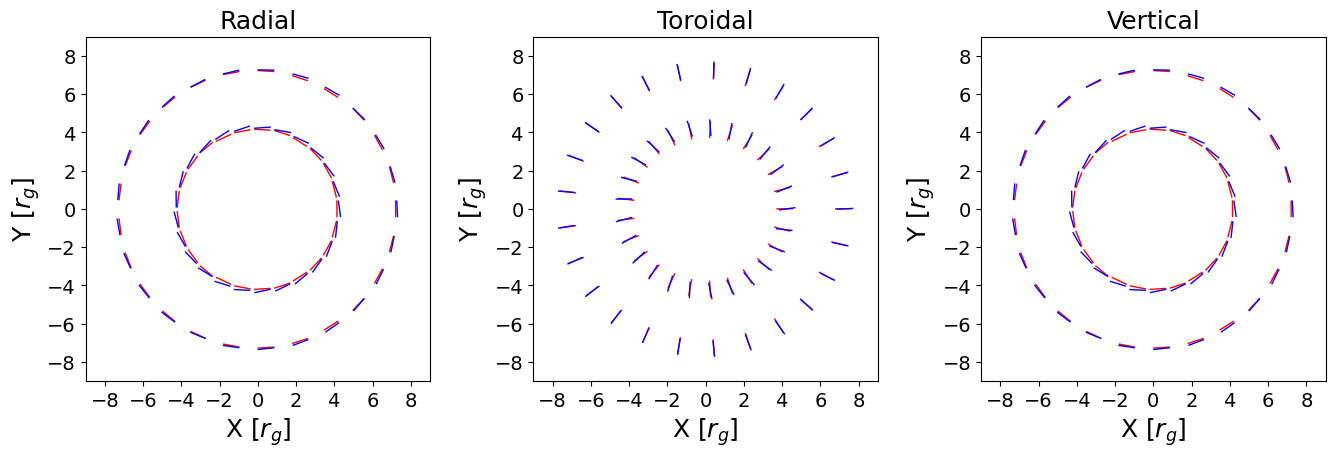}
    \caption{Polarized tick plots for three idealized magnetic field configurations: radial (\textbf{left}), toroidal (\textbf{middle}), and vertical (\textbf{right}) from a geometrically thin ring seen by an almost face-on observer $i = 0.1^\circ$. The fluid is comoving with the Zero Angular Momentum Observer frame.
          %modeled with a null 3D velocity (ZAMO).
          Each plot shows two spins ($a = 0$ and $a = -0.99$ in red and blue,
          respectively) as two emission radii ($r_1 = 3 \, r_g$ and $r_2 = 6 \, r_g$, corresponding to the inner and outer rings, respectively).}
    \label{fig:Gelles}
\end{figure}

We now want to make a similar test in curved spacetime using a Kerr metric. To validate the computation of EVPA in \gy, we compute the polarization from a geometrically thin ring as in~\citet{Gelles2021} taking as previously $I_\nu=1$, $Q_\nu =  I_\nu \, \cos\left(2*EVPA\right)$ and $U_\nu =  I_\nu \, \sin\left(2*EVPA\right)$. \citet{Gelles2021} consider a synchrotron emission that we discard here,
our only interest being in testing the EVPA. Here, we are only interested in the orientation of the polarization vector, that is, in the EVPA, at $r_1=3\, r_g$ and $r_2 = 6 \, r_g$ (the inner and outter radius of the ring). Figure~\ref{fig:Gelles} shows the tick plots for three magnetic field configurations : radial, toroidal and vertical with the same setup as in Fig.~1 of~\citet{Gelles2021},
% (the 3D velocity of the fluid is zero, see~\citet{Gelles2021} for more details).
the fluid being assumed to be comoving with the Zero Angular Momentum Observer frame.
The vertical configuration implemented in \gy reads
\begin{equation}
    \mathrm{\textbf{Vertical}} \:
    \label{eq:B_vertical}
    b^\alpha = \left\{
    \begin{array}{l}
        b^t = 0, \\
        b^r = \frac{\cos{\theta}}{\sqrt{g_{rr}}} , \\
        b^\theta = \frac{\sin{\theta}}{\sqrt{g_{\theta \theta}}} , \\
        b^\varphi = 0.
    \end{array} \right.
\end{equation}

The results of \gy shown in Fig.~\ref{fig:Gelles} are in perfect agreement with the ones in the Fig.~1 in~\citet{Gelles2021}. We note that~\citet{Gelles2021} scale the length of the ticks by the observed synchrotron intensity, while our ticks are all of unit length: we are only interested in the EVPA. This confirms that the calculation of the EVPA works correctly and we can now test the radiative transfer part and compare the results with another ray-tracing code.

\subsection{Comparison with \textsc{ipole}}
\label{sec:ipole}
To check that all parts of our code work correctly, we compare the results of \gy with the ones of another polarized ray-tracing code : \textsc{ipole}~\citep{ipole}. We will focus on polarized observables of a thick disk around a Schwarzschild black hole. We take power law profiles of the physical quantities following~\citet{Vos2022}. To compute the emission, absorption and rotation coefficients for the radiative transfer, we assume a thermal distribution of electrons~\citep[as in][]{Vos2022}, and, for this comparison only, we use the fitting formula of~\citet{Dexter2016} as used in \textsc{Ipole}~\citep[we remind that \gy implements the
formulae from][]{Pandya2021}.
%\footnote{These formulae have been implemented in \gy for these test. However, in regular use, \gy use the formulae from~\citet{Pandya2021}}.

We compared the three magnetic field configurations (toroidal, radial and vertical) described in~\citet{Vos2022} at two inclinations, close to face-on with $i=20$° and close to edge-on with $i=80$°. We define, as in~\citet{Prather2023}, the normalized mean squared error (NMSE) as
\begin{equation}
  NMSE(A, B) = \frac{\sum | A_j - B_j |^2}{\sum | A_j |^2}
\end{equation}
where $A_j$ and $B_j$ are the intensities of a particular Stokes parameter in two images at pixel $j$. The results of \gy are in perfect agreement with \textsc{Ipole} with a NMSE $< 10^{-4}$ for all configurations and Stokes parameters except for Stoke U in the radial cases at high inclination for which the NMSE is around $10^{-3}$. This can be compared to the worst NMSE of $\sim 0.01$ obtained in the code comparison made in~\citet{Prather2023} showing the perfect agreement between \gy and \textsc{Ipole}.

\begin{figure}[htbp]
    \centering
    \includegraphics[width=0.95\textwidth]{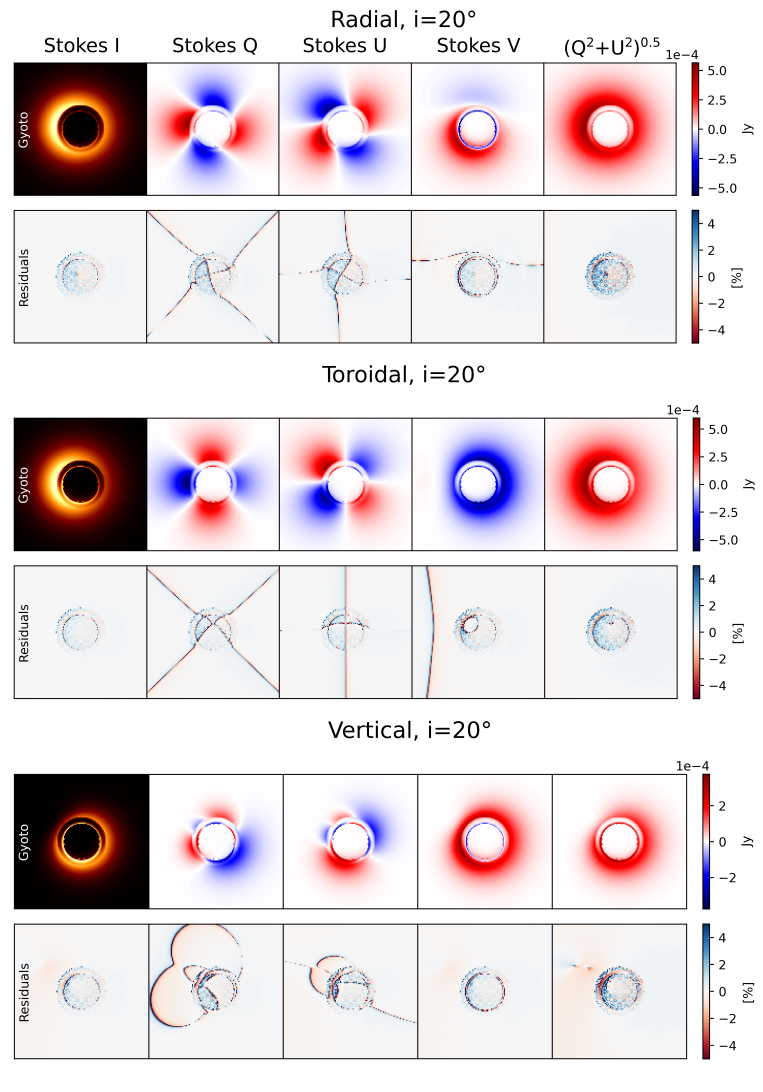}
    \caption{The first, third and fifth lines show images from left to right of the four Stokes parameters (I, Q, U, V) generated by \gy and $Q^2+U^2$ as the last column for the three magnetic field configuration at low inclination ($i=20$°). Their relative error maps with \textsc{ipole} images are shown in the second, fourth and sixth lines. The relative errors scale is linear between $\pm 5\%$.}
    \label{fig:Stokes_err_maps_20deg}
\end{figure}

We also performed pixel-to-pixel comparisons, not restricting our
comparison to integrated quantities like the NMSE.
Fig.~\ref{fig:Stokes_err_maps_20deg} illustrates this for
128x128 pixels images of the four Stokes parameters computed by \gy and their relative difference with \textsc{Ipole}, in a field of view of $40 \, r_g$,
for the three magnetic field configurations described above,
at low inclination $i=20^\circ$.
The relative error map is very close to zero for the vast majority of pixels,
with typical values $\lesssim 0.1 \%$.
Higher values are only reached on specific tracks, that correspond to the zeroes
of the corresponding Stokes parameters (as is clear by comparing with the
panels showing the maps of the corresponding Stokes parameters).
It is thus not surprising to get higher residuals there, and it does not
affect the radiative transfer, given that the Stokes parameters are anyway
close to zero in these regions.
Besides these tracks corresponding to the zeroes of the Stokes parameters,
the interior of the ``shadow'' region (i.e.~geodesics that asymptotically
approach the horizon when backward ray traced) lead to higher error.
This is due to the stop condition of the geodesic integration that differs
in the two codes. Given that this part of the image is anyway strongly
redshifted and leads to a very low flux, this has no impact on the
field-of-view integrated comparison of the NMSE.

%In most of the relative error maps, the error is close to zero. Some areas in the Stokes Q, U and V show a higher (but still low) relative error due to their inversion of sign. Indeed, for theses pixels their values are close to zero leading to a more important relative error compared to the rest of the field of view but do not affect the NMSE results and are thus negligible.

\section{Conclusion}

This article presents the polarized version of the
ray-tracing code \gy. After reviewing the formalism for polarized
GRRT, our main aim is to explain the details of our implementation,
and to provide tests of our code. In particular, we have shown that in the
framework of the GRRT computation of a geometrically thick, optically thin
accretion flow, we find results in perfect agreement with the
\textsc{ipole} code.

Polarized GRRT is of fundamental importance for interpreting current
and future observed data, in particular that of GRAVITY, the EHT,
the polarized loops of ALMA associated with Sgr~A* flares, or the
data of IXPE. Properly interpreting these data is key to better understanding
the properties of plasmas in the extreme environments of black holes,
and might offer new interesting probes of strong-field gravity.

The polarized \gy code is public,
actively maintained, and in constant development
for offering ever more diversed setups for relativistic astrophysics.
The recent polarized version of the code is accompanied by a
python notebook, available at \url{https://github.com/gyoto/Gyoto/blob/master/doc/examples/Gyoto_Polar_example.ipynb},
offering a quick and user-friendly first example
of the new environment.

%We have presented the new version of the publicly available ray-tracing code \gy which now solves the polarized radiative transfer equation for any given (analytic or numerically computed) metric. As our primary targets, like M87 and Sgr~A* emit synchrotron radiation we presented in this paper how the code compute polarized intensities from this emission mechanism. However, the computation of polarized intensities in the code is generic allowing the user to model any emission process (black body with scattering for example).

%The code, written in C++, already contain a large variety of astrophysical sources models and was built to be modular, so that any interested user can easily create its own model, and user-friendly with an XML and python interface. An example of polarized computation from a thick disk around a Kerr black hole is given in a notebook available at~\url{}{}.

%Ray-tracing codes like \gy is also of particular interest as post-processing of accretion-ejection simulations (GRMHD or GRPIC) to generate synthetic images, spectra, light curves or Q-U loops. \gy is already able to do so with the GRMHD code GR-AMRVAC~\citep{NOVAs}. This will be generalized for simulations code in the future.

\appendix

\section{Observer's screen polarization basis}
\label{app:observertriad}

\begin{figure}[htbp]
\centering
\includegraphics[width=0.5\textwidth]{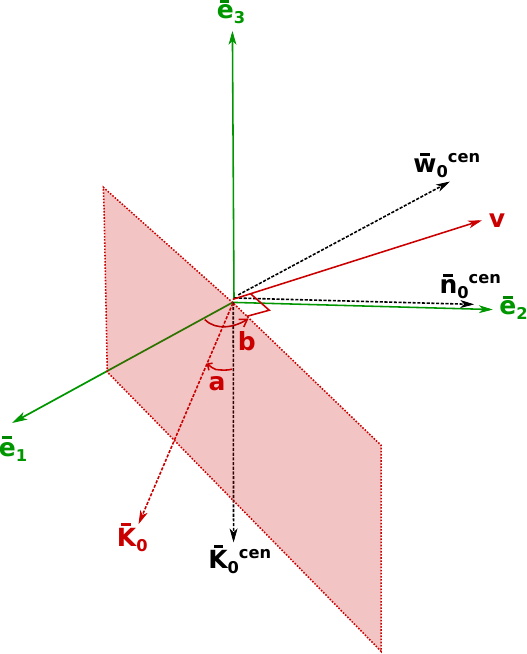}
\caption{Local observer's basis $(\mathbf{\bar{e}_1},\mathbf{\bar{e}_2},\mathbf{\bar{e}_3})$,
  in green, as represented in Fig.~\ref{fig:obsframe}.
  The polarization basis corresponding to the central pixel of the
  screen, that is, to a purely radial direction of incidence,
  is $(\mathbf{\bar{K}_0^\mathrm{cen}},\mathbf{\bar{w}_0^\mathrm{cen}},
  \mathbf{\bar{n}_0^\mathrm{cen}})$, and is aligned with the oberver's
  local basis vectors. When considering a non-central pixel, the direction
  of photon incidence is no longer purely radial. The corresponding vector
  $\mathbf{\bar{K}_0}$ is obtained from $\mathbf{\bar{K}_0^\mathrm{cen}}$ by a rotation
  of angle $a$ around the vector $\mathbf{v}$ (in red) normal to the red plane
  that contains $\mathbf{\bar{e}_3}$ and makes an angle $b$ with $\mathbf{\bar{e}_1}$.
  The same applies for $\mathbf{\bar{w}_0}$ and $\mathbf{\bar{n}_0}$.}
\label{fig:polarbasis}
\end{figure}
A general vector $\mathbf{\bar{K}_0}$, defined by the two spherical angles
$(a,b)$, see the right panel of Fig.~\ref{fig:obsframe},
is obtained by rotating
the vector $\mathbf{\bar{K}_0^\mathrm{cen}}$ by an angle $a$ in a plane
containing $\mathbf{\bar{e}_3}$ and making an angle $b$ with the
$\mathbf{\bar{e}_1}$ vector (see the red plane in Fig.~\ref{fig:polarbasis}).
This corresponds to a rotation of angle $a$ around the vector $\mathbf{v} = -\sin b \, \mathbf{\bar{e}_1} + \cos b \,\mathbf{\bar{e}_2} $. The corresponding rotation
matrix reads
\be
R(a,b) = \left( \begin{array}{cccc}
                  \sin^2 b \left(1 - \cos a \right) + \cos a        &   -\sin b \cos b \left(1-\cos a \right)  &  \cos b \sin a    \\
                  -\sin b \cos b \left(1-\cos a \right)         &   \cos^2 b \left(1 - \cos a \right) + \cos a  & \sin b \cos a  \\
                  -\cos b \sin a          &  -\sin b \sin a & \cos a  \end{array} \right).
\ee
The general vector $\mathbf{\bar{K}_0}$ then reads $\mathbf{\bar{K}_0} = R(a,b) \,\mathbf{\bar{K}_0^\mathrm{cen}}$, and similarly for the two other polarization vectors.
It is then easy to express the general local unit West and North directions
that complete the local incoming photon direction given by Eq.~\ref{eq:K0def}.
They read
\bea
\mathbf{\bar{w}_0} &=& \left[ - \sin^2 b \left(1 - \cos a \right) - \cos a\right] \mathbf{\bar{e}_1} + \sin b \cos b \left(1 - \cos a \right) \mathbf{\bar{e}_2} \\ \nn
&& + \cos b \sin a \, \mathbf{\bar{e}_3}, \\ \nn
\mathbf{\bar{n}_0} &=&  -\sin b \cos b \left(1 - \cos a \right)\mathbf{\bar{e}_1} + \left[  \cos^2 b \left(1 - \cos a \right) + \cos a\right] \mathbf{\bar{e}_2} \\ \nn
&&- \sin b \sin a \,\mathbf{\bar{e}_3}. \\ \nn
\eea

\section{Electron gyration and polarization vector direction}
\label{app:PolarVec}

Our discussion of the polarization vector in section~\ref{sec:parallel_transport_EVPA}
took place in the fluid frame. Here, we need to discuss a third natural frame (after the
observer frame and the fluid frame) naturally
associated with the GRRT problem,
namely the particle frame, the frame comoving with an individual
electron swirling around the magnetic field lines and emitting synchrotron radiation
as a consequence of this accelerated motion (see Fig.~\ref{fig:setup}). The description of this
infinitesimal
\footnote{Infinitesimal as compared to the natural scale of our problem coinciding
  with the gravitational radius.}
motion is the topic of this appendix. Our goal is
to express the radiation electric field of an individual electron, as measured in the
fluid frame. The direction of this vector should coincide with that of the polarization
vector, which was introduced in Eq.~\ref{eq:Femit} without any reference to the electron's motion,
nor to its radiation field.

\begin{figure}[htbp]
\centering
\includegraphics[width=0.5\textwidth]{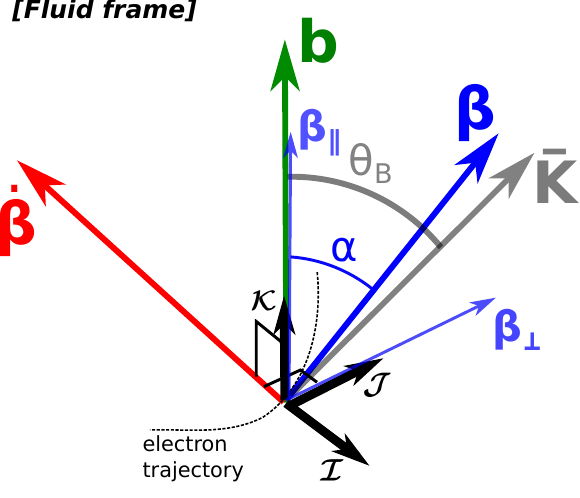}
\caption{Geometry of the synchrotron-emitting electron motion
  in the fluid frame. $\boldsymbol{\beta}$ is the velocity 3-vector, $\mathbf{\bar{K}}$ is the direction of emission, $\mathbf{b}$ is the magnetic vector, all these
  quantities defined in the fluid frame. The emission angle
  between the magnetic field direction and the direction of emission
  is called $\theta_B$ while $\alpha$ is the pitch angle between
  the magnetic field direction and the velocity of the electron.
}
\label{fig:PolarVec}
\end{figure}

Let us consider the standard picture of synchrotron emission by
a relativistic electron illustrated in
Fig.~\ref{fig:PolarVec}, using the same notation and following the
derivation of~\citet{westfold59}.
In the fluid frame, we consider a direct orthonormal triad,
%\textbf{CHANGE NOTATION k is the wave vector}.
$(\boldsymbol{\mathcal{I}},\boldsymbol{\mathcal{J}},\boldsymbol{\mathcal{K}})$, such that $\boldsymbol{\mathcal{I}}$
is antiparallel to the acceleration vector at some initial
time $t=0$ (coinciding with the proper time of the fluid frame),
$\boldsymbol{\mathcal{K}}$ is along the ambient magnetic field $\mathbf{b}$ (measured
in the fluid frame),
and $\boldsymbol{\mathcal{J}}$ completes the triad.
We call $\boldsymbol{\beta}$ the velocity 3-vector of the electron
in the fluid frame and $\boldsymbol{\dot{\beta}}$
the corresponding acceleration. A synchrotron
wave is emitted by the accelerated electron in the unit direction $\mathbf{\bar{K}}$
(measured in the fluid frame). The emission angle
$\theta_B$ and the pitch angle $\alpha$ are illustrated in Fig.~\ref{fig:PolarVec}.

The velocity and acceleration read
\bea
\boldsymbol{\beta} &=& \beta \sin \alpha \left(- \sin \omega_B t \, \boldsymbol{\mathcal{I}} + \cos \omega_B t \, \boldsymbol{\mathcal{J}}\right) + \beta \cos \alpha \,\boldsymbol{\mathcal{K}}, \\ \nn
\boldsymbol{\dot{\beta}} &=& - \omega_B \beta \sin \alpha  \left(\cos \omega_B t \, \boldsymbol{\mathcal{I}} + \sin \omega_B t \, \boldsymbol{\mathcal{J}}\right), \\ \nn
\eea
where $\beta$ is the velocity of the electron in units of the speed of light,
and $\omega_B$ is the cyclotron gyrofrequency.
It is easy to check that at $t=0$, the projection of the velocity vector
orthogonal to $\mathbf{b}$ is along $\boldsymbol{\mathcal{J}}$, and the acceleration
vector is along $-\boldsymbol{\mathcal{I}}$, which is the setup illustrated in Fig.~\ref{fig:PolarVec}.

The radiation field of the moving charge in the fluid frame satisfies
\be
\mathbf{E_\mathrm{rad}} \propto \mathbf{\bar{K}} \times \left[ \left(\mathbf{\bar{K}} - \boldsymbol{\beta}\right) \times \boldsymbol{\dot{\beta}}\right].
\ee

Let us consider the unit direction of emission of a synchrotron photon,
written in full generality as
\be
\mathbf{\bar{K}} = a \,\boldsymbol{\mathcal{I}} + b \, \boldsymbol{\mathcal{J}} + c \, \boldsymbol{\mathcal{K}},
\ee
where $(a,b,c)$ are arbitrary real numbers such that $\mathbf{\bar{K}}$
is a unit vector. It is easy to show that
\be
\mathbf{E_\mathrm{rad}} \cdot \boldsymbol{\mathcal{K}} \propto \beta \cos \alpha - c.
\ee
The lhs quantity represents the projection of the radiating electric field
along the ambient magnetic field direction.
It is well known that the beaming effect leads to the radiation being
confined within a narrow cone around the pitch angle of the relativistically
moving electron~\citep[see Fig.~6.5 of][]{RL79}. This exactly means
that
\be
\beta \cos \alpha - c \approx 0,
\ee
such that the radiating electric field is orthogonal to the ambient
magnetic field. It follows that
\be
\mathbf{E_\mathrm{rad}} \propto \mathbf{\bar{K}} \times \mathbf{b},
\ee
so that the rhs can be used to define the direction of the synchrotron
polarization vector, as done in Eq.~\ref{eq:Femit}.

\section{Solving the polarized radiative transfer equation}
\label{app:solve_RT}

Let us start with the unpolarized radiative transfer equation in the emitter frame
\be
\label{eq:radtransf}
\frac{\dd I}{\dd s} = - \alpha I + j
\ee
with obvious notations. The formal solution reads
\be
\label{eq:radtransfformal2}
I(\tau) = \int_{s_0}^s \mathrm{exp}\left( -(\tau - \tau')\right) S(\tau') \dd \tau'
\ee
where $S=j/\alpha$ is the source function, $\dd\tau=\alpha\, \delta s$ is the optical depth, $s$ is the proper length in the emitter frame
and $s_0$ is some initial value of $s$ where the intensity is assumed to be zero. This equation can equivalently be written
\be
\label{eq:radtransfformal3}
I(s) = \int_{s_0}^s \mathrm{exp}\left( -\alpha(s - s')\right) j(s') \dd s'
\ee
where $\alpha$ is assumed constant over the integration interval. Considering a small range
$\delta s = s - s'$ between some previous location $s'$ and some current location $s$,
over which interval $j$ and $\alpha$ can be considered constant in a realistic problem, the increment
of intensity reads
\be
\label{eq:deltaInopol}
\delta I(s) = j(s)\,\delta s\,\mathrm{exp}\left(-\alpha(s)\,\delta s\right).
\ee
Let us now come back to the polarized version of the radiative transfer equation
(Eq. \ref{eq:polareqgyoto}) which reads
\be
\label{eq:radtransfmat}
\frac{\dd \boldsymbol{\mathcal{I}}}{\dd s} = - \boldsymbol{\mathcal{K}} \boldsymbol{\mathcal{I}} + \boldsymbol{\mathcal{J}}
\ee
where $\boldsymbol{\mathcal{I}}$ is the vector of Stokes parameters and
\be
\boldsymbol{\mathcal{K}} =
\mathbf{R}\left(\chi\right) \left( \begin{array}{cccc}
     \alpha_I        &   \alpha_Q  &  \alpha_U  &  \alpha_V   \\
      \alpha_Q          &   \alpha_I  &r_V & -r_U  \\
     \alpha_U          &  -r_V & \alpha_I & r_Q  \\
    \alpha_V          &  r_U & -r_Q &  \alpha_I   \end{array} \right) \mathbf{R}\left(-\chi\right);
    \qquad
    \boldsymbol{\mathcal{J}} =
       \mathbf{R}\left(\chi\right)
   \left( \begin{array}{c}
     j_I    \\
     j_Q    \\
    j_U       \\
   j_V        \end{array} \right)
\ee
with $\mathbf{R}\left(\chi\right)$ being the rotation matrix given
in Eq.~\ref{eq:rotmatrix}.
Its formal solution is the direct generalization of Eq.~\ref{eq:radtransfformal2}, provided that
$\boldsymbol{\mathcal{K}}$ is constant with $s$:
\be
\label{eq:sol}
\boldsymbol{\mathcal{I}}(s) = \int_{s_0}^{s}  \mathrm{exp}\left(-\boldsymbol{\mathcal{K}}(s-s')\right) \boldsymbol{\mathcal{J}}(s') \dd s'.
\ee
Let us introduce the matrix
\be
\label{eq:Omat}
\mathbf{O}(s,s') = \mathrm{exp}\left(-\boldsymbol{\mathcal{K}}(s-s')\right).
\ee
Over a small interval of proper length $\delta s=s-s'$, over which the absorption
matrix and emission vector can be considered constant, the elementary increase of Stokes parameters is
\be
\delta \boldsymbol{\mathcal{I}} (s) = \mathbf{O}(\delta s) \boldsymbol{\mathcal{J}} (s) \delta s
\ee
which is the direct generalization of Eq.~\ref{eq:deltaInopol}.
This is the equation used in \gy to update the Stokes parameters
along the light ray.

We still need to compute the exponential of matrix appearing
in Eq.~\ref{eq:Omat}. \citet{LDI85}
have given an expression for this matrix.
It reads
\bea
\mathbf{O}(\delta s) = \mathrm{exp}\left(-\alpha_I \delta s\right) &&\left\{ \left[ \mathrm{cosh} \left(\Lambda_1 \delta s\right) + \cos \left( \Lambda_2 \delta s\right)\right] \mathbf{M_1}/2 \right.\\ \nn
 && \left. - \sin\left( \Lambda_2 \delta s\right)\mathbf{M_2} \right.\\ \nn
 &&\left.- \mathrm{sinh}\left( \Lambda_1 \delta s\right)\mathbf{M_3} \right. \\ \nn
 &&  \left. +  \left[ \mathrm{cosh} \left(\Lambda_1 \delta s\right) - \cos \left( \Lambda_2 \delta s\right)\right] \mathbf{M_4}/2 \right\} \\ \nn
\eea
with
\begin{landscape}
\bea
\mathbf{M_1} &=& \mathbf{1}, \\ \nn
\mathbf{M_2} &=& \frac{1}{\Theta}\left( \begin{array}{cccc}
     0       &   \Lambda_2\tilde{\alpha}_Q - \sigma \Lambda_1 \tilde{r}_Q  & \Lambda_2\tilde{\alpha}_U - \sigma \Lambda_1 \tilde{r}_U     &  \Lambda_2\tilde{\alpha}_V - \sigma \Lambda_1 \tilde{r}_V     \\
      \Lambda_2\tilde{\alpha}_Q - \sigma \Lambda_1 \tilde{r}_Q       &   0  & \sigma \Lambda_1 \tilde{\alpha}_V + \Lambda_2 \tilde{r}_V & -\sigma \Lambda_1 \tilde{\alpha}_U - \Lambda_2 \tilde{r}_U  \\
     \Lambda_2\tilde{\alpha}_U - \sigma \Lambda_1 \tilde{r}_U          &  -\sigma \Lambda_1 \tilde{\alpha}_V - \Lambda_2 \tilde{r}_V & 0 & \sigma \Lambda_1 \tilde{\alpha}_Q + \Lambda_2 \tilde{r}_Q \\
    \Lambda_2\tilde{\alpha}_V - \sigma \Lambda_1 \tilde{r}_V          &  \sigma \Lambda_1 \tilde{\alpha}_U + \Lambda_2 \tilde{r}_U &  -\sigma \Lambda_1 \tilde{\alpha}_Q - \Lambda_2 \tilde{r}_Q &  0  \end{array} \right),\\ \nn
\mathbf{M_3} &=& \frac{1}{\Theta}\left( \begin{array}{cccc}
     0       &   \Lambda_1\tilde{\alpha}_Q + \sigma \Lambda_2 \tilde{r}_Q  & \Lambda_1\tilde{\alpha}_U + \sigma \Lambda_2 \tilde{r}_U   &  \Lambda_1\tilde{\alpha}_V + \sigma \Lambda_2 \tilde{r}_V     \\
      \Lambda_1\tilde{\alpha}_Q + \sigma \Lambda_2 \tilde{r}_Q      &   0  & - \sigma \Lambda_2 \tilde{\alpha}_V + \Lambda_1 \tilde{r}_V & \sigma \Lambda_2 \tilde{\alpha}_U - \Lambda_1 \tilde{r}_U  \\
     \Lambda_1\tilde{\alpha}_U + \sigma \Lambda_2 \tilde{r}_U        &  \sigma \Lambda_2 \tilde{\alpha}_V - \Lambda_1 \tilde{r}_V & 0 & -\sigma \Lambda_2 \tilde{\alpha}_Q + \Lambda_1 \tilde{r}_Q \\
    \Lambda_1\tilde{\alpha}_V + \sigma \Lambda_2 \tilde{r}_V        &  -\sigma \Lambda_2 \tilde{\alpha}_U + \Lambda_1 \tilde{r}_U &  \sigma \Lambda_2 \tilde{\alpha}_Q - \Lambda_1 \tilde{r}_Q &  0  \end{array} \right),\\ \nn
\mathbf{M_4} &=& \frac{2}{\Theta} \\ \nn
&\times&\left( \begin{array}{cccc}
     (\tilde{\alpha}^2 + \tilde{r}^2)/2       &   \tilde{\alpha}_V \tilde{r}_U - \tilde{\alpha}_U \tilde{r}_V  & \tilde{\alpha}_Q \tilde{r}_V - \tilde{\alpha}_V \tilde{r}_Q  & \tilde{\alpha}_U\tilde{r}_Q - \tilde{\alpha}_Q \tilde{r}_U    \\
      \tilde{\alpha}_U\tilde{r}_V - \tilde{\alpha}_V\tilde{r}_U      &   \tilde{\alpha}_Q^2 + \tilde{r}_Q^2 -  (\tilde{\alpha}^2 + \tilde{r}^2)/2    & \tilde{\alpha}_Q\tilde{\alpha}_U + \tilde{r}_Q\tilde{r}_U & \tilde{\alpha}_V\tilde{\alpha}_Q + \tilde{r}_V\tilde{r}_Q   \\
     \tilde{\alpha}_V\tilde{r}_Q - \tilde{\alpha}_Q\tilde{r}_V      &  \tilde{\alpha}_Q\tilde{\alpha}_U + \tilde{r}_Q\tilde{r}_U & \tilde{\alpha}_U^2 + \tilde{r}_U^2 -  (\tilde{\alpha}^2 + \tilde{r}^2)/2  & \tilde{\alpha}_U\tilde{\alpha}_V + \tilde{r}_U\tilde{r}_V \\
    \tilde{\alpha}_Q\tilde{r}_U - \tilde{\alpha}_U\tilde{r}_Q        &  \tilde{\alpha}_V\tilde{\alpha}_Q + \tilde{r}_V\tilde{r}_Q &  \tilde{\alpha}_U\tilde{\alpha}_V+\tilde{r}_U\tilde{r}_V &  \tilde{\alpha}_V^2 + \tilde{r}_V^2 -  (\tilde{\alpha}^2 + \tilde{r}^2)/2  \end{array} \right)\\ \nn
\eea
\end{landscape}
where
\bea
\tilde{\alpha}^2 &=& \tilde{\alpha}_Q^2 + \tilde{\alpha}_U^2 + \tilde{\alpha}_V^2, \\ \nn
\tilde{r}^2 &=& \tilde{r}_Q^2 + \tilde{r}_U^2 + \tilde{r}_V^2, \\ \nn
\Lambda_1 &=& \sqrt{\sqrt{\frac{1}{4}(\tilde{\alpha}^2 - \tilde{r}^2)^2 + (\tilde{\alpha}_Q\tilde{r}_Q + \tilde{\alpha}_U\tilde{r}_U + \tilde{\alpha}_V\tilde{r}_V)^2} + \frac{1}{2}(\tilde{\alpha}^2 - \tilde{r}^2)}, \\ \nn
\Lambda_2 &=& \sqrt{\sqrt{\frac{1}{4}(\tilde{\alpha}^2 - \tilde{r}^2)^2 + (\tilde{\alpha}_Q\tilde{r}_Q + \tilde{\alpha}_U\tilde{r}_U + \tilde{\alpha}_V\tilde{r}_V)^2} - \frac{1}{2}(\tilde{\alpha}^2 - \tilde{r}^2)}, \\ \nn
\Theta &=& \Lambda_1^2 + \Lambda_2^2, \\ \nn
\sigma &=& \mathrm{sign}\left(\tilde{\alpha}_Q\tilde{r}_Q + \tilde{\alpha}_U\tilde{r}_U + \tilde{\alpha}_V\tilde{r}_V \right), \\ \nn
\eea
and where the tilde quantities $\tilde{\alpha}_X$, $\tilde{r}_X$
take into account the basis rotation by an angle $\chi$ and read
\be
 \left( \begin{array}{c}
     \tilde{\alpha}_Q    \\
     \tilde{\alpha}_U    \end{array} \right)
   =
   \left( \begin{array}{cc}
     \cos 2\chi        &   -\sin 2\chi   \\
     \sin 2\chi          &   \cos 2\chi  \end{array} \right)
  \left( \begin{array}{c}
     \alpha_Q    \\
     \alpha_U      \end{array} \right),
 \ee
 and similarly for $\tilde{r}_Q$ and $\tilde{r}_U$,
 while $\tilde{\alpha}_{I,V}$ and $\tilde{r}_{V}$
 are the same as their counterparts without a tilde,
 given that $I$ and $V$ are not affected by the rotation.

%---------------------------------------------------------------------
%---------------------------------------------------------------------
\section*{Acknowledgements}
{
  It is a pleasure for the authors to acknowledge continuous
  discussions with Maciek Wielgus on the topic of the paper.
  The authors gratefully acknowledge fruitful discussions with
  B. Cerutti and B. Crinquand, as well as very useful exchanges with J. Vos
  for the code comparison of section~\ref{sec:ipole}.
  E. G. acknowledges funding by l'Agence Nationale de la Recherche, Project StronG
  ANR-22-CE31-0015.
  }
%---------------------------------------------------------------------
%-------------------------------
 % --------------------------------------

\bibliography{GyotoPolar}

@ARTICLE{Dauser2010,
       author = {{Dauser}, T. and {Wilms}, J. and {Reynolds}, C.~S. and {Brenneman}, L.~W.},
        title = "{Broad emission lines for a negatively spinning black hole}",
      journal = {\mnras},
         year = 2010,
        month = dec,
       volume = {409},
       number = {4},
        pages = {1534-1540},
          doi = {10.1111/j.1365-2966.2010.17393.x},
archivePrefix = {arXiv},
       eprint = {1007.4937},
 primaryClass = {astro-ph.HE},
       adsurl = {https://ui.adsabs.harvard.edu/abs/2010MNRAS.409.1534D}
}

@ARTICLE{Dovciak2008,
       author = {{Dov{\v{c}}iak}, M. and {Muleri}, F. and {Goosmann}, R.~W. and {Karas}, V. and {Matt}, G.},
        title = "{Thermal disc emission from a rotating black hole: X-ray polarization signatures}",
      journal = {\mnras},
         year = 2008,
        month = nov,
       volume = {391},
       number = {1},
        pages = {32-38},
          doi = {10.1111/j.1365-2966.2008.13872.x},
archivePrefix = {arXiv},
       eprint = {0809.0418},
 primaryClass = {astro-ph},
       adsurl = {https://ui.adsabs.harvard.edu/abs/2008MNRAS.391...32D}
}

@ARTICLE{Pihajoki18,
       author = {{Pihajoki}, Pauli and {Mannerkoski}, Matias and {N{\"a}ttil{\"a}}, Joonas and {Johansson}, Peter H.},
        title = "{General Purpose Ray Tracing and Polarized Radiative Transfer in General Relativity}",
      journal = {\apj},
         year = 2018,
        month = aug,
       volume = {863},
       number = {1},
          eid = {8},
        pages = {8},
          doi = {10.3847/1538-4357/aacea0},
archivePrefix = {arXiv},
       eprint = {1804.04670},
 primaryClass = {astro-ph.HE},
       adsurl = {https://ui.adsabs.harvard.edu/abs/2018ApJ...863....8P}
}

@ARTICLE{Yang21,
       author = {{Xiao-lin}, Yang and {Jian-cheng}, Wang and {Chu-yuan}, Yang and {Zun-li}, Yuan},
        title = "{A New Fast Monte Carlo Code for Solving Radiative Transfer Equations Based on the Neumann Solution}",
      journal = {\apjs},
         year = 2021,
        month = jun,
       volume = {254},
       number = {2},
          eid = {29},
        pages = {29},
          doi = {10.3847/1538-4365/abec73},
archivePrefix = {arXiv},
       eprint = {2104.07007},
 primaryClass = {physics.comp-ph},
       adsurl = {https://ui.adsabs.harvard.edu/abs/2021ApJS..254...29X}
}

@ARTICLE{Noble2007,
       author = {{Noble}, Scott C. and {Leung}, Po Kin and {Gammie}, Charles F. and {Book}, Laura G.},
        title = "{Simulating the emission and outflows from accretion discs}",
      journal = {Classical and Quantum Gravity},
         year = 2007,
        month = jun,
       volume = {24},
       number = {12},
        pages = {S259-S274},
          doi = {10.1088/0264-9381/24/12/S17},
archivePrefix = {arXiv},
       eprint = {astro-ph/0701778},
 primaryClass = {astro-ph},
       adsurl = {https://ui.adsabs.harvard.edu/abs/2007CQGra..24S.259N}
}

@ARTICLE{Dexter2009,
       author = {{Dexter}, Jason and {Agol}, Eric},
        title = "{A Fast New Public Code for Computing Photon Orbits in a Kerr Spacetime}",
      journal = {\apj},
         year = 2009,
        month = may,
       volume = {696},
       number = {2},
        pages = {1616-1629},
          doi = {10.1088/0004-637X/696/2/1616},
archivePrefix = {arXiv},
       eprint = {0903.0620},
 primaryClass = {astro-ph.HE},
       adsurl = {https://ui.adsabs.harvard.edu/abs/2009ApJ...696.1616D}
}

@ARTICLE{Moscibrodzka2016,
       author = {{Mo{\'s}cibrodzka}, Monika and {Falcke}, Heino and {Shiokawa}, Hotaka},
        title = "{General relativistic magnetohydrodynamical simulations of the jet in M 87}",
      journal = {\aap},
         year = 2016,
        month = feb,
       volume = {586},
          eid = {A38},
        pages = {A38},
          doi = {10.1051/0004-6361/201526630},
archivePrefix = {arXiv},
       eprint = {1510.07243},
 primaryClass = {astro-ph.HE},
       adsurl = {https://ui.adsabs.harvard.edu/abs/2016A&A...586A..38M}
}

@ARTICLE{Nalewajko2020,
       author = {{Nalewajko}, Krzysztof and {Sikora}, Marek and {R{\'o}{\.z}a{\'n}ska}, Agata},
        title = "{Orientation of the crescent image of M 87*}",
      journal = {\aap},
         year = 2020,
        month = feb,
       volume = {634},
          eid = {A38},
        pages = {A38},
          doi = {10.1051/0004-6361/201936586},
archivePrefix = {arXiv},
       eprint = {1908.10376},
 primaryClass = {astro-ph.HE},
       adsurl = {https://ui.adsabs.harvard.edu/abs/2020A&A...634A..38N}
}

@ARTICLE{Dovciak2022,
       author = {{Dov{\v{c}}iak}, M. and {Papadakis}, I.~E. and {Kammoun}, E.~S. and {Zhang}, W.},
        title = "{Physical model for the broadband energy spectrum of X-ray illuminated accretion discs: Fitting the spectral energy distribution of NGC 5548}",
      journal = {\aap},
         year = 2022,
        month = may,
       volume = {661},
          eid = {A135},
        pages = {A135},
          doi = {10.1051/0004-6361/202142358},
archivePrefix = {arXiv},
       eprint = {2110.01249},
 primaryClass = {astro-ph.HE},
       adsurl = {https://ui.adsabs.harvard.edu/abs/2022A&A...661A.135D}
}

@ARTICLE{Broderick2016,
       author = {{Broderick}, Avery E. and {Fish}, Vincent L. and {Johnson}, Michael D. and {Rosenfeld}, Katherine and {Wang}, Carlos and {Doeleman}, Sheperd S. and {Akiyama}, Kazunori and {Johannsen}, Tim and {Roy}, Alan L.},
        title = "{Modeling Seven Years of Event Horizon Telescope Observations with Radiatively Inefficient Accretion Flow Models}",
      journal = {\apj},
         year = 2016,
        month = apr,
       volume = {820},
       number = {2},
          eid = {137},
        pages = {137},
          doi = {10.3847/0004-637X/820/2/137},
archivePrefix = {arXiv},
       eprint = {1602.07701},
 primaryClass = {astro-ph.HE},
       adsurl = {https://ui.adsabs.harvard.edu/abs/2016ApJ...820..137B}
}

@ARTICLE{Aimar2023,
       author = {{Aimar}, N. and {Dmytriiev}, A. and {Vincent}, F.~H. and {El Mellah}, I. and {Paumard}, T. and {Perrin}, G. and {Zech}, A.},
        title = "{Magnetic reconnection plasmoid model for Sagittarius A* flares}",
      journal = {\aap},
         year = 2023,
        month = apr,
       volume = {672},
          eid = {A62},
        pages = {A62},
          doi = {10.1051/0004-6361/202244936},
archivePrefix = {arXiv},
       eprint = {2301.11874},
 primaryClass = {astro-ph.HE},
       adsurl = {https://ui.adsabs.harvard.edu/abs/2023A&A...672A..62A}
}

@ARTICLE{Vos2022,
       author = {{Vos}, J. and {Mo{\'s}cibrodzka}, M.~A. and {Wielgus}, M.},
        title = "{Polarimetric signatures of hot spots in black hole accretion flows}",
      journal = {\aap},
         year = 2022,
        month = dec,
       volume = {668},
          eid = {A185},
        pages = {A185},
          doi = {10.1051/0004-6361/202244840},
archivePrefix = {arXiv},
       eprint = {2209.09931},
 primaryClass = {astro-ph.HE},
       adsurl = {https://ui.adsabs.harvard.edu/abs/2022A&A...668A.185V}
}

@ARTICLE{krawczynski22,
       author = {{Krawczynski}, Henric and {Muleri}, Fabio and {Dov{\v{c}}iak}, Michal and {Veledina}, Alexandra and {Rodriguez Cavero}, Nicole and {Svoboda}, Jiri and {Ingram}, Adam and {Matt}, Giorgio and {Garcia}, Javier A. and {Loktev}, Vladislav and {Negro}, Michela and {Poutanen}, Juri and {Kitaguchi}, Takao and {Podgorn{\'y}}, Jakub and {Rankin}, John and {Zhang}, Wenda and {Berdyugin}, Andrei and {Berdyugina}, Svetlana V. and {Bianchi}, Stefano and {Blinov}, Dmitry and {Capitanio}, Fiamma and {Di Lalla}, Niccol{\`o} and {Draghis}, Paul and {Fabiani}, Sergio and {Kagitani}, Masato and {Kravtsov}, Vadim and {Kiehlmann}, Sebastian and {Latronico}, Luca and {Lutovinov}, Alexander A. and {Mandarakas}, Nikos and {Marin}, Fr{\'e}d{\'e}ric and {Marinucci}, Andrea and {Miller}, Jon M. and {Mizuno}, Tsunefumi and {Molkov}, Sergey V. and {Omodei}, Nicola and {Petrucci}, Pierre-Olivier and {Ratheesh}, Ajay and {Sakanoi}, Takeshi and {Semena}, Andrei N. and {Skalidis}, Raphael and {Soffitta}, Paolo and {Tennant}, Allyn F. and {Thalhammer}, Phillipp and {Tombesi}, Francesco and {Weisskopf}, Martin C. and {Wilms}, Joern and {Zhang}, Sixuan and {Agudo}, Iv{\'a}n and {Antonelli}, Lucio A. and {Bachetti}, Matteo and {Baldini}, Luca and {Baumgartner}, Wayne H. and {Bellazzini}, Ronaldo and {Bongiorno}, Stephen D. and {Bonino}, Raffaella and {Brez}, Alessandro and {Bucciantini}, Niccol{\`o} and {Castellano}, Simone and {Cavazzuti}, Elisabetta and {Ciprini}, Stefano and {Costa}, Enrico and {De Rosa}, Alessandra and {Del Monte}, Ettore and {Di Gesu}, Laura and {Di Marco}, Alessandro and {Donnarumma}, Immacolata and {Doroshenko}, Victor and {Ehlert}, Steven R. and {Enoto}, Teruaki and {Evangelista}, Yuri and {Ferrazzoli}, Riccardo and {Gunji}, Shuichi and {Hayashida}, Kiyoshi and {Heyl}, Jeremy and {Iwakiri}, Wataru and {Jorstad}, Svetlana G. and {Karas}, Vladimir and {Kolodziejczak}, Jeffery J. and {La Monaca}, Fabio and {Liodakis}, Ioannis and {Maldera}, Simone and {Manfreda}, Alberto and {Marscher}, Alan P. and {Marshall}, Herman L. and {Mitsuishi}, Ikuyuki and {Ng}, Chi-Yung and {O{\textquoteright}Dell}, Stephen L. and {Oppedisano}, Chiara and {Papitto}, Alessandro and {Pavlov}, George G. and {Peirson}, Abel L. and {Perri}, Matteo and {Pesce-Rollins}, Melissa and {Pilia}, Maura and {Possenti}, Andrea and {Puccetti}, Simonetta and {Ramsey}, Brian D. and {Romani}, Roger W. and {Sgr{\`o}}, Carmelo and {Slane}, Patrick and {Spandre}, Gloria and {Tamagawa}, Toru and {Tavecchio}, Fabrizio and {Taverna}, Roberto and {Tawara}, Yuzuru and {Thomas}, Nicholas E. and {Trois}, Alessio and {Tsygankov}, Sergey and {Turolla}, Roberto and {Vink}, Jacco and {Wu}, Kinwah and {Xie}, Fei and {Zane}, Silvia},
        title = "{Polarized x-rays constrain the disk-jet geometry in the black hole x-ray binary Cygnus X-1}",
      journal = {Science},
         year = 2022,
        month = nov,
       volume = {378},
       number = {6620},
        pages = {650-654},
          doi = {10.1126/science.add5399},
archivePrefix = {arXiv},
       eprint = {2206.09972},
 primaryClass = {astro-ph.HE},
       adsurl = {https://ui.adsabs.harvard.edu/abs/2022Sci...378..650K}
}

@INPROCEEDINGS{varniere18,
       author = {{Varniere}, P. and {Casse}, F. and {Vincent}, F.~H.},
        title = "{NOVAs: a Numerical Observatory of Violent Accreting systems}",
    booktitle = {SF2A-2018: Proceedings of the Annual meeting of the French Society of Astronomy and Astrophysics},
         year = 2018,
       editor = {{Di Matteo}, P. and {Billebaud}, F. and {Herpin}, F. and {Lagarde}, N. and {Marquette}, J. -B. and {Robin}, A. and {Venot}, O.},
        month = dec,
        pages = {Di},
       adsurl = {https://ui.adsabs.harvard.edu/abs/2018sf2a.conf..233V}
}

@ARTICLE{casse17,
       author = {{Casse}, F. and {Varniere}, P. and {Meliani}, Z.},
        title = "{Impact of the gravity of a Schwarzschild black hole upon the Rossby wave instability}",
      journal = {\mnras},
         year = 2017,
        month = jan,
       volume = {464},
       number = {3},
        pages = {3704-3712},
          doi = {10.1093/mnras/stw2572},
archivePrefix = {arXiv},
       eprint = {1610.00887},
 primaryClass = {astro-ph.HE},
       adsurl = {https://ui.adsabs.harvard.edu/abs/2017MNRAS.464.3704C}
}

@BOOK{MTW,
       author = {{Misner}, Charles W. and {Thorne}, Kip S. and {Wheeler}, John Archibald},
        title = "{Gravitation}",
         year = 1973,
    publisher = {Freeman},
     address  = {New York},
       adsurl = {https://ui.adsabs.harvard.edu/abs/1973grav.book.....M}
}

@ARTICLE{grould17,
       author = {{Grould}, M. and {Vincent}, F.~H. and {Paumard}, T. and {Perrin}, G.},
        title = "{General relativistic effects on the orbit of the S2 star with GRAVITY}",
      journal = {\aap},
         year = 2017,
        month = dec,
       volume = {608},
          eid = {A60},
        pages = {A60},
          doi = {10.1051/0004-6361/201731148},
archivePrefix = {arXiv},
       eprint = {1709.04492},
 primaryClass = {astro-ph.HE},
       adsurl = {https://ui.adsabs.harvard.edu/abs/2017A&A...608A..60G}
}

@ARTICLE{himwich20,
       author = {{Himwich}, Elizabeth and {Johnson}, Michael D. and {Lupsasca}, Alexandru and {Strominger}, Andrew},
        title = "{Universal polarimetric signatures of the black hole photon ring}",
      journal = {\prd},
         year = 2020,
        month = apr,
       volume = {101},
       number = {8},
          eid = {084020},
        pages = {084020},
          doi = {10.1103/PhysRevD.101.084020},
archivePrefix = {arXiv},
       eprint = {2001.08750},
 primaryClass = {gr-qc},
       adsurl = {https://ui.adsabs.harvard.edu/abs/2020PhRvD.101h4020H}
}

@ARTICLE{vincent23,
       author = {{Vincent}, F.~H. and {Wielgus}, M. and {Aimar}, N. and {Paumard}, T. and {Perrin}, G.},
        title = "{Polarized signatures of orbiting hot spots: special relativity impact and probe of spacetime curvature}",
      journal = {arXiv e-prints},
         year = 2023,
        month = sep,
          eid = {arXiv:2309.10053},
        pages = {arXiv:2309.10053},
          doi = {10.48550/arXiv.2309.10053},
archivePrefix = {arXiv},
       eprint = {2309.10053},
 primaryClass = {astro-ph.HE},
       adsurl = {https://ui.adsabs.harvard.edu/abs/2023arXiv230910053V}
}

@ARTICLE{gravity18,
       author = {{GRAVITY Collaboration} and {Abuter}, R. and {Amorim}, A. and {Baub{\"o}ck}, M. and {Berger}, J.~P. and {Bonnet}, H. and {Brandner}, W. and {Cl{\'e}net}, Y. and {Coud{\'e} Du Foresto}, V. and {de Zeeuw}, P.~T. and {Deen}, C. and {Dexter}, J. and {Duvert}, G. and {Eckart}, A. and {Eisenhauer}, F. and {F{\"o}rster Schreiber}, N.~M. and {Garcia}, P. and {Gao}, F. and {Gendron}, E. and {Genzel}, R. and {Gillessen}, S. and {Guajardo}, P. and {Habibi}, M. and {Haubois}, X. and {Henning}, Th. and {Hippler}, S. and {Horrobin}, M. and {Huber}, A. and {Jim{\'e}nez-Rosales}, A. and {Jocou}, L. and {Kervella}, P. and {Lacour}, S. and {Lapeyr{\`e}re}, V. and {Lazareff}, B. and {Le Bouquin}, J. -B. and {L{\'e}na}, P. and {Lippa}, M. and {Ott}, T. and {Panduro}, J. and {Paumard}, T. and {Perraut}, K. and {Perrin}, G. and {Pfuhl}, O. and {Plewa}, P.~M. and {Rabien}, S. and {Rodr{\'\i}guez-Coira}, G. and {Rousset}, G. and {Sternberg}, A. and {Straub}, O. and {Straubmeier}, C. and {Sturm}, E. and {Tacconi}, L.~J. and {Vincent}, F. and {von Fellenberg}, S. and {Waisberg}, I. and {Widmann}, F. and {Wieprecht}, E. and {Wiezorrek}, E. and {Woillez}, J. and {Yazici}, S.},
        title = "{Detection of orbital motions near the last stable circular orbit of the massive black hole SgrA*}",
      journal = {\aap},
         year = 2018,
        month = oct,
       volume = {618},
          eid = {L10},
        pages = {L10},
          doi = {10.1051/0004-6361/201834294},
archivePrefix = {arXiv},
       eprint = {1810.12641},
 primaryClass = {astro-ph.GA},
       adsurl = {https://ui.adsabs.harvard.edu/abs/2018A&A...618L..10G}
}

@ARTICLE{WalkerPenrose70,
       author = {{Walker}, Martin and {Penrose}, Roger},
        title = "{On quadratic first integrals of the geodesic equations for type \{ 22\} spacetimes}",
      journal = {Communications in Mathematical Physics},
         year = 1970,
        month = dec,
       volume = {18},
       number = {4},
        pages = {265-274},
          doi = {10.1007/BF01649445},
       adsurl = {https://ui.adsabs.harvard.edu/abs/1970CMaPh..18..265W}
}

@ARTICLE{Huang09,
       author = {{Huang}, Lei and {Liu}, Siming and {Shen}, Zhi-Qiang and {Yuan}, Ye-Fei and {Cai}, Mike J. and {Li}, Hui and {Fryer}, Christopher L.},
        title = "{Polarized Emission of Sagittarius A*}",
      journal = {\apj},
         year = 2009,
        month = sep,
       volume = {703},
       number = {1},
        pages = {557-568},
          doi = {10.1088/0004-637X/703/1/557},
archivePrefix = {arXiv},
       eprint = {0907.5463},
 primaryClass = {astro-ph.GA},
       adsurl = {https://ui.adsabs.harvard.edu/abs/2009ApJ...703..557H}
}

@ARTICLE{Shcherbakov08,
       author = {{Shcherbakov}, Roman V.},
        title = "{Propagation Effects in Magnetized Transrelativistic Plasmas}",
      journal = {\apj},
         year = 2008,
        month = nov,
       volume = {688},
       number = {1},
        pages = {695-700},
          doi = {10.1086/592326},
archivePrefix = {arXiv},
       eprint = {0809.0012},
 primaryClass = {astro-ph},
       adsurl = {https://ui.adsabs.harvard.edu/abs/2008ApJ...688..695S}
}

@ARTICLE{Westfold59,
       author = {{Westfold}, K.~C.},
        title = "{The Polarization of Synchrotron Radiation.}",
      journal = {\apj},
         year = 1959,
        month = jul,
       volume = {130},
        pages = {241},
          doi = {10.1086/146713},
       adsurl = {https://ui.adsabs.harvard.edu/abs/1959ApJ...130..241W}
}

@article{IAU74,
title={Transactions of the International Astronomical Union},
volume={15},
DOI={10.1017/S0251107X00031606},
number={2},
journal={Transactions of the International Astronomical Union},
publisher={Cambridge University Press},
year={1973},
pages={165–167}}

@ARTICLE{LDI85,
       author = {{Landi Degl'Innocenti}, E. and {Landi Degl'Innocenti}, M.},
        title = "{On the solution of the radiative transfer equations for polarized radiation}",
      journal = {\solphys},
         year = 1985,
        month = jun,
       volume = {97},
        pages = {239-250},
          doi = {10.1007/BF00165988},
       adsurl = {https://ui.adsabs.harvard.edu/abs/1985SoPh...97..239L}
}

@BOOK{RL79,
       author = {{Rybicki}, George B. and {Lightman}, Alan P.},
        title = "{Radiative processes in astrophysics}",
         year = 1979,
       adsurl = {https://ui.adsabs.harvard.edu/abs/1979rpa..book.....R}
}

@ARTICLE{vincent11,
   author = {{Vincent}, F.~H. and {Paumard}, T. and {Gourgoulhon}, E. and
  {Perrin}, G.},
    title = "{GYOTO: a new general relativistic ray-tracing code}",
  journal = {Classical and Quantum Gravity},
archivePrefix = "arXiv",
   eprint = {1109.4769},
 primaryClass = "gr-qc",
     year = 2011,
    month = nov,
   volume = 28,
   number = 22,
      eid = {225011},
    pages = {225011},
      doi = {10.1088/0264-9381/28/22/225011},
   adsurl = {http://adsabs.harvard.edu/abs/2011CQGra..28v5011V}
}

@ARTICLE{hamaker96,
       author = {{Hamaker}, J.~P. and {Bregman}, J.~D.},
        title = "{Understanding radio polarimetry. III. Interpreting the IAU/IEEE definitions of the Stokes parameters.}",
      journal = {\aaps},
         year = 1996,
        month = may,
       volume = {117},
        pages = {161-165},
       adsurl = {https://ui.adsabs.harvard.edu/abs/1996A&AS..117..161H}
}

@ARTICLE{Pandya2021,
    author = {{Marszewski}, Andrew and {Prather}, Ben S. and {Joshi}, Abhishek V. and {Pandya}, Alex and {Gammie}, Charles F.},
    title = "{Updated Transfer Coefficients for Magnetized Plasmas}",
    journal = {\apj},
    year = 2021,
    month = nov,
    volume = {921},
    number = {1},
    eid = {17},
    pages = {17},
    doi = {10.3847/1538-4357/ac1b28},
    archivePrefix = {arXiv},
    eprint = {2108.10359},
    primaryClass = {astro-ph.HE},
    adsurl = {https://ui.adsabs.harvard.edu/abs/2021ApJ...921...17M}
}

@ARTICLE{Huang2011,
    author = {{Huang}, Lei and {Shcherbakov}, Roman V.},
    title = "{Faraday conversion and rotation in uniformly magnetized relativistic plasmas}",
    journal = {\mnras},
    year = 2011,
    month = oct,
    volume = {416},
    number = {4},
    pages = {2574-2592},
    doi = {10.1111/j.1365-2966.2011.19207.x},
    archivePrefix = {arXiv},
    eprint = {1106.1630},
    primaryClass = {astro-ph.HE},
    adsurl = {https://ui.adsabs.harvard.edu/abs/2011MNRAS.416.2574H}
}

@ARTICLE{Dexter2016,
    author = {{Dexter}, Jason},
    title = "{A public code for general relativistic, polarised radiative transfer around spinning black holes}",
    journal = {\mnras},
    year = 2016,
    month = oct,
    volume = {462},
    number = {1},
    pages = {115-136},
    doi = {10.1093/mnras/stw1526},
    archivePrefix = {arXiv},
    eprint = {1602.03184},
    primaryClass = {astro-ph.HE},
    adsurl = {https://ui.adsabs.harvard.edu/abs/2016MNRAS.462..115D}
}

@ARTICLE{PT1974,
    author = {{Page}, Don N. and {Thorne}, Kip S.},
    title = "{Disk-Accretion onto a Black Hole. Time-Averaged Structure of Accretion Disk}",
    journal = {\apj},
    year = 1974,
    month = jul,
    volume = {191},
    pages = {499-506},
    doi = {10.1086/152990},
    adsurl = {https://ui.adsabs.harvard.edu/abs/1974ApJ...191..499P}
}

@ARTICLE{ipole,
    author = {{Mo{\'s}cibrodzka}, M. and {Gammie}, C.~F.},
    title = "{IPOLE - semi-analytic scheme for relativistic polarized radiative transport}",
    journal = {\mnras},
    year = 2018,
    month = mar,
    volume = {475},
    number = {1},
    pages = {43-54},
    doi = {10.1093/mnras/stx3162},
    archivePrefix = {arXiv},
    eprint = {1712.03057},
    primaryClass = {astro-ph.HE},
    adsurl = {https://ui.adsabs.harvard.edu/abs/2018MNRAS.475...43M}
}

@ARTICLE{Gravity2023,
    author = {{GRAVITY Collaboration} and {Abuter}, R. and {Aimar}, N. and {Amaro Seoane}, P. and {Amorim}, A. and {Baub{\"o}ck}, M. and {Berger}, J.~P. and {Bonnet}, H. and {Bourdarot}, G. and {Brandner}, W. and {Cardoso}, V. and {Cl{\'e}net}, Y. and {Davies}, R. and {de Zeeuw}, P.~T. and {Dexter}, J. and {Drescher}, A. and {Eckart}, A. and {Eisenhauer}, F. and {Feuchtgruber}, H. and {Finger}, G. and {F{\"o}rster Schreiber}, N.~M. and {Foschi}, A. and {Garcia}, P. and {Gao}, F. and {Gelles}, Z. and {Gendron}, E. and {Genzel}, R. and {Gillessen}, S. and {Hartl}, M. and {Haubois}, X. and {Haussmann}, F. and {Hei{\ss}el}, G. and {Henning}, T. and {Hippler}, S. and {Horrobin}, M. and {Jochum}, L. and {Jocou}, L. and {Kaufer}, A. and {Kervella}, P. and {Lacour}, S. and {Lapeyr{\`e}re}, V. and {Le Bouquin}, J. -B. and {L{\'e}na}, P. and {Lutz}, D. and {Mang}, F. and {More}, N. and {Ott}, T. and {Paumard}, T. and {Perraut}, K. and {Perrin}, G. and {Pfuhl}, O. and {Rabien}, S. and {Ribeiro}, D.~C. and {Sadun Bordoni}, M. and {Scheithauer}, S. and {Shangguan}, J. and {Shimizu}, T. and {Stadler}, J. and {Straub}, O. and {Straubmeier}, C. and {Sturm}, E. and {Tacconi}, L.~J. and {Vincent}, F. and {von Fellenberg}, S. and {Widmann}, F. and {Wielgus}, M. and {Wieprecht}, E. and {Wiezorrek}, E. and {Woillez}, J.},
    title = "{Polarimetry and astrometry of NIR flares as event horizon scale, dynamical probes for the mass of Sgr A*}",
    journal = {\aap},
    year = 2023,
    month = sep,
    volume = {677},
    eid = {L10},
    pages = {L10},
    doi = {10.1051/0004-6361/202347416},
    archivePrefix = {arXiv},
    eprint = {2307.11821},
    primaryClass = {astro-ph.GA},
    adsurl = {https://ui.adsabs.harvard.edu/abs/2023A&A...677L..10G}
}

@ARTICLE{EHT2022,
    author = {{Event Horizon Telescope Collaboration} and {Akiyama}, Kazunori and {Alberdi}, Antxon and {Alef}, Walter and {Algaba}, Juan Carlos and {Anantua}, Richard and {Asada}, Keiichi and {Azulay}, Rebecca and {Bach}, Uwe and {Baczko}, Anne-Kathrin and {Ball}, David and {Balokovi{\'c}}, Mislav and {Barrett}, John and {Baub{\"o}ck}, Michi and {Benson}, Bradford A. and {Bintley}, Dan and {Blackburn}, Lindy and {Blundell}, Raymond and {Bouman}, Katherine L. and {Bower}, Geoffrey C. and {Boyce}, Hope and {Bremer}, Michael and {Brinkerink}, Christiaan D. and {Brissenden}, Roger and {Britzen}, Silke and {Broderick}, Avery E. and {Broguiere}, Dominique and {Bronzwaer}, Thomas and {Bustamante}, Sandra and {Byun}, Do-Young and {Carlstrom}, John E. and {Ceccobello}, Chiara and {Chael}, Andrew and {Chan}, Chi-kwan and {Chatterjee}, Koushik and {Chatterjee}, Shami and {Chen}, Ming-Tang and {Chen}, Yongjun and {Cheng}, Xiaopeng and {Cho}, Ilje and {Christian}, Pierre and {Conroy}, Nicholas S. and {Conway}, John E. and {Cordes}, James M. and {Crawford}, Thomas M. and {Crew}, Geoffrey B. and {Cruz-Osorio}, Alejandro and {Cui}, Yuzhu and {Davelaar}, Jordy and {De Laurentis}, Mariafelicia and {Deane}, Roger and {Dempsey}, Jessica and {Desvignes}, Gregory and {Dexter}, Jason and {Dhruv}, Vedant and {Doeleman}, Sheperd S. and {Dougal}, Sean and {Dzib}, Sergio A. and {Eatough}, Ralph P. and {Emami}, Razieh and {Falcke}, Heino and {Farah}, Joseph and {Fish}, Vincent L. and {Fomalont}, Ed and {Ford}, H. Alyson and {Fraga-Encinas}, Raquel and {Freeman}, William T. and {Friberg}, Per and {Fromm}, Christian M. and {Fuentes}, Antonio and {Galison}, Peter and {Gammie}, Charles F. and {Garc{\'\i}a}, Roberto and {Gentaz}, Olivier and {Georgiev}, Boris and {Goddi}, Ciriaco and {Gold}, Roman and {G{\'o}mez-Ruiz}, Arturo I. and {G{\'o}mez}, Jos{\'e} L. and {Gu}, Minfeng and {Gurwell}, Mark and {Hada}, Kazuhiro and {Haggard}, Daryl and {Haworth}, Kari and {Hecht}, Michael H. and {Hesper}, Ronald and {Heumann}, Dirk and {Ho}, Luis C. and {Ho}, Paul and {Honma}, Mareki and {Huang}, Chih-Wei L. and {Huang}, Lei and {Hughes}, David H. and {Ikeda}, Shiro and {Impellizzeri}, C.~M. Violette and {Inoue}, Makoto and {Issaoun}, Sara and {James}, David J. and {Jannuzi}, Buell T. and {Janssen}, Michael and {Jeter}, Britton and {Jiang}, Wu and {Jim{\'e}nez-Rosales}, Alejandra and {Johnson}, Michael D. and {Jorstad}, Svetlana and {Joshi}, Abhishek V. and {Jung}, Taehyun and {Karami}, Mansour and {Karuppusamy}, Ramesh and {Kawashima}, Tomohisa and {Keating}, Garrett K. and {Kettenis}, Mark and {Kim}, Dong-Jin and {Kim}, Jae-Young and {Kim}, Jongsoo and {Kim}, Junhan and {Kino}, Motoki and {Koay}, Jun Yi and {Kocherlakota}, Prashant and {Kofuji}, Yutaro and {Koch}, Patrick M. and {Koyama}, Shoko and {Kramer}, Carsten and {Kramer}, Michael and {Krichbaum}, Thomas P. and {Kuo}, Cheng-Yu and {La Bella}, Noemi and {Lauer}, Tod R. and {Lee}, Daeyoung and {Lee}, Sang-Sung and {Leung}, Po Kin and {Levis}, Aviad and {Li}, Zhiyuan and {Lico}, Rocco and {Lindahl}, Greg and {Lindqvist}, Michael and {Lisakov}, Mikhail and {Liu}, Jun and {Liu}, Kuo and {Liuzzo}, Elisabetta and {Lo}, Wen-Ping and {Lobanov}, Andrei P. and {Loinard}, Laurent and {Lonsdale}, Colin J. and {Lu}, Ru-Sen and {Mao}, Jirong and {Marchili}, Nicola and {Markoff}, Sera and {Marrone}, Daniel P. and {Marscher}, Alan P. and {Mart{\'\i}-Vidal}, Iv{\'a}n and {Matsushita}, Satoki and {Matthews}, Lynn D. and {Medeiros}, Lia and {Menten}, Karl M. and {Michalik}, Daniel and {Mizuno}, Izumi and {Mizuno}, Yosuke and {Moran}, James M. and {Moriyama}, Kotaro and {Moscibrodzka}, Monika and {M{\"u}ller}, Cornelia and {Mus}, Alejandro and {Musoke}, Gibwa and {Myserlis}, Ioannis and {Nadolski}, Andrew and {Nagai}, Hiroshi and {Nagar}, Neil M. and {Nakamura}, Masanori and {Narayan}, Ramesh and {Narayanan}, Gopal and {Natarajan}, Iniyan and {Nathanail}, Antonios and {Fuentes}, Santiago Navarro and {Neilsen}, Joey and {Neri}, Roberto and {Ni}, Chunchong and {Noutsos}, Aristeidis and {Nowak}, Michael A. and {Oh}, Junghwan and {Okino}, Hiroki and {Olivares}, H{\'e}ctor and {Ortiz-Le{\'o}n}, Gisela N. and {Oyama}, Tomoaki and {{\"O}zel}, Feryal and {Palumbo}, Daniel C.~M. and {Paraschos}, Georgios Filippos and {Park}, Jongho and {Parsons}, Harriet and {Patel}, Nimesh and {Pen}, Ue-Li and {Pesce}, Dominic W. and {Pi{\'e}tu}, Vincent and {Plambeck}, Richard and {PopStefanija}, Aleksandar and {Porth}, Oliver and {P{\"o}tzl}, Felix M. and {Prather}, Ben and {Preciado-L{\'o}pez}, Jorge A. and {Psaltis}, Dimitrios and {Pu}, Hung-Yi and {Ramakrishnan}, Venkatessh and {Rao}, Ramprasad and {Rawlings}, Mark G. and {Raymond}, Alexander W. and {Rezzolla}, Luciano and {Ricarte}, Angelo and {Ripperda}, Bart and {Roelofs}, Freek and {Rogers}, Alan and {Ros}, Eduardo and {Romero-Ca{\~n}izales}, Cristina and {Roshanineshat}, Arash and {Rottmann}, Helge and {Roy}, Alan L. and {Ruiz}, Ignacio and {Ruszczyk}, Chet and {Rygl}, Kazi L.~J. and {S{\'a}nchez}, Salvador and {S{\'a}nchez-Arg{\"u}elles}, David and {S{\'a}nchez-Portal}, Miguel and {Sasada}, Mahito and {Satapathy}, Kaushik and {Savolainen}, Tuomas and {Schloerb}, F. Peter and {Schonfeld}, Jonathan and {Schuster}, Karl-Friedrich and {Shao}, Lijing and {Shen}, Zhiqiang and {Small}, Des and {Sohn}, Bong Won and {SooHoo}, Jason and {Souccar}, Kamal and {Sun}, He and {Tazaki}, Fumie and {Tetarenko}, Alexandra J. and {Tiede}, Paul and {Tilanus}, Remo P.~J. and {Titus}, Michael and {Torne}, Pablo and {Traianou}, Efthalia and {Trent}, Tyler and {Trippe}, Sascha and {Turk}, Matthew and {van Bemmel}, Ilse and {van Langevelde}, Huib Jan and {van Rossum}, Daniel R. and {Vos}, Jesse and {Wagner}, Jan and {Ward-Thompson}, Derek and {Wardle}, John and {Weintroub}, Jonathan and {Wex}, Norbert and {Wharton}, Robert and {Wielgus}, Maciek and {Wiik}, Kaj and {Witzel}, Gunther and {Wondrak}, Michael F. and {Wong}, George N. and {Wu}, Qingwen and {Yamaguchi}, Paul and {Yoon}, Doosoo and {Young}, Andr{\'e} and {Young}, Ken and {Younsi}, Ziri and {Yuan}, Feng and {Yuan}, Ye-Fei and {Zensus}, J. Anton and {Zhang}, Shuo and {Zhao}, Guang-Yao and {Zhao}, Shan-Shan and {Agurto}, Claudio and {Allardi}, Alexander and {Amestica}, Rodrigo and {Araneda}, Juan Pablo and {Arriagada}, Oriel and {Berghuis}, Jennie L. and {Bertarini}, Alessandra and {Berthold}, Ryan and {Blanchard}, Jay and {Brown}, Ken and {C{\'a}rdenas}, Mauricio and {Cantzler}, Michael and {Caro}, Patricio and {Castillo-Dom{\'\i}nguez}, Edgar and {Chan}, Tin Lok and {Chang}, Chih-Cheng and {Chang}, Dominic O. and {Chang}, Shu-Hao and {Chang}, Song-Chu and {Chen}, Chung-Chen and {Chilson}, Ryan and {Chuter}, Tim C. and {Ciechanowicz}, Miroslaw and {Colin-Beltran}, Edgar and {Coulson}, Iain M. and {Crowley}, Joseph and {Degenaar}, Nathalie and {Dornbusch}, Sven and {Dur{\'a}n}, Carlos A. and {Everett}, Wendeline B. and {Faber}, Aaron and {Forster}, Karl and {Fuchs}, Miriam M. and {Gale}, David M. and {Geertsema}, Gertie and {Gonz{\'a}lez}, Edouard and {Graham}, Dave and {Gueth}, Fr{\'e}d{\'e}ric and {Halverson}, Nils W. and {Han}, Chih-Chiang and {Han}, Kuo-Chang and {Hasegawa}, Yutaka and {Hern{\'a}ndez-Rebollar}, Jos{\'e} Luis and {Herrera}, Cristian and {Herrero-Illana}, Ruben and {Heyminck}, Stefan and {Hirota}, Akihiko and {Hoge}, James and {Hostler Schimpf}, Shelbi R. and {Howie}, Ryan E. and {Huang}, Yau-De and {Jiang}, Homin and {Jinchi}, Hao and {John}, David and {Kimura}, Kimihiro and {Klein}, Thomas and {Kubo}, Derek and {Kuroda}, John and {Kwon}, Caleb and {Lacasse}, Richard and {Laing}, Robert and {Leitch}, Erik M. and {Li}, Chao-Te and {Liu}, Ching-Tang and {Liu}, Kuan-Yu and {Lin}, Lupin C. -C. and {Lu}, Li-Ming and {Mac-Auliffe}, Felipe and {Martin-Cocher}, Pierre and {Matulonis}, Callie and {Maute}, John K. and {Messias}, Hugo and {Meyer-Zhao}, Zheng and {Monta{\~n}a}, Alfredo and {Montenegro-Montes}, Francisco and {Montgomerie}, William and {Moreno Nolasco}, Marcos Emir and {Muders}, Dirk and {Nishioka}, Hiroaki and {Norton}, Timothy J. and {Nystrom}, George and {Ogawa}, Hideo and {Olivares}, Rodrigo and {Oshiro}, Peter and {P{\'e}rez-Beaupuits}, Juan Pablo and {Parra}, Rodrigo and {Phillips}, Neil M. and {Poirier}, Michael and {Pradel}, Nicolas and {Qiu}, Richard and {Raffin}, Philippe A. and {Rahlin}, Alexandra S. and {Ram{\'\i}rez}, Jorge and {Ressler}, Sean and {Reynolds}, Mark and {Rodr{\'\i}guez-Montoya}, Iv{\'a}n and {Saez-Madain}, Alejandro F. and {Santana}, Jorge and {Shaw}, Paul and {Shirkey}, Leslie E. and {Silva}, Kevin M. and {Snow}, William and {Sousa}, Don and {Sridharan}, T.~K. and {Stahm}, William and {Stark}, Anthony A. and {Test}, John and {Torstensson}, Karl and {Venegas}, Paulina and {Walther}, Craig and {Wei}, Ta-Shun and {White}, Chris and {Wieching}, Gundolf and {Wijnands}, Rudy and {Wouterloot}, Jan G.~A. and {Yu}, Chen-Yu and {Yu}, Wei and {Zeballos}, Milagros},
    title = "{First Sagittarius A* Event Horizon Telescope Results. I. The Shadow of the Supermassive Black Hole in the Center of the Milky Way}",
    journal = {\apjl},
    year = 2022,
    month = may,
    volume = {930},
    number = {2},
    eid = {L12},
    pages = {L12},
    doi = {10.3847/2041-8213/ac6674},
    adsurl = {https://ui.adsabs.harvard.edu/abs/2022ApJ...930L..12E}
}

@ARTICLE{Wielgus2022,
    author = {{Wielgus}, M. and {Moscibrodzka}, M. and {Vos}, J. and {Gelles}, Z. and {Mart{\'\i}-Vidal}, I. and {Farah}, J. and {Marchili}, N. and {Goddi}, C. and {Messias}, H.},
    title = "{Orbital motion near Sagittarius A$^{*}$ . Constraints from polarimetric ALMA observations}",
    journal = {\aap},
    year = 2022,
    month = sep,
    volume = {665},
    eid = {L6},
    pages = {L6},
    doi = {10.1051/0004-6361/202244493},
    archivePrefix = {arXiv},
    eprint = {2209.09926},
    primaryClass = {astro-ph.HE},
    adsurl = {https://ui.adsabs.harvard.edu/abs/2022A&A...665L...6W}
}

@ARTICLE{Jimenez2021,
    author = {{Jim{\'e}nez-Rosales}, A. and {Dexter}, J. and {Ressler}, S.~M. and {Tchekhovskoy}, A. and {Baub{\"o}ck}, M. and {Dallilar}, Y. and {de Zeeuw}, P.~T. and {Drescher}, A. and {Eisenhauer}, F. and {von Fellenberg}, S. and {Gao}, F. and {Genzel}, R. and {Gillessen}, S. and {Habibi}, M. and {Ott}, T. and {Stadler}, J. and {Straub}, O. and {Widmann}, F.},
    title = "{Relative depolarization of the black hole photon ring in GRMHD models of Sgr A* and M87*}",
    journal = {\mnras},
    year = 2021,
    month = may,
    volume = {503},
    number = {3},
    pages = {4563-4575},
    doi = {10.1093/mnras/stab784},
    archivePrefix = {arXiv},
    eprint = {2103.06292},
    primaryClass = {astro-ph.HE},
    adsurl = {https://ui.adsabs.harvard.edu/abs/2021MNRAS.503.4563J}
}

@ARTICLE{Pu2016,
    author = {{Pu}, Hung-Yi and {Yun}, Kiyun and {Younsi}, Ziri and {Yoon}, Suk-Jin},
    title = "{Odyssey: A Public GPU-based Code for General Relativistic Radiative Transfer in Kerr Spacetime}",
    journal = {\apj},
    year = 2016,
    month = apr,
    volume = {820},
    number = {2},
    eid = {105},
    pages = {105},
    doi = {10.3847/0004-637X/820/2/105},
    archivePrefix = {arXiv},
    eprint = {1601.02063},
    primaryClass = {astro-ph.HE},
    adsurl = {https://ui.adsabs.harvard.edu/abs/2016ApJ...820..105P}
}

@INPROCEEDINGS{Younsi2020,
    author = {{Younsi}, Ziri and {Porth}, Oliver and {Mizuno}, Yosuke and {Fromm}, Christian M. and {Olivares}, Hector},
    title = "{Modelling the polarised emission from black holes on event horizon-scales}",
    booktitle = {Perseus in Sicily: From Black Hole to Cluster Outskirts},
    year = 2020,
    editor = {{Asada}, Keiichi and {de Gouveia Dal Pino}, Elisabete and {Giroletti}, Marcello and {Nagai}, Hiroshi and {Nemmen}, Rodrigo},
    volume = {342},
    month = jan,
    pages = {9-12},
    doi = {10.1017/S1743921318007263},
    archivePrefix = {arXiv},
    eprint = {1907.09196},
    primaryClass = {astro-ph.HE},
    adsurl = {https://ui.adsabs.harvard.edu/abs/2020IAUS..342....9Y}
}

@ARTICLE{Bronzwaer2018,
       author = {{Bronzwaer}, T. and {Davelaar}, J. and {Younsi}, Z. and {Mo{\'s}cibrodzka}, M. and {Falcke}, H. and {Kramer}, M. and {Rezzolla}, L.},
        title = "{RAPTOR. I. Time-dependent radiative transfer in arbitrary spacetimes}",
      journal = {\aap},
         year = 2018,
        month = may,
       volume = {613},
          eid = {A2},
        pages = {A2},
          doi = {10.1051/0004-6361/201732149},
archivePrefix = {arXiv},
       eprint = {1801.10452},
 primaryClass = {astro-ph.HE},
       adsurl = {https://ui.adsabs.harvard.edu/abs/2018A&A...613A...2B}
}

@ARTICLE{Bronzwaer2020,
    author = {{Bronzwaer}, T. and {Younsi}, Z. and {Davelaar}, J. and {Falcke}, H.},
    title = "{RAPTOR. II. Polarized radiative transfer in curved spacetime}",
    journal = {\aap},
    year = 2020,
    month = sep,
    volume = {641},
    eid = {A126},
    pages = {A126},
    doi = {10.1051/0004-6361/202038573},
    archivePrefix = {arXiv},
    eprint = {2007.03045},
    primaryClass = {astro-ph.HE},
    adsurl = {https://ui.adsabs.harvard.edu/abs/2020A&A...641A.126B}
}

@ARTICLE{Moscibrodzka2018,
    author = {{Mo{\'s}cibrodzka}, M. and {Gammie}, C.~F.},
    title = "{IPOLE - semi-analytic scheme for relativistic polarized radiative transport}",
    journal = {\mnras},
    year = 2018,
    month = mar,
    volume = {475},
    number = {1},
    pages = {43-54},
    doi = {10.1093/mnras/stx3162},
    archivePrefix = {arXiv},
    eprint = {1712.03057},
    primaryClass = {astro-ph.HE},
    adsurl = {https://ui.adsabs.harvard.edu/abs/2018MNRAS.475...43M}
}

@ARTICLE{Prather2023,
    author = {{Prather}, Ben S. and {Dexter}, Jason and {Moscibrodzka}, Monika and {Pu}, Hung-Yi and {Bronzwaer}, Thomas and {Davelaar}, Jordy and {Younsi}, Ziri and {Gammie}, Charles F. and {Gold}, Roman and {Wong}, George N. and {Akiyama}, Kazunori and {Alberdi}, Antxon and {Alef}, Walter and {Algaba}, Juan Carlos and {Anantua}, Richard and {Asada}, Keiichi and {Azulay}, Rebecca and {Bach}, Uwe and {Baczko}, Anne-Kathrin and {Ball}, David and {Balokovi{\'c}}, Mislav and {Barrett}, John and {Baub{\"o}ck}, Michi and {Benson}, Bradford A. and {Bintley}, Dan and {Blackburn}, Lindy and {Blundell}, Raymond and {Bouman}, Katherine L. and {Bower}, Geoffrey C. and {Boyce}, Hope and {Bremer}, Michael and {Brinkerink}, Christiaan D. and {Brissenden}, Roger and {Britzen}, Silke and {Broderick}, Avery E. and {Broguiere}, Dominique and {Bustamante}, Sandra and {Byun}, Do-Young and {Carlstrom}, John E. and {Ceccobello}, Chiara and {Chael}, Andrew and {Chan}, Chi-kwan and {Chang}, Dominic O. and {Chatterjee}, Koushik and {Chatterjee}, Shami and {Chen}, Ming-Tang and {Chen}, Yongjun and {Cheng}, Xiaopeng and {Cho}, Ilje and {Christian}, Pierre and {Conroy}, Nicholas S. and {Conway}, John E. and {Cordes}, James M. and {Crawford}, Thomas M. and {Crew}, Geoffrey B. and {Cruz-Osorio}, Alejandro and {Cui}, Yuzhu and {De Laurentis}, Mariafelicia and {Deane}, Roger and {Dempsey}, Jessica and {Desvignes}, Gregory and {Dhruv}, Vedant and {Doeleman}, Sheperd S. and {Dougal}, Sean and {Dzib}, Sergio A. and {Eatough}, Ralph P. and {Emami}, Razieh and {Falcke}, Heino and {Farah}, Joseph and {Fish}, Vincent L. and {Fomalont}, Ed and {Ford}, H. Alyson and {Fraga-Encinas}, Raquel and {Freeman}, William T. and {Friberg}, Per and {Fromm}, Christian M. and {Fuentes}, Antonio and {Galison}, Peter and {Garc{\'\i}a}, Roberto and {Gentaz}, Olivier and {Georgiev}, Boris and {Goddi}, Ciriaco and {G{\'o}mez-Ruiz}, Arturo I. and {G{\'o}mez}, Jos{\'e} L. and {Gu}, Minfeng and {Gurwell}, Mark and {Hada}, Kazuhiro and {Haggard}, Daryl and {Haworth}, Kari and {Hecht}, Michael H. and {Hesper}, Ronald and {Heumann}, Dirk and {Ho}, Luis C. and {Ho}, Paul and {Honma}, Mareki and {Huang}, Chih-Wei L. and {Huang}, Lei and {Hughes}, David H. and {Ikeda}, Shiro and {Impellizzeri}, C.~M. Violette and {Inoue}, Makoto and {Issaoun}, Sara and {James}, David J. and {Jannuzi}, Buell T. and {Janssen}, Michael and {Jeter}, Britton and {Jiang}, Wu and {Jim{\'e}nez-Rosales}, Alejandra and {Johnson}, Michael D. and {Jorstad}, Svetlana and {Joshi}, Abhishek V. and {Jung}, Taehyun and {Karami}, Mansour and {Karuppusamy}, Ramesh and {Kawashima}, Tomohisa and {Keating}, Garrett K. and {Kettenis}, Mark and {Kim}, Dong-Jin and {Kim}, Jae-Young and {Kim}, Jongsoo and {Kim}, Junhan and {Kino}, Motoki and {Koay}, Jun Yi and {Kocherlakota}, Prashant and {Kofuji}, Yutaro and {Koyama}, Shoko and {Kramer}, Carsten and {Kramer}, Michael and {Krichbaum}, Thomas P. and {Kuo}, Cheng-Yu and {La Bella}, Noemi and {Lauer}, Tod R. and {Lee}, Daeyoung and {Lee}, Sang-Sung and {Leung}, Po Kin and {Levis}, Aviad and {Li}, Zhiyuan and {Lico}, Rocco and {Lindahl}, Greg and {Lindqvist}, Michael and {Lisakov}, Mikhail and {Liu}, Jun and {Liu}, Kuo and {Liuzzo}, Elisabetta and {Lo}, Wen-Ping and {Lobanov}, Andrei P. and {Loinard}, Laurent and {Lonsdale}, Colin J. and {Lu}, Ru-Sen and {MacDonald}, Nicholas R. and {Mao}, Jirong and {Marchili}, Nicola and {Markoff}, Sera and {Marrone}, Daniel P. and {Marscher}, Alan P. and {Mart{\'\i}-Vidal}, Iv{\'a}n and {Matsushita}, Satoki and {Matthews}, Lynn D. and {Medeiros}, Lia and {Menten}, Karl M. and {Michalik}, Daniel and {Mizuno}, Izumi and {Mizuno}, Yosuke and {Moran}, James M. and {Moriyama}, Kotaro and {M{\"u}ller}, Cornelia and {Mus}, Alejandro and {Musoke}, Gibwa and {Myserlis}, Ioannis and {Nadolski}, Andrew and {Nagai}, Hiroshi and {Nagar}, Neil M. and {Nakamura}, Masanori and {Narayan}, Ramesh and {Narayanan}, Gopal and {Natarajan}, Iniyan and {Nathanail}, Antonios and {Fuentes}, Santiago Navarro and {Neilsen}, Joey and {Neri}, Roberto and {Ni}, Chunchong and {Noutsos}, Aristeidis and {Nowak}, Michael A. and {Oh}, Junghwan and {Okino}, Hiroki and {Olivares}, H{\'e}ctor and {Ortiz-Le{\'o}n}, Gisela N. and {Oyama}, Tomoaki and {{\"O}zel}, Feryal and {Palumbo}, Daniel C.~M. and {Paraschos}, Georgios Filippos and {Park}, Jongho and {Parsons}, Harriet and {Patel}, Nimesh and {Pen}, Ue-Li and {Pesce}, Dominic W. and {Pi{\'e}tu}, Vincent and {Plambeck}, Richard and {PopStefanija}, Aleksandar and {Porth}, Oliver and {P{\"o}tzl}, Felix M. and {Preciado-L{\'o}pez}, Jorge A. and {Psaltis}, Dimitrios and {Ramakrishnan}, Venkatessh and {Rao}, Ramprasad and {Rawlings}, Mark G. and {Raymond}, Alexander W. and {Rezzolla}, Luciano and {Ricarte}, Angelo and {Ripperda}, Bart and {Roelofs}, Freek and {Rogers}, Alan and {Ros}, Eduardo and {Romero-Ca{\~n}izales}, Cristina and {Roshanineshat}, Arash and {Rottmann}, Helge and {Roy}, Alan L. and {Ruiz}, Ignacio and {Ruszczyk}, Chet and {Rygl}, Kazi L.~J. and {S{\'a}nchez}, Salvador and {S{\'a}nchez-Arg{\"u}elles}, David and {S{\'a}nchez-Portal}, Miguel and {Sasada}, Mahito and {Satapathy}, Kaushik and {Savolainen}, Tuomas and {Schloerb}, F. Peter and {Schonfeld}, Jonathan and {Schuster}, Karl-Friedrich and {Shao}, Lijing and {Shen}, Zhiqiang and {Small}, Des and {Sohn}, Bong Won and {SooHoo}, Jason and {Souccar}, Kamal and {Sun}, He and {Tazaki}, Fumie and {Tetarenko}, Alexandra J. and {Tiede}, Paul and {Tilanus}, Remo P.~J. and {Titus}, Michael and {Torne}, Pablo and {Traianou}, Efthalia and {Trent}, Tyler and {Trippe}, Sascha and {Turk}, Matthew and {van Bemmel}, Ilse and {van Langevelde}, Huib Jan and {van Rossum}, Daniel R. and {Vos}, Jesse and {Wagner}, Jan and {Ward-Thompson}, Derek and {Wardle}, John and {Weintroub}, Jonathan and {Wex}, Norbert and {Wharton}, Robert and {Wielgus}, Maciek and {Wiik}, Kaj and {Witzel}, Gunther and {Wondrak}, Michael F. and {Wu}, Qingwen and {Yamaguchi}, Paul and {Yfantis}, Aristomenis and {Yoon}, Doosoo and {Young}, Andr{\'e} and {Young}, Ken and {Yu}, Wei and {Yuan}, Feng and {Yuan}, Ye-Fei and {Zensus}, J. Anton and {Zhang}, Shuo and {Zhao}, Guang-Yao and {Zhao}, Shan-Shan and {Event Horizon Telescope Collaboration}},
    title = "{Comparison of Polarized Radiative Transfer Codes Used by the EHT Collaboration}",
    journal = {\apj},
    year = 2023,
    month = jun,
    volume = {950},
    number = {1},
    eid = {35},
    pages = {35},
    doi = {10.3847/1538-4357/acc586},
    archivePrefix = {arXiv},
    eprint = {2303.12004},
    primaryClass = {astro-ph.HE},
    adsurl = {https://ui.adsabs.harvard.edu/abs/2023ApJ...950...35P}
}

@INPROCEEDINGS{NOVAs,
    author = {{Mignon-Risse}, R. and {Aimar}, N. and {Varniere}, P. and {Casse}, F. and {Vincent}, F.},
    title = "{A Possible Instability Origin for the Flares in Sagittarius A*: Linking Simulations and Observations}",
    booktitle = {SF2A-2021: Proceedings of the Annual meeting of the French Society of Astronomy and Astrophysics. Eds.: A. Siebert},
    year = 2021,
    editor = {{Siebert}, A. and {Bailli{\'e}}, K. and {Lagadec}, E. and {Lagarde}, N. and {Malzac}, J. and {Marquette}, J. -B. and {N'Diaye}, M. and {Richard}, J. and {Venot}, O.},
    month = dec,
    pages = {113-116},
    adsurl = {https://ui.adsabs.harvard.edu/abs/2021sf2a.conf..113M}
}

@ARTICLE{Porth2021,
    author = {{Porth}, O. and {Mizuno}, Y. and {Younsi}, Z. and {Fromm}, C.~M.},
    title = "{Flares in the Galactic Centre - I. Orbiting flux tubes in magnetically arrested black hole accretion discs}",
    journal = {\mnras},
    year = 2021,
    month = apr,
    volume = {502},
    number = {2},
    pages = {2023-2032},
    doi = {10.1093/mnras/stab163},
    archivePrefix = {arXiv},
    eprint = {2006.03658},
    primaryClass = {astro-ph.HE},
    adsurl = {https://ui.adsabs.harvard.edu/abs/2021MNRAS.502.2023P}
}

@ARTICLE{Anantua2020,
    author = {{Anantua}, Richard and {Ressler}, Sean and {Quataert}, Eliot},
    title = "{On the comparison of AGN with GRMHD simulations: I. Sgr A*}",
    journal = {\mnras},
    year = 2020,
    month = mar,
    volume = {493},
    number = {1},
    pages = {1404-1418},
    doi = {10.1093/mnras/staa318},
    archivePrefix = {arXiv},
    eprint = {2001.11556},
    primaryClass = {astro-ph.HE},
    adsurl = {https://ui.adsabs.harvard.edu/abs/2020MNRAS.493.1404A}
}

@ARTICLE{Chael2019,
    author = {{Chael}, Andrew and {Narayan}, Ramesh and {Johnson}, Michael D.},
    title = "{Two-temperature, Magnetically Arrested Disc simulations of the jet from the supermassive black hole in M87}",
    journal = {\mnras},
    year = 2019,
    month = jun,
    volume = {486},
    number = {2},
    pages = {2873-2895},
    doi = {10.1093/mnras/stz988},
    archivePrefix = {arXiv},
    eprint = {1810.01983},
    primaryClass = {astro-ph.HE},
    adsurl = {https://ui.adsabs.harvard.edu/abs/2019MNRAS.486.2873C}
}

@ARTICLE{Chael2018,
    author = {{Chael}, Andrew and {Rowan}, Michael and {Narayan}, Ramesh and {Johnson}, Michael and {Sironi}, Lorenzo},
    title = "{The role of electron heating physics in images and variability of the Galactic Centre black hole Sagittarius A*}",
    journal = {\mnras},
    year = 2018,
    month = aug,
    volume = {478},
    number = {4},
    pages = {5209-5229},
    doi = {10.1093/mnras/sty1261},
    archivePrefix = {arXiv},
    eprint = {1804.06416},
    primaryClass = {astro-ph.HE},
    adsurl = {https://ui.adsabs.harvard.edu/abs/2018MNRAS.478.5209C}
}

@ARTICLE{Chan2015,
    author = {{Chan}, Chi-kwan and {Psaltis}, Dimitrios and {{\"O}zel}, Feryal and {Medeiros}, Lia and {Marrone}, Daniel and {Sadowski}, Aleksander and {Narayan}, Ramesh},
    title = "{Fast Variability and Millimeter/IR Flares in GRMHD Models of Sgr A* from Strong-field Gravitational Lensing}",
    journal = {\apj},
    year = 2015,
    month = oct,
    volume = {812},
    number = {2},
    eid = {103},
    pages = {103},
    doi = {10.1088/0004-637X/812/2/103},
    archivePrefix = {arXiv},
    eprint = {1505.01500},
    primaryClass = {astro-ph.HE},
    adsurl = {https://ui.adsabs.harvard.edu/abs/2015ApJ...812..103C}
}

@ARTICLE{Davelaar2018,
    author = {{Davelaar}, J. and {Mo{\'s}cibrodzka}, M. and {Bronzwaer}, T. and {Falcke}, H.},
    title = "{General relativistic magnetohydrodynamical {\ensuremath{\kappa}}-jet models for Sagittarius A*}",
    journal = {\aap},
    year = 2018,
    month = apr,
    volume = {612},
    eid = {A34},
    pages = {A34},
    doi = {10.1051/0004-6361/201732025},
    archivePrefix = {arXiv},
    eprint = {1712.02266},
    primaryClass = {astro-ph.HE},
    adsurl = {https://ui.adsabs.harvard.edu/abs/2018A&A...612A..34D}
}

@ARTICLE{Dexter2020,
    author = {{Dexter}, J. and {Tchekhovskoy}, A. and {Jim{\'e}nez-Rosales}, A. and {Ressler}, S.~M. and {Baub{\"o}ck}, M. and {Dallilar}, Y. and {de Zeeuw}, P.~T. and {Eisenhauer}, F. and {von Fellenberg}, S. and {Gao}, F. and {Genzel}, R. and {Gillessen}, S. and {Habibi}, M. and {Ott}, T. and {Stadler}, J. and {Straub}, O. and {Widmann}, F.},
    title = "{Sgr A* near-infrared flares from reconnection events in a magnetically arrested disc}",
    journal = {\mnras},
    year = 2020,
    month = oct,
    volume = {497},
    number = {4},
    pages = {4999-5007},
    doi = {10.1093/mnras/staa2288},
    archivePrefix = {arXiv},
    eprint = {2006.03657},
    primaryClass = {astro-ph.HE},
    adsurl = {https://ui.adsabs.harvard.edu/abs/2020MNRAS.497.4999D}
}

@ARTICLE{EHT2019,
    author = {{Event Horizon Telescope Collaboration} and {Akiyama}, Kazunori and {Alberdi}, Antxon and {Alef}, Walter and {Asada}, Keiichi and {Azulay}, Rebecca and {Baczko}, Anne-Kathrin and {Ball}, David and {Balokovi{\'c}}, Mislav and {Barrett}, John and {Bintley}, Dan and {Blackburn}, Lindy and {Boland}, Wilfred and {Bouman}, Katherine L. and {Bower}, Geoffrey C. and {Bremer}, Michael and {Brinkerink}, Christiaan D. and {Brissenden}, Roger and {Britzen}, Silke and {Broderick}, Avery E. and {Broguiere}, Dominique and {Bronzwaer}, Thomas and {Byun}, Do-Young and {Carlstrom}, John E. and {Chael}, Andrew and {Chan}, Chi-kwan and {Chatterjee}, Shami and {Chatterjee}, Koushik and {Chen}, Ming-Tang and {Chen}, Yongjun and {Cho}, Ilje and {Christian}, Pierre and {Conway}, John E. and {Cordes}, James M. and {Crew}, Geoffrey B. and {Cui}, Yuzhu and {Davelaar}, Jordy and {De Laurentis}, Mariafelicia and {Deane}, Roger and {Dempsey}, Jessica and {Desvignes}, Gregory and {Dexter}, Jason and {Doeleman}, Sheperd S. and {Eatough}, Ralph P. and {Falcke}, Heino and {Fish}, Vincent L. and {Fomalont}, Ed and {Fraga-Encinas}, Raquel and {Friberg}, Per and {Fromm}, Christian M. and {G{\'o}mez}, Jos{\'e} L. and {Galison}, Peter and {Gammie}, Charles F. and {Garc{\'\i}a}, Roberto and {Gentaz}, Olivier and {Georgiev}, Boris and {Goddi}, Ciriaco and {Gold}, Roman and {Gu}, Minfeng and {Gurwell}, Mark and {Hada}, Kazuhiro and {Hecht}, Michael H. and {Hesper}, Ronald and {Ho}, Luis C. and {Ho}, Paul and {Honma}, Mareki and {Huang}, Chih-Wei L. and {Huang}, Lei and {Hughes}, David H. and {Ikeda}, Shiro and {Inoue}, Makoto and {Issaoun}, Sara and {James}, David J. and {Jannuzi}, Buell T. and {Janssen}, Michael and {Jeter}, Britton and {Jiang}, Wu and {Johnson}, Michael D. and {Jorstad}, Svetlana and {Jung}, Taehyun and {Karami}, Mansour and {Karuppusamy}, Ramesh and {Kawashima}, Tomohisa and {Keating}, Garrett K. and {Kettenis}, Mark and {Kim}, Jae-Young and {Kim}, Junhan and {Kim}, Jongsoo and {Kino}, Motoki and {Koay}, Jun Yi and {Koch}, Patrick M. and {Koyama}, Shoko and {Kramer}, Michael and {Kramer}, Carsten and {Krichbaum}, Thomas P. and {Kuo}, Cheng-Yu and {Lauer}, Tod R. and {Lee}, Sang-Sung and {Li}, Yan-Rong and {Li}, Zhiyuan and {Lindqvist}, Michael and {Liu}, Kuo and {Liuzzo}, Elisabetta and {Lo}, Wen-Ping and {Lobanov}, Andrei P. and {Loinard}, Laurent and {Lonsdale}, Colin and {Lu}, Ru-Sen and {MacDonald}, Nicholas R. and {Mao}, Jirong and {Markoff}, Sera and {Marrone}, Daniel P. and {Marscher}, Alan P. and {Mart{\'\i}-Vidal}, Iv{\'a}n and {Matsushita}, Satoki and {Matthews}, Lynn D. and {Medeiros}, Lia and {Menten}, Karl M. and {Mizuno}, Yosuke and {Mizuno}, Izumi and {Moran}, James M. and {Moriyama}, Kotaro and {Moscibrodzka}, Monika and {Mul{\ensuremath{\ddot{}}}ler}, Cornelia and {Nagai}, Hiroshi and {Nagar}, Neil M. and {Nakamura}, Masanori and {Narayan}, Ramesh and {Narayanan}, Gopal and {Natarajan}, Iniyan and {Neri}, Roberto and {Ni}, Chunchong and {Noutsos}, Aristeidis and {Okino}, Hiroki and {Olivares}, H{\'e}ctor and {Oyama}, Tomoaki and {{\"O}zel}, Feryal and {Palumbo}, Daniel C.~M. and {Patel}, Nimesh and {Pen}, Ue-Li and {Pesce}, Dominic W. and {Pi{\'e}tu}, Vincent and {Plambeck}, Richard and {PopStefanija}, Aleksandar and {Porth}, Oliver and {Prather}, Ben and {Preciado-L{\'o}pez}, Jorge A. and {Psaltis}, Dimitrios and {Pu}, Hung-Yi and {Ramakrishnan}, Venkatessh and {Rao}, Ramprasad and {Rawlings}, Mark G. and {Raymond}, Alexander W. and {Rezzolla}, Luciano and {Ripperda}, Bart and {Roelofs}, Freek and {Rogers}, Alan and {Ros}, Eduardo and {Rose}, Mel and {Roshanineshat}, Arash and {Rottmann}, Helge and {Roy}, Alan L. and {Ruszczyk}, Chet and {Ryan}, Benjamin R. and {Rygl}, Kazi L.~J. and {S{\'a}nchez}, Salvador and {S{\'a}nchez-Arguelles}, David and {Sasada}, Mahito and {Savolainen}, Tuomas and {Schloerb}, F. Peter and {Schuster}, Karl-Friedrich and {Shao}, Lijing and {Shen}, Zhiqiang and {Small}, Des and {Sohn}, Bong Won and {SooHoo}, Jason and {Tazaki}, Fumie and {Tiede}, Paul and {Tilanus}, Remo P.~J. and {Titus}, Michael and {Toma}, Kenji and {Torne}, Pablo and {Trent}, Tyler and {Trippe}, Sascha and {Tsuda}, Shuichiro and {van Bemmel}, Ilse and {van Langevelde}, Huib Jan and {van Rossum}, Daniel R. and {Wagner}, Jan and {Wardle}, John and {Weintroub}, Jonathan and {Wex}, Norbert and {Wharton}, Robert and {Wielgus}, Maciek and {Wong}, George N. and {Wu}, Qingwen and {Young}, Andr{\'e} and {Young}, Ken and {Younsi}, Ziri and {Yuan}, Feng and {Yuan}, Ye-Fei and {Zensus}, J. Anton and {Zhao}, Guangyao and {Zhao}, Shan-Shan and {Zhu}, Ziyan and {Anczarski}, Jadyn and {Baganoff}, Frederick K. and {Eckart}, Andreas and {Farah}, Joseph R. and {Haggard}, Daryl and {Meyer-Zhao}, Zheng and {Michalik}, Daniel and {Nadolski}, Andrew and {Neilsen}, Joseph and {Nishioka}, Hiroaki and {Nowak}, Michael A. and {Pradel}, Nicolas and {Primiani}, Rurik A. and {Souccar}, Kamal and {Vertatschitsch}, Laura and {Yamaguchi}, Paul and {Zhang}, Shuo},
title = "{First M87 Event Horizon Telescope Results. V. Physical Origin of the Asymmetric Ring}",
journal = {\apjl},
year = 2019,
month = apr,
volume = {875},
number = {1},
eid = {L5},
pages = {L5},
doi = {10.3847/2041-8213/ab0f43},
archivePrefix = {arXiv},
eprint = {1906.11242},
primaryClass = {astro-ph.GA},
adsurl = {https://ui.adsabs.harvard.edu/abs/2019ApJ...875L...5E}
}

@ARTICLE{Gravity2020,
    author = {{GRAVITY Collaboration} and {Baub{\"o}ck}, M. and {Dexter}, J. and {Abuter}, R. and {Amorim}, A. and {Berger}, J.~P. and {Bonnet}, H. and {Brandner}, W. and {Cl{\'e}net}, Y. and {Coud{\'e} Du Foresto}, V. and {de Zeeuw}, P.~T. and {Duvert}, G. and {Eckart}, A. and {Eisenhauer}, F. and {F{\"o}rster Schreiber}, N.~M. and {Gao}, F. and {Garcia}, P. and {Gendron}, E. and {Genzel}, R. and {Gerhard}, O. and {Gillessen}, S. and {Habibi}, M. and {Haubois}, X. and {Henning}, T. and {Hippler}, S. and {Horrobin}, M. and {Jim{\'e}nez-Rosales}, A. and {Jocou}, L. and {Kervella}, P. and {Lacour}, S. and {Lapeyr{\`e}re}, V. and {Le Bouquin}, J. -B. and {L{\'e}na}, P. and {Ott}, T. and {Paumard}, T. and {Perraut}, K. and {Perrin}, G. and {Pfuhl}, O. and {Rabien}, S. and {Rodriguez Coira}, G. and {Rousset}, G. and {Scheithauer}, S. and {Stadler}, J. and {Sternberg}, A. and {Straub}, O. and {Straubmeier}, C. and {Sturm}, E. and {Tacconi}, L.~J. and {Vincent}, F. and {von Fellenberg}, S. and {Waisberg}, I. and {Widmann}, F. and {Wieprecht}, E. and {Wiezorrek}, E. and {Woillez}, J. and {Yazici}, S.},
    title = "{Modeling the orbital motion of Sgr A*'s near-infrared flares}",
    journal = {\aap},
    year = 2020,
    month = mar,
    volume = {635},
    eid = {A143},
    pages = {A143},
    doi = {10.1051/0004-6361/201937233},
    archivePrefix = {arXiv},
    eprint = {2002.08374},
    primaryClass = {astro-ph.HE},
    adsurl = {https://ui.adsabs.harvard.edu/abs/2020A&A...635A.143G}
}

@ARTICLE{Gralla2020,
    author = {{Gralla}, Samuel E. and {Lupsasca}, Alexandru and {Marrone}, Daniel P.},
    title = "{The shape of the black hole photon ring: A precise test of strong-field general relativity}",
    journal = {\prd},
    year = 2020,
    month = dec,
    volume = {102},
    number = {12},
    eid = {124004},
    pages = {124004},
    doi = {10.1103/PhysRevD.102.124004},
    archivePrefix = {arXiv},
    eprint = {2008.03879},
    primaryClass = {gr-qc},
    adsurl = {https://ui.adsabs.harvard.edu/abs/2020PhRvD.102l4004G}
}

@ARTICLE{Gralla2019,
    author = {{Gralla}, Samuel E. and {Holz}, Daniel E. and {Wald}, Robert M.},
    title = "{Black hole shadows, photon rings, and lensing rings}",
    journal = {\prd},
    year = 2019,
    month = jul,
    volume = {100},
    number = {2},
    eid = {024018},
    pages = {024018},
    doi = {10.1103/PhysRevD.100.024018},
    archivePrefix = {arXiv},
    eprint = {1906.00873},
    primaryClass = {astro-ph.HE},
    adsurl = {https://ui.adsabs.harvard.edu/abs/2019PhRvD.100b4018G}
}

@ARTICLE{Gralla2018,
    author = {{Gralla}, Samuel E. and {Lupsasca}, Alexandru and {Strominger}, Andrew},
    title = "{Observational signature of high spin at the Event Horizon Telescope}",
    journal = {\mnras},
    year = 2018,
    month = apr,
    volume = {475},
    number = {3},
    pages = {3829-3853},
    doi = {10.1093/mnras/sty039},
    archivePrefix = {arXiv},
    eprint = {1710.11112},
    primaryClass = {astro-ph.HE},
    adsurl = {https://ui.adsabs.harvard.edu/abs/2018MNRAS.475.3829G}
}

@ARTICLE{CardenasAvendano2023,
    author = {{C{\'a}rdenas-Avenda{\~n}o}, Alejandro and {Lupsasca}, Alexandru},
    title = "{Prediction for the interferometric shape of the first black hole photon ring}",
    journal = {\prd},
    year = 2023,
    month = sep,
    volume = {108},
    number = {6},
    eid = {064043},
    pages = {064043},
    doi = {10.1103/PhysRevD.108.064043},
    archivePrefix = {arXiv},
    eprint = {2305.12956},
    primaryClass = {gr-qc},
    adsurl = {https://ui.adsabs.harvard.edu/abs/2023PhRvD.108f4043C}
}

@ARTICLE{AART,
    author = {{C{\'a}rdenas-Avenda{\~n}o}, Alejandro and {Lupsasca}, Alexandru and {Zhu}, Hengrui},
    title = "{Adaptive analytical ray tracing of black hole photon rings}",
    journal = {\prd},
    year = 2023,
    month = feb,
    volume = {107},
    number = {4},
    eid = {043030},
    pages = {043030},
    doi = {10.1103/PhysRevD.107.043030},
    archivePrefix = {arXiv},
    eprint = {2211.07469},
    primaryClass = {gr-qc},
    adsurl = {https://ui.adsabs.harvard.edu/abs/2023PhRvD.107d3030C}
}

@ARTICLE{Vincent2019,
    author = {{Vincent}, F.~H. and {Abramowicz}, M.~A. and {Zdziarski}, A.~A. and {Wielgus}, M. and {Paumard}, T. and {Perrin}, G. and {Straub}, O.},
    title = "{Multi-wavelength torus-jet model for Sagittarius A*}",
    journal = {\aap},
    year = 2019,
    month = apr,
    volume = {624},
    eid = {A52},
    pages = {A52},
    doi = {10.1051/0004-6361/201834946},
    archivePrefix = {arXiv},
    eprint = {1902.01175},
    primaryClass = {astro-ph.HE},
    adsurl = {https://ui.adsabs.harvard.edu/abs/2019A&A...624A..52V}
}

@ARTICLE{White2022,
    author = {{White}, Christopher J.},
    title = "{Blacklight: A General-relativistic Ray-tracing and Analysis Tool}",
    journal = {\apjs},
    year = 2022,
    month = sep,
    volume = {262},
    number = {1},
    eid = {28},
    pages = {28},
    doi = {10.3847/1538-4365/ac77ef},
    archivePrefix = {arXiv},
    eprint = {2203.15963},
    primaryClass = {astro-ph.HE},
    adsurl = {https://ui.adsabs.harvard.edu/abs/2022ApJS..262...28W}
}

@ARTICLE{Chan2018,
    author = {{Chan}, Chi-kwan and {Medeiros}, Lia and {{\"O}zel}, Feryal and {Psaltis}, Dimitrios},
    title = "{GRay2: A General Purpose Geodesic Integrator for Kerr Spacetimes}",
    journal = {\apj},
    year = 2018,
    month = nov,
    volume = {867},
    number = {1},
    eid = {59},
    pages = {59},
    doi = {10.3847/1538-4357/aadfe5},
    archivePrefix = {arXiv},
    eprint = {1706.07062},
    primaryClass = {astro-ph.HE},
    adsurl = {https://ui.adsabs.harvard.edu/abs/2018ApJ...867...59C}
}

@ARTICLE{Gelles2021,
    author = {{Gelles}, Zachary and {Himwich}, Elizabeth and {Johnson}, Michael D. and {Palumbo}, Daniel C.~M.},
    title = "{Polarized image of equatorial emission in the Kerr geometry}",
    journal = {\prd},
    year = 2021,
    month = aug,
    volume = {104},
    number = {4},
    eid = {044060},
    pages = {044060},
    doi = {10.1103/PhysRevD.104.044060},
    archivePrefix = {arXiv},
    eprint = {2105.09440},
    primaryClass = {gr-qc},
    adsurl = {https://ui.adsabs.harvard.edu/abs/2021PhRvD.104d4060G}
}

\end{document}